%% file: aa60445-26.tex
\documentclass[longauth]{aa}

\usepackage{graphicx}
\usepackage{natbib}
\usepackage{scalerel}
\usepackage{comment}
\usepackage{ulem}
\usepackage[table]{xcolor}
\usepackage{subcaption}
\usepackage{cancel}
\usepackage{txfonts}
\usepackage[pdfencoding=auto,psdextra]{hyperref}
\hypersetup{
    colorlinks=true,
    linkcolor=blue,
    filecolor=magenta, 
    urlcolor=blue,
    citecolor=blue}

\usepackage{lastpage}
\makeatletter
\renewcommand*\aa@pageof{, page \thepage{} of \pageref*{LastPage}}
\makeatother

\usepackage[utf8]{inputenc}

\usepackage[switch, modulo]{lineno}
              
\renewcommand{\linenumbers}[0]{}

\usepackage{euclid}

\defcitealias{Castander2024}{CF25}
 
\newif\iffirstcite
\firstcitetrue

\newcommand{\firstciteCF}{%
  \iffirstcite
    \hyperlink{cite.Castander2024}{Euclid Collaboration: Castander et al.~2025} (hereafter CF25)%
    \firstcitefalse
  \else
    \citetalias{Castander2024}%
  \fi
}

\RequirePackage{etex}
\begin{document}

%
%
\title{\Euclid}
  \subtitle{VIII. Populating a dark universe using SciPIC}

\include{authors} 

%
\abstract{High-fidelity galaxy mocks are crucial for validating analysis pipelines and for cosmological inference. In this context, the Science Pipeline at PIC (\texttt{SciPIC}) is a pipeline specifically designed for the fast generation of synthetic galaxy catalogues from the halo properties identified in cosmological simulations. \texttt{SciPIC} delivers galaxy catalogues that aim to reproduce the observed luminosity function and clustering above a given flux detection limit over a wide redshift range. In this work, we introduce \texttt{SciPICal}, an automated pipeline that calibrates the parameters that set the main mock galaxy properties, namely number density, luminosities, colours, and positions. The pipeline was applied to the Euclid Flagship 2 Wide and Deep halo catalogues, specifically built to support the \textit{Euclid} wide and deep surveys. Compared to the recently released Flagship 2 Wide mock, our calibrated version improves the clustering predictions by approximately 50\% based on chi-squared values. Furthermore, we produced the Euclid Deep mock catalogue, which reaches up to $z = 10$ by populating a light cone and a complementary snapshot at $z = 0$. We validated these catalogues using measurements from spectroscopic and photometric galaxy surveys, as well as with results from a hydrodynamical simulation. The obtained good agreement (within $15\%$ for most of the samples) in the clustering predictions across the different galaxy samples considered, validates our calibration strategy and demonstrates the strong predictive power of the generated mocks. This pipeline will allow us to improve the methodology applied for assigning the galaxy properties and will ensure that the galaxy mocks remain up-to-date by incorporating constraints from upcoming observational data into the calibration procedure.}

%
%
%
%
   \titlerunning{\Euclid\/: Populating a dark universe}
   \authorrunning{Euclid Collaboration: E. J. Gonzalez et al.}
   
   \maketitle
%
%
%
%
   
\section{Introduction}
\label{sc:Intro}

The formation and evolution of the large-scale structure of the Universe are well described within the standard cosmological model. However, the nature of its dark components, dark matter and dark energy, remains unknown. One strategy for studying these components involves improving the precision of cosmological parameter constraints by combining a diverse set of complementary cosmological probes. Large-scale galaxy surveys are then designed following this approach. To achieve this goal, a new generation of photometric and spectroscopic surveys, such as LSST at the NSF \textit{Vera Rubin} Observatory \citep{LSST2009,Ivezic2019}, DESI \citep{DESI2016,DESI2022}, the \textit{Nancy Grace Roman} Space Telescope \citep{Spergel2015}, and \Euclid \citep{Laureijs2011, Mellier2025}, will map the galaxy distribution out to high redshifts, providing unprecedented galaxy number densities across vast cosmological volumes. 
The large amount of data and the exquisite imaging quality expected from these surveys will enable measurements of data vectors with a significantly lower statistical uncertainty than previous determinations, positioning cosmology in a new era of high precision. In this new era, statistical uncertainties are subdominant, while systematic effects and modelling assumptions determine the ultimate precision of the measurements. We can identify the systematic effects related to the way in which we model galaxy formation and its interplay with the dark content as one of the main drawbacks in current cosmological analyses.

In this context, cosmological galaxy simulations emerge as a powerful tool, both in the planning and the analysis of galaxy surveys. A strategy for understanding the connection between galaxies and the dark matter content emerges from the use of hydrodynamical simulations that attempt to directly model relevant physical processes that shape galaxy formation and evolution \citep[e.g.][]{Vogelsberger2020,Villaescusa2021,Nelson2024}. 
In this way, the simulated data provide information on the dark matter content and the galaxies that would be observed under certain conditions \citep{Bose2019,Perez2024}. 
However, producing these simulations in cosmological volumes is numerically expensive. 
A less computationally expensive approach is provided by semi-analytic models that combine numerical dark-matter-only simulations and analytic models for the prescription of baryonic physics. 
These models rely on simplified flow equations to describe the key physical processes of galaxy formation and evolution. Nevertheless, both semi-analytic and hydrodynamic simulations depend on approximations to describe the physics below their respective resolution scales. 
Different yet equally reasonable sub-resolution prescriptions can therefore lead to significant variations in the predicted galaxy properties \citep{Kim2016,Lu2014}. An alternative and more computationally efficient approach for generating synthetic galaxy catalogues in cosmological volumes involves populating haloes identified in dark matter-only simulations with galaxies in a way that reproduces a range of observational constraints. 
This empirical approach is applied by setting the halo occupation distribution (HOD), which statistically describes the number of galaxies residing in haloes of a given mass \citep{Zheng2007}. 
Other properties such as luminosity or stellar mass can also be assigned using the abundance matching (AM) technique \citep{Peacock2000,Kravtsov2004,Vale2004}. 
In this approach, galaxies are ranked and matched to haloes by mass, assuming that galaxies with a given observed abundance occupy dark matter haloes with the same spatial number density. 
Given that they enable a fast generation of mocks, they can be used to constrain cosmological information by populating a large number of simulations with different cosmologies and comparing them with the observed properties of spectroscopic galaxy samples \citep[e.g.][]{Reid2014,Zhai2019,Yuan2022}. 
More sophisticated versions of this approach have been developed to offer a more accurate description of the observed constraints by incorporating information on subhalo and halo evolution history \citep{Hearin2016,Contreras2019,Behroozi2019}, thereby enabling higher precision in determining cosmological parameters \citep{Mahony2025}. 

Science Pipeline at PIC \citep[\texttt{SciPIC},][]{Carretero:17} is a galaxy simulation pipeline developed under a collaborative effort with multiple contributors. 
The main goal of \texttt{SciPIC} is to create synthetic galaxy catalogues that accurately mimic the observed properties of galaxy populations, which can be selected with different selection criteria. The pipeline assigns a large number of galaxy parameters, including fluxes, lensing, and morphological properties. To manage this complexity, its design follows a modular scheme in which each module assigns properties that depend on those already assigned by preceding modules. This enables the efficient generation of comprehensive galaxy mocks from any set of haloes identified in a dark matter simulation. In particular, \texttt{SciPIC} was applied to generate the MICE mock catalogue \citep{Carretero:15} and to populate the Euclid Flagship 2 Wide (FS2-Wide) halo catalogue. 
The FS2-Wide mock presented in \firstciteCF\,\,is the official simulation of the Euclid Consortium and was designed to support the Euclid Wide Survey \citep[EWS,][]{Scaramella2022,Mellier2025}. 
It constitutes the largest virtual galaxy catalogue ever built to this day and is publicly available through the \texttt{CosmoHub} platform \citep{Tallada2024,Carretero:17}.\footnote{\url{https://cosmohub.pic.es/home}} 
The parameters involved in generating the FS2-Wide mock were set to reproduce clustering at low redshifts \citep{Carretero:15}. 
This simulation has been widely used to produce, test, and validate analysis pipelines with the aim of applying them to observational data with different applications. Its applications include, for example, predicting redshift distributions to generate synthetic data vectors \citep[e.g.][]{Cloe}, determining photometric redshifts \citep[e.g.][]{Pocino2021,Euclid_Photoz_SOM,Euclid_Photoz_Optimiz}, and producing pixel-level simulations for the development of image reduction pipelines \citep{Serrano2024}. 
Moreover, the resulting HOD of H\,$\alpha$ emitters from this simulation was used to generate simulated datasets, which in turn can be used to assess the impact of sample variance on cosmological inference \citep{Monaco2025}. 
 
While the Flagship 2 Wide mock has been successfully used for various purposes, its clustering predictions for certain galaxy samples still require improvement \citep{Hoffmann2026}. Among the various modules in \texttt{SciPIC}, the initial one assigns number densities, colours, and positions -- the properties with the greatest impact on the resulting galaxy clustering. A dedicated pipeline capable of fully exploring the parameter space involved in this crucial step for mock generation is therefore necessary to test the underlying methodology. This is essential for obtaining clustering predictions in better agreement with observations and meeting the \textit{Euclid} requirements for cosmological analyses. Moreover, an automated procedure for calibrating the galaxy properties assigned by \texttt{SciPIC} is mandatory to keep the mock catalogues updated as new observational data become available.

In this work, we present the \texttt{SciPIC} calibration pipeline (\texttt{SciPICal}), developed to optimise the parameters of the initial module of \texttt{SciPIC} using observed clustering constraints at low redshift, $z < 0.1$. We applied this pipeline to the FS2-Wide and the Euclid Flagship 2 Deep (FS2-Deep) halo catalogues. The latter has a mass resolution one order of magnitude higher than FS2-Wide and is specifically designed to support the science analysis pipelines for the Euclid Deep Survey (EDS). The EDS complements the EWS by reaching approximately two magnitudes deeper across a total area of $53\,\mathrm{deg}^2$. Using the optimised parameters from \texttt{SciPICal}, we ran \texttt{SciPIC} to produce three galaxy mock catalogues for the \textit{Euclid} survey: an upgraded version of the existing FS2-Wide catalogue, the first release of the FS2-Deep light cone reaching $z=10$, and a complementary snapshot at $z=0$. Moreover, we validated the calibration strategy and the resulting mocks by comparing the two-point correlations with measurements from spectroscopic and photometric galaxy surveys and from a hydrodynamical simulation. The outline of this paper is as follows. First, we describe the dark-matter-halo catalogues, FS2-Wide and FS2-Deep, in Sect. \ref{sec:data}. 
We then present the mock production and calibration strategy in Sect. \ref{sec:scipical}. 
We present the resulting optimised parameters and the validated calibration in Sect. \ref{sec:calFS}. 
Then, we discuss validating the mocks generated with the calibrated parameters using different sets of observational and simulated constraints over a wide redshift range in Sect. \ref{sec:validation}. 
Finally, we summarise and discuss our results in Sect. \ref{sec:conclusion}.

\section{Dark matter halo catalogues}
\label{sec:data} 

The calibration pipeline was built taking as input the halo catalogues from the FS2-Wide and FS2-Deep simulations. 
These catalogues were specifically designed for the \textit{Euclid} Wide and Deep surveys, respectively. 
We describe the details of these catalogues below.

\subsection{FS2-Wide}
\label{subsec:halo_catalogue_wide}

 The FS2-Wide dark matter-only simulation has a box side of $3600\,h^{-1}\,\mathrm{Mpc}$ with a particle mass resolution of $10^9\,h^{-1}\,M_\odot$. 
It was generated using the \texttt{PKDGRAV3} $N$-body code \citep{Potter2016}. 
The simulation was run with a softening length of $4.5\,h^{-1}\,\mathrm{kpc}$, using the \Euclid reference cosmology, with matter density $\Omega_{\mathrm{m}} = 0.319$, baryon density $\Omega_{\mathrm{b}} = 0.049$, and dark energy density $\Omega_\Lambda = 0.681 - \Omega_{\mathrm{r}} - \Omega_\nu$, where $\Omega_{\mathrm{r}} = 5.509 \times 10^{-5}$ and $\Omega_\nu = 1.40343 \times 10^{-3}$ are the radiation and neutrino density, respectively, $h = 0.67$, and the primordial power spectrum amplitude $A_{\mathrm{s}} = 2.1 \times 10^{-9}$. 
The main data product consists of a full-sky particle light cone up to $z=3$ that was produced on the fly while running the simulation. 
Haloes were identified using the \texttt{ROCKSTAR} algorithm \citep{Behroozi2013} with a minimum of ten particles ($\approx\,10^{10}\,h^{-1}\,M_\odot$), although some haloes ended up with lower mass after discarding unbound particles. 
The halo catalogue contains 126 billion main haloes with several properties provided by the \texttt{ROCKSTAR} halo identifier. 
For a more detailed description and characterisation of this simulation, see Sect. 4 of CF25.

Since the calibration process is performed at low redshifts ($z < 0.1$; see Sect. \ref{subsec:cal}), we selected a subset of haloes from the light cone. Specifically, we defined a cubic volume of side length $300\,\hMpc$ with the observer placed at one corner and selected all haloes whose comoving coordinates are each less than $300\,\hMpc$. In total, we used 8\,075\,637 haloes to compute the calibration. We adopted the virial mass computed by \texttt{ROCKSTAR} using the bound particles as the reference halo mass, $M_{\mathrm{h}}$.

\subsection{FS2-Deep}
\label{subsec:halo_catalogue_deep}

The FS2-Deep dark matter-only simulation consists of a box with a side length of $1000\,\hMpc$ and a particle mass resolution of $10^8\,h^{-1}\,M_\odot$. 
It was also generated using the \texttt{PKDGRAV3} $N$-body code. 
The simulation was run with the same softening length and cosmology as FS2-Wide. 
The main data product consists of two light cones pointing in opposite directions on the sky, each covering $50\,\mathrm{deg}^2$ up to $z=10$. 
They were produced on the fly during the simulation generation. 
Similar to FS2-Wide, haloes were identified using the \texttt{ROCKSTAR} algorithm with a minimum of ten particles ($\approx\,1.0\times 10^{9}\,h^{-1}\,M_\odot$), although again some haloes ended up with lower mass after discarding unbound particles. 
Besides the light cones, halo catalogues for 100 different redshift snapshots were stored.

\begin{figure}
    \captionsetup{font={small},labelfont=bf}
    \centering
    \includegraphics[scale=0.85]{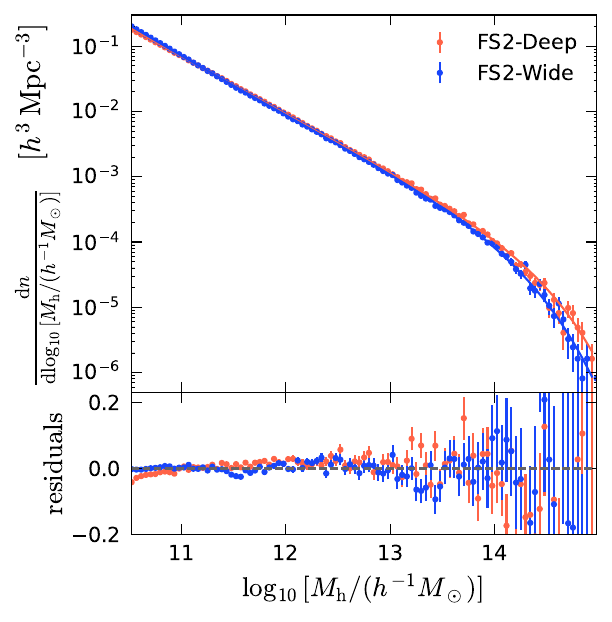}
    \caption{HMF, defined as the number density of haloes in logarithmic mass bins, for the haloes used for calibration from the FS2-Deep and FS2-Wide halo catalogues (red and blue, respectively). For FS2-Deep, haloes were selected from the snapshot at $z=0$ within a sub-box of a side length of $300\,h^{-1}\,\rm Mpc$. For FS2-Wide we selected the haloes from the light cone within a cubic volume of a side length of $300\,h^{-1}\,\rm Mpc$ with the observed placed at one corner. The solid lines indicate the Schechter fits. \textit{Bottom panel}: Dimensionless residuals, defined as the difference between the measurement and the Schechter fit relative to the fit.}
    \label{fig:HMF}
    
\end{figure}

For the calibration, we selected a subset of haloes within a box of side length $300\,\hMpc$ in the snapshot at $z=0$. 
This subset includes more than 63 million haloes. 
To reduce the data volume and speed up the calibration process, we further considered a halo mass cut of $\logten[M_{\mathrm{h}}/(h^{-1}\,M_\odot)] > 10.5$. 
This cut was applied to exclude very low-mass haloes, taking into account the minimum luminosity threshold of the galaxies used as reference for the calibration (see Sect. \ref{subsec:cal}). 
In total, we used 2\,539\,365 haloes to perform the calibration. 
Following the same approach as for FS2-Wide, we used the halo virial mass computed by \texttt{ROCKSTAR} using the bound particles for the calibration. 
In Fig. \ref{fig:HMF}, we show the halo mass functions (HMFs) of the subset of haloes considered for the FS2-Wide and FS2-Deep halo catalogues, together with the Schechter fit model \citep{Schechter1976} used to perform the AM (see Sect. \ref{subsubsec:AM}). 
Both HMFs agree well at the low-mass end but show some discrepancy at the high-mass end, where FS2-Deep presents a higher abundance. 
This behaviour is expected as the FS2-Deep HMF is for a snapshot at $z=0$, while the FS2-Wide HMF is measured in the light cone and covers a range of nearby redshifts with a median redshift around $0.1$.

\section{SciPICal}
\label{sec:scipical}

The \texttt{SciPIC} pipeline is composed of multiple modules that perform different steps in the assignment of galaxy properties. 
The pipeline starts with a halo catalogue as input, with a set of required halo properties: the halo mass, the comoving coordinates that describe position and velocity, the halo shape represented by an ellipsoid, and a concentration parameter that characterises the inner slope of the radial density profile. 
The modules then compute various classes of galaxy properties, such as lensing estimators, fluxes, and morphological parameters. 
For a more detailed description of these modules, we refer the reader to CF25.

The calibration pipeline developed in this work aims to obtain the parameters involved in the assignment of the main properties modelled: number of galaxies in each halo, positions, luminosities, and colours. It builds upon and extends the framework introduced by \citet{Tutusaus2025}. The main improvements include updated prescriptions for assigning colours and positions, as well as the flexibility to process both light cones and snapshots as input halo catalogues. In addition, we implemented an improved calibration strategy that yields posterior probability distributions (PDFs) for the optimised parameters. This allows us to examine correlations between them and ensures more robust results.

In this section we describe in detail how these properties were assigned and the calibration strategy applied to constrain the parameters involved in the assignment process. We first provide a more detailed description of the initial part of \texttt{SciPIC}, which we aimed to calibrate, highlighting the parameters optimised in the procedure. Then, we describe the calibration process.

\subsection{Mock generation}

\subsubsection{Assigning the number of galaxies}
\label{subsec:hod}

The first step of this module involves assigning the number of galaxies that each halo will host, for which we use a HOD model. Since we aim to reproduce the whole set of galaxy populations observed for a flux-limited survey, we populate all the haloes by placing a central galaxy in each of them. With this approach, we reach the observed flux limit set by the halo catalogue resolution. For the FS2-Wide and FS2-Deep halo catalogues, we set $\logten[M_\mathrm{min}/(h^{-1} \, M_\odot)] = 9.865$ and $\logten[M_\mathrm{min}/(h^{-1} \, M_\odot)] = 8.8$, respectively, which correspond to the minimum mass of each halo catalogue after applying completeness and discreteness corrections (see Sect.\,4.2 in \citetalias{Castander2024}).
The number of satellites is given by a Poisson realisation with mean $\langle N_\mathrm{sats} \rangle$ that depends on the halo mass,
\begin{equation}
    \langle N_\mathrm{sats} \rangle = \left( \frac{M_\mathrm{h}}{f \, M_\mathrm{min}} \right)^\alpha \, ,
\end{equation}
where the parameters $f$ and $\alpha$ are optimised in the calibration process. In general, the HOD of central galaxies and $f \, M_\mathrm{min}$ depend on the target galaxy population for which the mock is constructed. For example, for star-forming galaxies the HOD of centrals can be modelled with a Gaussian-like distribution, while for luminous red galaxies this can be modelled considering a smoothed step function \citep{Zheng2005,Avila2020}. In our case the full diversity of the different populations  is introduced a posteriori when assigning the luminosities and colours, and later the spectral energy distribution (SED).

\subsubsection{Central galaxy luminosity assignment}
\label{subsubsec:AM}
Once the number of galaxies is assigned, we compute each galaxy's luminosity. We assign central galaxy luminosities by applying an AM approach. To this aim, and following \citet{Carretero:15} and \citetalias{Castander2024}, we computed the cumulative galaxy number density function defined as
\begin{equation}
\label{eq:cum_gal}
    n_\mathrm{gal}(> M_\mathrm{min}) = \int^{\infty}_{M_\mathrm{min}} n_\text{h}(M') \left[ 1 + \left( \frac{M'}{f \, M_\mathrm{min}}\right)^\alpha \right] \diff M' \, ,
\end{equation}
where $n_\text{h}(M)$ is given by the HMF. We computed an AM relation between halo mass and luminosity comparing Eq.~(\ref{eq:cum_gal}) to the observed luminosity function (LF), $\Phi_\mathrm{obs}(L)$, in its cumulative form. We took the same LF as in \citetalias{Castander2024}, which is based on the Sloan Digital Sky Survey \citep[SDSS,][]{Johnston2019} local LF \citep{Blanton2003} and the Cosmic Assembly Near-Infrared Deep Extragalactic Legacy Survey \citep[CANDELS,][]{Koekemoer2011,Grogin2012} LF up to around redshift $2$ \citep{Dahlen2005}.

We also consider scatter in the LF to account for the stochastic character of the galaxy-formation process. This scatter is also necessary to reduce the predicted clustering for the most luminous galaxies. The observed LF, $\Phi_\mathrm{obs}$, can be expressed as the result of the convolution of the unscattered LF, $\Phi_\mathrm{unscat}$, with a convolution kernel
\begin{equation}
\begin{aligned}
    \Phi_\mathrm{obs}\{\logten[L/(h^{-2}\,L_\odot)]\} = & \int^{\infty}_{0} \Phi_\mathrm{unscat}\{\logten[L'/(h^{-2}\,L_\odot)]\} \\
    & \times \mathcal{N}[\logten(L/L');\sigma_L]\, \\
    & \times \diff \logten[L'/(h^{-2}\,L_\odot)] \, ,
\end{aligned}
\end{equation}
where $\mathcal{N}$ is the convolution kernel that is implemented as a Gaussian scatter in the logarithm of the luminosity, $\sigma_L$.\footnote{Note that this parameter represents the standard deviation of the introduced scatter in logarithmic luminosity space.} This last parameter is optimised in the calibration procedure. 

\subsubsection{Satellite galaxy luminosity assignment}

Satellite luminosities are assigned for the mock galaxies to reproduce $\Phi_\mathrm{obs}$, considering the cumulative LF for the central galaxies that were already assigned. This procedure is performed by assuming that the cumulative LF for the satellite galaxies in each halo follows a Schechter function given by
\begin{equation}
\label{eq:CLF_halo_satellites}
    \Phi_\mathrm{sat}^\mathrm{halo} (>L; L_\mathrm{cen}, a, \beta) = \frac{A}{V} \, \left( \frac{L}{a L_\mathrm{cen}} \right)^{\alpha} \exp \left[ - \left( \frac{L}{a L_\mathrm{cen}} \right)^{\beta} \right] \, ,
\end{equation}
where the cumulative function depends on the luminosity of the halo central galaxy, $L_\mathrm{cen}$, $V$ is the volume of the subset of the halo catalogue used for the calibration, $A$ is the normalisation that ensures that the halo contains the number of satellites for that halo at the minimum luminosity, and $\alpha$ is set to $-0.5$ as in \citetalias{Castander2024}. The parameters $\beta$ and $a$ are optimised by requiring that the cumulative LF for all the satellite galaxies, plus that for the centrals, equals the total observed cumulative $\Phi_\mathrm{obs}$:
\begin{equation}
\label{eq:CLF_satellites}
\begin{aligned}
    \Phi_\mathrm{sat}(>L) & =  \Phi_\mathrm{obs}(>L) - \Phi_\mathrm{cen}(>L) \\ &  = \sum_{i=0}^{N_\mathrm{h}} N_\mathrm{sat}^\mathrm{HOD} \, \Phi_\mathrm{sat}^\mathrm{halo} (>L; L_\mathrm{cen}, a, \beta) \, .
\end{aligned}
\end{equation}
It is possible to obtain this last equation once we set the HOD parameters, $f$ and $\alpha$, and the introduced scatter in the luminosity assignment for central galaxies, $\sigma_L$. With this aim, we first compute the differential LF of central galaxies, $\phi_\mathrm{cen}(L)$, taking into account the HMF, $\diff n / \diff M_\text{h}$, and the relation between $M_\mathrm{h}$ and the luminosity of the central galaxy given by the AM, $M^\text{AM}_\mathrm{h}(L)$,
\begin{equation}
    \phi_\mathrm{cen}(L) = \frac{\diff n}{\diff M_\text{h}} \frac{\diff M^\text{AM}_\mathrm{h}(L)}{\diff L} \, .
\end{equation}
We can also compute the luminosity of the central galaxy using the AM relation and the scatter introduced, $L^\mathrm{AM}_\mathrm{cen}$ and $\delta L$, respectively. Namely, $L_\mathrm{cen}(M_\text{h},\delta L) = L_\mathrm{cen}^\mathrm{AM}(M_\mathrm{h}) + \delta L$. Then, we can rewrite Eq.~(\ref{eq:CLF_satellites}) as
\begin{equation}
\begin{aligned}
    \Phi_\mathrm{sat}(>L)
    = & \int^{\infty}_{-\infty} \int_{M_\mathrm{min}}^{\infty} \Phi_\mathrm{sat}^\mathrm{halo}[>L; L_\mathrm{cen}(M_\text{h},\delta L), a, \beta] \\
      & \times \mathcal{H}[L_\mathrm{cen}(M_\text{h},\delta L)-L]  \,\mathcal{N}(\delta L; \sigma_L)  \\
      & \times \langle N_\mathrm{sats} \rangle(M_\text{h};\alpha,f) \frac{\diff n}{\diff M_\mathrm{h}} \diff M_\mathrm{h} \, \diff \delta L \,,
\end{aligned}
\end{equation}
where $\mathcal{H}$ is the Heaviside function that accounts for the fact that the satellite luminosity has to be lower than $L_\mathrm{cen}$. The Gaussian factor, $\mathcal{N}(\delta L; \sigma_L)$, accounts for the scatter introduced when assigning the luminosity to central galaxies and, for simplicity, is integrated over $\pm 3.5 \, \sigma_L$. Using this equation, we fit $\beta$ and $a$ on the fly for each set of optimised parameters in the calibration procedure.

\subsubsection{Galaxy colours}
\label{subsubsec:colors}

After assigning luminosities to all galaxies, we compute the $(g01-r01)_\text{HOD}$ colour, defined as the rest-frame colour obtained from the SDSS $g$ and $r$ filters redshifted to $z=0.1$. To assign colours, we use the observed colour-magnitude diagram from the low-redshift NYU-VAGC SDSS galaxy catalogue \citep{Blanton2005} as reference. From this diagram we identified three distinct galaxy populations: red, green, and blue. In bins of the $r$-band absolute magnitude at $z=0.1$, $M^{0.1}_r$, we then determined the fraction of galaxies belonging to each population together with the mean and standard deviation of their colour distributions.

For each galaxy, the colour is assigned in two steps. First, a colour class (red, green, or blue) is drawn such that the observed fractions of the colour-magnitude diagram are reproduced. The probability of assigning a given class depends on both the galaxy type (central or satellite) and $M^{0.1}_r$. The class is assigned by sampling a random number and comparing it with the probability $f^\mathrm{type}_\mathrm{class}(M^{0.1}_r)$, which gives the likelihood that a galaxy with magnitude $M^{0.1}_r$ and of a given type belongs to each colour population. We adopted the superscripts $\mathrm{cen}$, $\mathrm{sat}$, and $\mathrm{tot}$ to denote central, satellite, and all galaxies combined, respectively. 

These probability functions are set considering the fit fractions from the colour-magnitude diagram, i.e. $f^\mathrm{tot}_\mathrm{red}$, $f^\mathrm{tot}_\mathrm{green}$, and $f^\mathrm{tot}_\mathrm{blue}$, and the fraction of central galaxies as a function of $M^{0.1}_r$, $f_\mathrm{cen}(M^{0.1}_r)$. This last function is fitted on the fly in the calibration process.
In \citetalias{Castander2024}, $f^\mathrm{sat}_\mathrm{red}$ and $f^\mathrm{sat}_\mathrm{green}$ were modelled with the sum of two sigmoid functions.
Here, we simplified these relations by using one sigmoid function and adopted
\begin{equation}
\label{eq:sat_colors}
\begin{split}
\begin{aligned}
 f^\mathrm{sat}_\mathrm{red}(M^{0.1}_r) = 
 & 1.05 - \frac{1}{1 + \exp[-0.8\,(M^{0.1}_r-M_r^\mathrm{red})
]} \, ;\\
  f^\mathrm{sat}_\mathrm{green}(M^{0.1}_r) =  
  & \frac{0.25}{1 + \exp[-1.5\,(M^{0.1}_r-M_r^\mathrm{green})]} \, ,\\
\end{aligned}
\end{split}
\end{equation}
where $M_r^\mathrm{red} = -19.5$, and $M_r^\mathrm{green} = -20.0$ are the values adopted in \citetalias{Castander2024}. The rest of the probabilities, $f^\mathrm{sat}_\mathrm{blue}$ $f^\mathrm{cen}_\mathrm{red}$, $f^\mathrm{cen}_\mathrm{green}$, and $f^\mathrm{cen}_\mathrm{blue}$, were obtained according to the total observed probabilities and $f_\mathrm{cen}$.
 Once the colour class is assigned, the $(g01-r01)_\mathrm{HOD}$ value was drawn from a Gaussian distribution whose mean and standard deviation are taken from the observed colour-magnitude diagram, evaluated for the corresponding galaxy population and luminosity bin. In the calibration process, we optimised $M_r^\mathrm{red}$ and $M_r^\mathrm{green}$.
 
\subsubsection{Galaxy positions and velocities}
\label{subsubsec:positions}

The galaxy positions are assigned taking into account the halo shape provided by \texttt{ROCKSTAR}. Central galaxies are placed at the halo centre, while satellites  follow a triaxial NFW profile \citep{Navarro97,Jing2002} with concentrations that depend on the satellite colour class and the halo concentration, following the procedure of \citetalias{Castander2024}. The halo concentrations for high-mass haloes are those determined by \texttt{ROCKSTAR}, and for low-mass haloes they were assigned based on the \citet{Diemer2019} relation (see Sect.~5.2 in \citetalias{Castander2024}). We populate the haloes with satellite galaxies up to a maximum radius given by $r_\mathrm{vir} \, f^\mathrm{cut}_\mathrm{r}$, where $r_\mathrm{vir}$ is the halo virial radius and $f^\mathrm{cut}_\mathrm{r}$ is optimised in the calibration procedure.

To assign galaxy velocities, \texttt{SciPIC} assumes that central galaxies are placed at rest at the halo centre. Thus, central galaxies are assigned the velocity of the halo, i.e. we assumed no velocity bias. For the satellite galaxies, the velocities are assigned by solving the spherical, stationary Jeans equation of local dynamical equilibrium using the parameters found for massive clusters at $z=0.05$ by \citet{Mamon2019}. Further details are given in Sect.~5.3 in \citetalias{Castander2024}. During calibration we neglected this step, and we did not model the galaxy peculiar velocities in order to simplify the procedure. Since we used the projected two-point correlation function to optimise the parameters (see Sect.~\ref{subsec:cal}), the impact of the peculiar velocities in this estimator is negligible. On the other hand, to validate the calibration, we computed clustering using the redshifts that include the galaxy peculiar velocities for the observed samples (see Sect.~\ref{sec:validation}).

\subsection{Calibration process}
\label{subsec:cal}

To perform the calibration, we used as constraints the observed clustering, $w^\mathrm{obs}_\mathrm{p}$, provided by \citet{Zehavi2011} obtained using the Seventh Data Release \citep[DR7;][]{Abazajian2009} of the Sloan Digital Sky Survey \citep[SDSS;][]{York2000}. We used the two-point projected correlation functions from \citet{Zehavi2011}, computed in luminosity bins, using the total sample of galaxies and divided by colour, red and blue. For the calibration we discarded the measurements for the highest luminosity bin, $-23 < M^{0.1}_r - 5 \logten (h) < -22$, given the lack of statistics for this particular subset. The red and blue galaxy subsamples were defined using a magnitude-dependent colour cut as the one adopted in \citet{Zehavi2011},
\begin{equation} \label{eq:colour_cut}
    (g-r)_\mathrm{cut} = 0.21 - 0.03 \, M_r - 5 \logten( h)\, .
\end{equation}
The red galaxies are those that have $(g01-r01)_\mathrm{HOD} > 0.21 - 0.03 \, M^{0.1}_r - 5 \logten( h)$, while the rest are classified as blue galaxies. In total, we considered the observed clustering of total, red, and blue galaxies, each split into four luminosity bins, resulting in 12 galaxy samples for the calibration. We include in Appendix~\ref{app:2pt} the definitions and strategy adopted to compute the projected two-point correlation functions for the mock catalogues. We used ten logarithmically spaced projected radial bins between $0.1 \, \hMpc$ up to $20 \, \hMpc$ and five linearly spaced bins in $\pi$.

The calibration process starts with the halo catalogue, and we also provide the Schechter parameters that characterise the HMF to perform the AM (Fig.~\ref{fig:HMF}). As described, the set of six parameters to be optimised are $\Theta = \{ \alpha, f, \sigma_L, M_r^\mathrm{red}, M_r^\mathrm{green}, f^\mathrm{cut}_\mathrm{r} \}$. For each set of parameters, we generated a mock catalogue and computed the clustering in luminosity and colour bins following the same procedure used for the measurements that serve as constraints. In the case of the FS2-Wide light cone, we further considered an upper redshift limit of $z=0.1$ for the catalogue to be complete in volume. Additionally, to increase the statistical significance of the lowest-luminosity bin sample, $-19 < M^{0.1}_r - 5 \logten (h) < -18$, we extended the redshift range from $z=0.026167$ \citep[as used by][]{Zehavi2011} to $z=0.06$. 

To perform the calibration, we needed to produce a mock catalogue for each set of parameters $\Theta$ and to measure the clustering for the 12 samples divided by colour and luminosity considered as constraints. To achieve this in a reasonable amount of time, we ran the code on top of \texttt{Apache Spark}, an engine for scalable computing. \texttt{SciPICal} is executed in the PIC Big Data platform, based on \texttt{Hadoop}. The code uses two different \texttt{Spark} configurations: one for mock catalogue generation and another for clustering measurements. The first stage is run using 600 CPU cores. Subsequently, clustering is measured in parallel by assigning one node with 40 CPU cores to each galaxy sample, totalling 480 CPU cores. This parallelisation is achieved by combining \texttt{Apache Spark} for sample distribution and \texttt{TreeCorr}\footnote{\url{https://rmjarvis.github.io/TreeCorr/}} for the clustering measurements.

For optimisation we tested two methodologies. One is a minimisation algorithm, which considers a target function defined as
\begin{equation}
\label{eq:chi}
\chi^2 = \sum^{N_\text{bin}}_{i} \frac{\brackets{ w_{\mathrm{p},i}(\Theta) - w^\mathrm{obs}_{\mathrm{p},i} }^2}{\sigma^2_{2\mathrm{p},i}}\;,
\end{equation}
where the sum runs over each radial bin, $i$, included in all the luminosity and colour-selected subsamples. Here, $\Theta$ is the set of parameters to be optimised, $w_{\mathrm{p},i}$ and $w^\mathrm{obs}_{\mathrm{p},i}$ are the two-point correlation functions at the radial bin $i$, obtained for each mock realisation and the SDSS observations, respectively. The considered uncertainty $\sigma^2_{2\mathrm{p}}$ is the variance of the projected correlation function computed as the quadratic sum of the uncertainties on the model and the observations: $\sigma^2_{2\mathrm{p},i} = \epsilon_{2\mathrm{p},i}^2 +\epsilon_{2\mathrm{p,obs},i}^2$. We estimated $\epsilon_{2\mathrm{p},i}^2$ from the diagonal elements of the covariance matrix, computed using jackknife resampling with 16 patches. The $\epsilon_{2\mathrm{p,obs},i}$ values correspond to the uncertainties reported by \citet{Zehavi2011}, which were also derived using jackknife resampling with 144 spatially contiguous subsamples.

To constrain the set of parameters, $\Theta$, we first implemented the \texttt{scipy.optimize.minimize} package, using the Nelder--Mead method. This method allows us to obtain a first guess for the optimised parameters in around $4 \, \mathrm{h}$. We implemented this method by including boundary conditions taking as reference the values adopted to generate the FS2-Wide (\citetalias{Castander2024}). The adopted boundaries are specified in Table~\ref{tab:initial_mcmc}. With the aim of obtaining more information regarding the parameter space and to look for possible degeneracies, we also implemented a Markov chain Monte Carlo (MCMC) method using the \texttt{emcee} \texttt{Python} package \citep{Foreman2013}. The adopted initial values to explore the parameter space were generated using 15 walkers, taking into account the optimised parameters with the minimisation method $\Theta_0 = \{ \alpha_0, f_0, \sigma_{L,0}, M_{r,0}^\mathrm{red}, M_{r,0}^\mathrm{green}, f^\mathrm{cut}_{\mathrm{r},0} \}$. Details of the distributions used to generate the random initial values are detailed in Appendix~\ref{sec:mcmc}. The calibration process consists of 500 steps to explore the parameter space. With the configuration described above, each iteration takes on average 2.4 minutes, requiring approximately 10 days of total execution time. Optimised parameters were determined as the median values of the posterior distributions after discarding the first 300 steps of each chain, while the errors enclose $68\%$ of these distributions.

\section{Calibrating Euclid Flagship galaxy mocks}
\label{sec:calFS}
\begin{figure}
    \captionsetup{font={small},labelfont=bf} 
    \centering
    \includegraphics[scale=0.9]{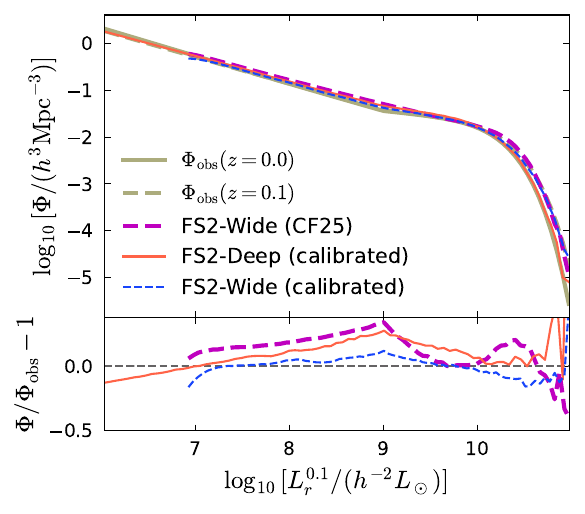}
    \caption{Differential LF distributions for the generated mocks in this work (FS2-Wide and FS2-Deep), the FS2-Wide mock release (\citetalias{Castander2024}), and the observed LF, $\Phi_\mathrm{obs}$, used to perform the AM at $z=0.1$ and $z=0$. \textit{Bottom panel}: Residuals of the LF computed for the mock catalogues, compared with $\Phi_\mathrm{obs}$ at $z=0$ for the FS2-Deep snapshot and at $z=0.1$ for the FS2-Wide catalogues.}
    \label{fig:fs2_LF}
\end{figure}

\begin{figure*}[h!]
        \captionsetup{font={small},labelfont=bf} 
    \centering
    \includegraphics[scale=0.7]{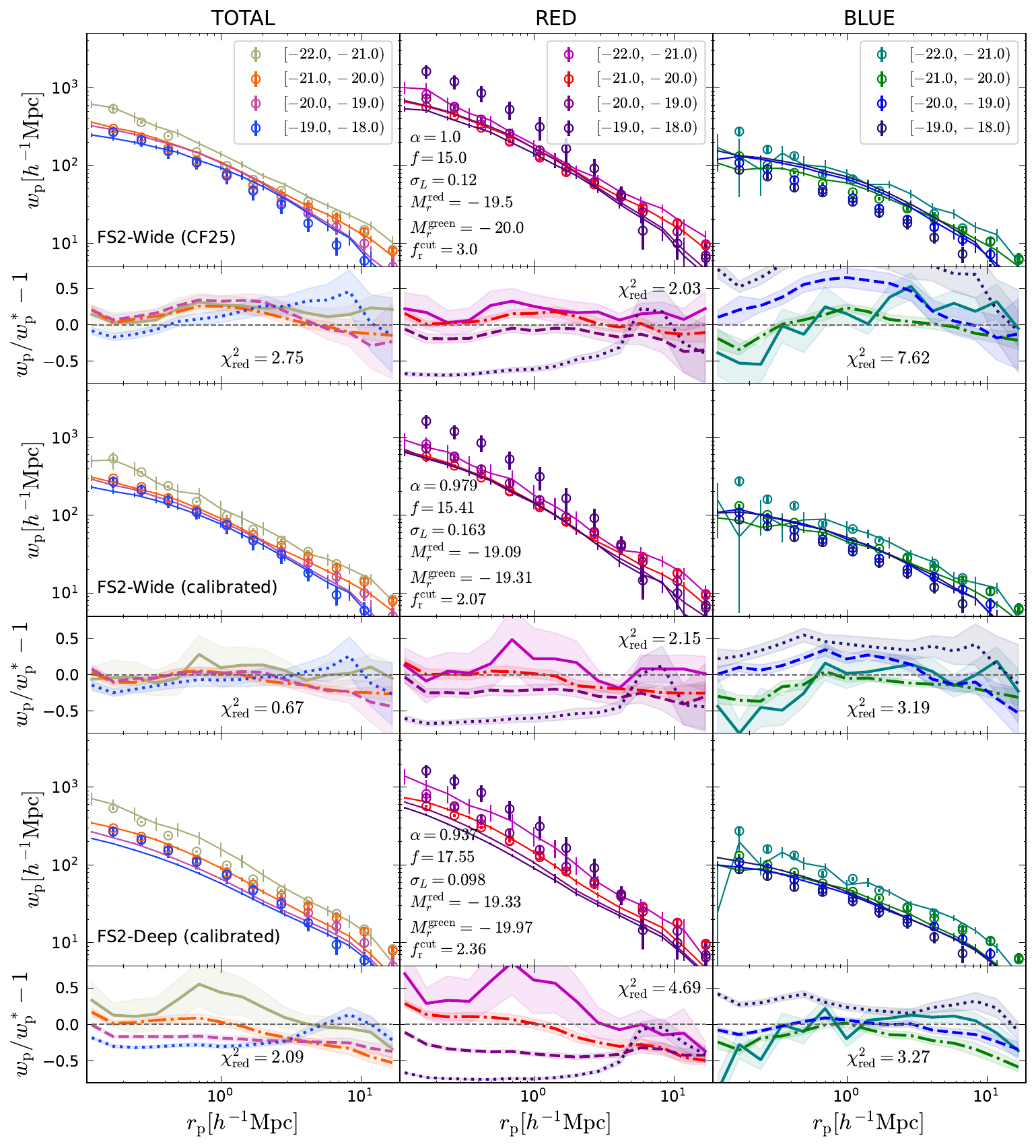}
    \caption{Two-point correlations for the galaxies split according to luminosity (\textit{left panels}) and considering the colour cut defined in Eq.~(\ref{eq:colour_cut}). \textit{Middle panels}: Red galaxies. \textit{Right panels}: Blue galaxies. The dots represent the \citet{Zehavi2011} measurements taken as reference for the calibration. The solid lines indicate the measured clustering obtained for the mocks: FS2-Wide (\citetalias{Castander2024}) release (\textit{top panels}), FS2-Wide (\textit{middle panels}), and FS2-Deep (\textit{bottom panels}) obtained by adopting the optimised parameters from the calibration. The colour-coding indicates the luminosity bin in $M^{0.1}_r - 5 \logten (h)$ specified in the legends. Below the correlations, the residuals are shown relative to the reference clustering together with the reduced chi-square values ($\chi^2_\mathrm{red}$). For clarity, the following line styles indicate the luminosity bins: solid, dot-dashed, dashed, and dotted, from brightest to faintest.}
    \label{fig:FS2_corr}
\end{figure*}

To evaluate the calibration performance of our pipeline, we created a test mock catalogue. This test mock was generated by populating the FS2-Wide halo catalogue with an adopted set of parameters, $\Theta_\mathrm{test}$. We then ran the calibration pipeline, using the measured clustering from the test mock as a constraint. With this approach, we can verify if the pipeline can recover the adopted $\Theta_\mathrm{test}$ parameters correctly. The details are presented in Appendix~\ref{sec:test}. From this test we can notice that the minimiser methodology implemented is not adequate since it tends to converge in local minima. On the other hand, the MCMC optimisation successfully recovers the parameters with high precision, yielding clustering measurements that agree within the uncertainties with those of the test mock catalogue used as a calibration reference.

Given the calibration results from the test mock, our discussion focuses on the results obtained with the MCMC optimisation approach after running the minimiser. The results of the MCMC run are presented and discussed in Appendix \ref{sec:mcmc_output}. In general, there is good agreement between the parameters adopted by \citetalias{Castander2024} and those obtained from the optimisation in this work. A parameter where a considerable difference is obtained is the limiting radius up to which the galaxies are placed ($f^\mathrm{cut}_\mathrm{r}$), where in this case the optimised parameter is considerably lower.
In the case of the FS2-Deep, the main differences are obtained for the HOD parameters. This simulation prefers lower values for $\alpha$ and $f$. Given that both simulations were produced adopting the same cosmological parameters, the differences are mainly related to the redshift range used to determine the HMF for the AM technique and the halo mass resolution. While the FS2-Wide calibration was performed in the light cone, in the case of the FS2-Deep we used a $300 \, \hMpc$ length box subsample, of a snapshot at $z=0$. Therefore, evolutionary effects could explain the obtained difference between both calibrations.

The main target of the FS2 simulations is to jointly reproduce the observed number density of galaxies and their clustering. The luminosity assignment process involves fitting the luminosity function distribution of satellite galaxies within each halo defined in Eq.~(\ref{eq:CLF_halo_satellites}). Inadequate fitting due to limitations in the adopted model can produce substantial discrepancies between the total luminosity distribution obtained in the mock catalogue and that observed. Then, to evaluate the performance of our mocks, we compared the obtained LF with the adopted observed relations. To perform this comparison, in the case of the FS2-Deep, we populated a subset of haloes in the snapshot at $z=0$ within a sub-box with length $200 \, \hMpc$ without a cut in mass to reach up to the mock magnitude limit. In Fig.~\ref{fig:fs2_LF}  we show the comparison between the LF obtained from the mock catalogues and the observed LF taken as reference, $\Phi_\mathrm{obs}$. The residuals are associated with the parameter fitting involved in assigning satellite luminosities. In particular, the discontinuity at $\logten [L_{r}^{0.1} /(h^{-2} \, L_{\odot})] = 9$ arises from the way the SDSS luminosity function is parametrised, where a power-law component is added below that luminosity to the standard Schechter function, introducing a discontinuity in its derivative. We notice a substantial improvement in reproducing the expected number density of galaxies for the mocks produced in this work in comparison with the previous FS2-Wide mock version (\citetalias{Castander2024}).

In Fig.~\ref{fig:FS2_corr} we show clustering for the mocks computed using the released FS2-Wide mock catalogue (\citetalias{Castander2024}) and from the mocks generated with the optimised parameters obtained using the calibration pipeline, for FS2-Wide and FS2-Deep. We compare the measured two-point correlation with the \citet{Zehavi2011} measurements, used as constraints. Reduced $\chi^2$, $\chi^2_\mathrm{red}$, values are also included in the figure. These were computed by dividing Eq.~(\ref{eq:chi}) by the number of radial bins multiplied by the number of galaxy samples considered, $N_\text{bin} = 120$. Overall, we obtain an improvement for the FS2-Wide calibrated version; the total reduced $\chi^2_\mathrm{red} \approx 2$, which, compared to $\chi^2_\mathrm{red} \approx 4$ for the \citetalias{Castander2024} catalogue, represents an improvement of approximately 50\%. Although we obtain a slightly larger $\chi^2_\mathrm{red}$ value for the red galaxy sample, this result arises because the calibration is performed using all samples simultaneously. Consequently, the fit for specific subsamples may show less agreement while achieving a lower total $\chi^2_\mathrm{red}$ value. For the FS2-Deep catalogue we obtain a larger mismatch between the observed clustering and the one obtained from the mock as indicated by the resulting reduced $\chi^2_\mathrm{red}$ values. Differences with the FS2-Wide catalogue are mainly related to the higher resolution of this catalogue, which modifies $M_\text{min}$, and, in turn, the number of satellites, and the clustering evolution. Comparing the clustering of the haloes in mass bins, we obtain differences of around $10\%$ between FS2-Wide and FS2-Deep, with the clustering in FS2-Deep being systematically lower.

 For both calibrated versions, we obtain total $\chi^2_\mathrm{red}$ values significantly larger than unity, particularly for the colour-selected samples. Taking into account the test presented in Appendix~\ref{sec:test}, we can also conclude that the differences obtained in the clustering between the observed constraints and those from the mocks obtained after the calibration evince the limitations of the modelling adopted by \texttt{SciPIC}. In particular, our strategy cannot properly reproduce the observed clustering for low-brightness red galaxies and tends to underpredict it. The development of \texttt{SciPICal} allows us to explore different strategies to improve the results obtained.

\section{Validating Flagship mocks}
\label{sec:validation}

In this section we validate the galaxy mocks by comparing the clustering measurements from the FS2-Wide and FS2-Deep catalogues with those from various observational and simulated datasets. Our goal is to assess whether the low-redshift calibration strategy can produce galaxy mocks that accurately reproduce the clustering of different galaxy selections across a broad redshift range.
 
 To this end, we first applied the full \texttt{SciPIC} pipeline to obtain the galaxy mocks for the light cones. Additionally, due to the limited sky coverage of the FS2-Deep light cone, we also populated the halo catalogue from the snapshot at $z=0$ to ensure sufficient statistics for the low-redshift comparison. Once we obtained the mocks, we compared the obtained clustering with two observational datasets: a spectroscopic sample from the VIMOS Public Extragalactic Redshift Survey \citep{Guzzo:2013} and a photometric sample from the Dark Energy Survey \citep[DES;][]{DES2005}. Moreover, to have higher signal-to-noise constraints at higher redshifts, we validated the agreement between the clustering predictions of both calibrated galaxy mocks using the Horizon-AGN hydrodynamical simulation \citep{Dubois2014, Kaviraj2017}. We also included in the comparison the clustering measurements from the FS2-Wide \citetalias{Castander2024} release to evaluate the improvements introduced by the calibration procedure.

Below, we first describe generating the galaxy mocks using the optimised parameters from \texttt{SciPICal}, and then present the comparison results of the two-point correlation measurements with the reference datasets. To quantify the agreement between the mock clustering and the reference data, we used the $\chi^2_\mathrm{red}$ computed from Eq.~(\ref{eq:chi}) divided by the total number of bins across all galaxy subsamples used in the comparison.

\subsection{Producing Flagship galaxy mock catalogues}

Taking into account the results from the calibration presented in the previous section, we implemented the optimised parameters in \texttt{SciPIC} and produced galaxy mock catalogues for the FS2-Wide and FS2-Deep light cones and for the FS2-Deep snapshot at $z=0$ by executing the full pipeline. To update the pipeline with these new results, we computed the AM for the whole redshift range considered. This was performed by first characterising the HMF for the light-cone halo catalogues. Then, the cumulative galaxy number density function (Eq.~\ref{eq:cum_gal}) was computed according to the HMF, the HOD optimised parameters ($\alpha$, $f$), and the minimum halo mass of the catalogues, $M_\mathrm{min}$. In the case of the FS2-Wide light cone, since it is designed for the EWS, our mock catalogue needs to reach an apparent magnitude of $\IE = 24.5$, which corresponds approximately to an absolute magnitude of $M_{\IE} - 5 \log_{10}(h) = -13.0$ at $z=0.1$. Since the $\IE$ band mainly encompasses the SDSS $r$ band used in the luminosity assignment, we can estimate the minimum halo mass needed according to the AM relation to be $1.5 \times 10^{10} \, h^{-1} M_\odot$. This mass corresponds to a halo identified with 15 particles and is two times larger than the FS2-Wide minimum mass, $M_{\mathrm{min}}$, indicated in Sect.~\ref{subsec:hod} after the corrections described in Sect.~4.2 of \citetalias{Castander2024}.

The EDS is expected to be approximately two magnitudes fainter than the EWS. In this case, we estimate a minimum halo mass of $5 \times 10^{9} \, h^{-1} M_\odot$, which in the FS2-Deep simulation corresponds to a halo of approximately 50 particles. Although this mass limit is higher than the minimum mass of our halo catalogue, these haloes are affected by incompleteness and discreteness effects. To obtain a mass-complete halo catalogue, we applied a procedure similar to that described in \citetalias{Castander2024}, which is detailed in Appendix~\ref{sec:deep_halo}. Then, to run \texttt{SciPIC} on the FS2-Deep light cone
 we extended the pipeline to populate haloes down to a minimum mass of $10^9 \, h^{-1} \, M_{\odot}$ and up to $z=10$. Details on the HMF computation and corrections to the halo mass and concentrations are presented in Appendix~\ref{sec:deep_halo}. The AM relations computed for the FS2-Deep light cone are presented in Appendix~\ref{sec:AM_deep}, where we describe using the COSMOS2020 catalogue \citep{COSMOS2020} to compute the LF for the whole redshift range. To assign an SED to each galaxy, we modified this module to reach the limiting redshift by transforming the $g-r$ colour distribution to distributions of redshifted colours in COSMOS2020. Further details regarding the extension of the pipeline for the luminosity and the SED assignment can be found in~\cite{Ramakrishnan2025b}. In the case of FS2-Wide, we performed the same procedure as that described in \citetalias{Castander2024} and obtained the galaxy fluxes in several bands for all galaxies up to $z=3$. The main modification in this case is the update of the AM relations with redshift, using the calibrated HOD parameters. Details on these measurements are presented in Appendix~\ref{sec:AM_deep}.

For both simulations, we did not assign morphological properties and intrinsic shapes, since a new calibration for these modules is going to be implemented in a future work (Gonzalez et al., in prep.). It is important to highlight that we do not consider a dependence on redshift of the optimised parameters. The redshift dependence of the assigned galaxy properties is indirectly introduced by the intrinsic evolution of the HMF and the observed LFs considered for the computation of the AM. Taking into account the minimum masses of the catalogues after corrections for incompleteness and discreteness, the generated FS2-Wide and FS2-Deep galaxy mocks are complete up to a limiting $r$-band apparent magnitudes of approximately 26 and 30, respectively, at $z=0.1$.

\subsection{Comparison with clustering measurements}

\subsubsection{\textsc{Y3 MagLim} DES sample}

To evaluate the mock clustering predictions for a broad redshift range, we considered as references the observed clustering from the \textsc{Y3 MagLim} DES galaxy sample \citep{Porredon2021}. In particular, we took into account the measurements\footnote{\url{https://des.ncsa.illinois.edu/releases/y3a2/Y3key-products}} presented in \citet{DES2022}. The angular clustering measurements used for the comparison were obtained by dividing the galaxy sample into six tomographic bins, with edges defined by the photometric redshift estimates: $z_\mathrm{phot}\in\{0.20, 0.40, 0.55, 0.70, 0.85, 0.95, 1.05\}$, corresponding to bins 1 through 6.

We performed the comparison with the FS2-Wide mock catalogues by selecting the galaxies with the same magnitude cuts as for the \textsc{Y3 MagLim} sample, $i > 17.5$ and $i < 4z + 18$, where we adopted $z$ as the galaxy redshift computed according to its position and peculiar velocity. To obtain the tomographic subsamples, we assigned a weight to each galaxy in the mocks to transform the observed number density distribution as a function of the redshift, $n(z)$, to be similar to the redshift distributions in the \textsc{Y3 MagLim} tomographic subsamples.
Then, we randomly sampled $A\,n(z)$ galaxies using these weights, where $A$ is the total area coverage of the mock light cone, corresponding to one octant of the sky. With this approach, we ensured that the tomographic subsamples from the mock match the same number density and redshift distributions as the observed subsamples. In this comparison, we did not include the clustering measurements from the FS2-Deep mock, since it has a small sky coverage that strongly affects the statistics for the lower tomographic redshift bins.

\begin{figure}
    \captionsetup{font={small},labelfont=bf}
    \centering
    \includegraphics[scale=0.7]{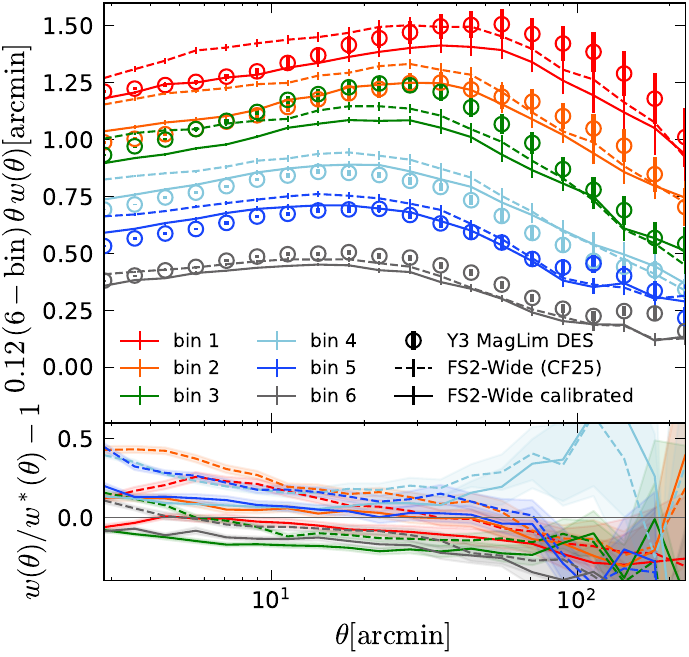}
    \caption{Clustering measurements from the \textsc{Y3 MagLim} DES galaxy sample compared to the clustering obtained from the galaxies of the FS2-Wide mocks for six tomographic bins. For a better visualisation of the results, we shifted the $\theta \, w(\theta)$ measurements by $0.12\,(6 - \mathrm{bin\,number})$. \textit{Bottom panel}: Residuals between the mock clustering and the observed one, where the shaded region represents the uncertainties computed for the mock clustering.}
    \label{fig:corr_DES}
\end{figure}

The clustering measurement comparison is shown in Fig.~\ref{fig:corr_DES}. We obtain good agreement between the observed clustering and that obtained from the galaxy mock catalogues. In particular, we obtain a tighter agreement for the calibrated catalogue, especially at smaller scales where the FS2-Wide (\citetalias{Castander2024}) mock tends to overpredict the clustering signal. For this comparison, we obtain $\chi^2_\mathrm{red}$ values of ten for the calibrated version and 30 for the \citetalias{Castander2024} mock, representing a substantial improvement in matching the clustering measurements.

\subsubsection{VIPERS}

\begin{figure*}
    \captionsetup{font={small},labelfont=bf} 
    \centering
    \includegraphics[scale=0.65]{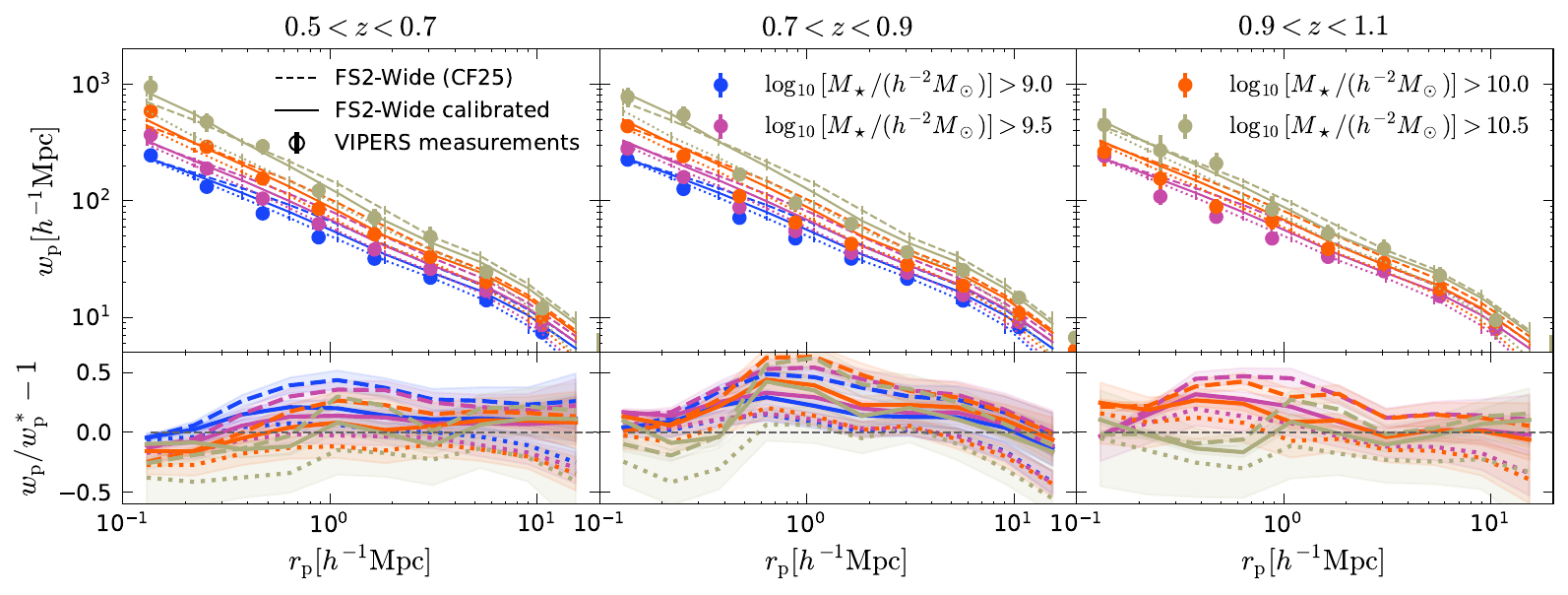}
    \caption{Comparison of the projected two-point correlation functions measured in the FS2 catalogues (solid lines) to those from the VIPERS galaxy samples (dots) for the redshift bins and stellar mass cuts indicated in the legend. Below the correlations, the residuals are shown relative to the VIPERS clustering, with the shaded region indicating the uncertainties of the VIPERS measurements.}
    \label{fig:corr_vipers}
\end{figure*}

The VIMOS Public Extragalactic Redshift Survey \citep[VIPERS;][]{Guzzo:2013} is a spectroscopic survey aimed at measuring redshifts in a redshift range of $0.5 < z \lesssim 1.2$ and covering about $24\,\mathrm{deg}^2$. The target galaxy sample was selected from the Canada-France-Hawaii Telescope Legacy Survey Wide optical photometric catalogue \citep{Erben2013} with a flux cut of $i_\mathrm{AB} < 22.5$ and a $gri$ colour pre-selection applied to remove objects at $z < 0.5$. To perform the comparison with the galaxy mock catalogues, we used the clustering measurements presented in \citet{Marulli2013} obtained using stellar-mass thresholds and stellar-mass binning to define the galaxy samples. Stellar masses in \texttt{SciPIC} were computed by first assigning a stellar mass-to-light ratio from the galaxy colours and the SED, taking as reference the $L^{0.1}_r$ luminosity (see Sect.~5.7 in \citetalias{Castander2024}). Therefore, this comparison simultaneously tests the validity of the extrapolation applied when generating the catalogue to redshifts beyond those used for calibration, and the subsequent stellar-mass assignment based on the SEDs.

In Fig.~\ref{fig:corr_vipers}, we compare the VIPERS clustering measurements with those from the FS2-Wide (\citetalias{Castander2024}) catalogue and from the mocks generated using the calibrated parameters introduced in this work. The comparison was performed using galaxy samples selected via stellar mass thresholds, as these yield higher signal-to-noise ratios than stellar mass binning. For completeness, the results based on stellar mass bins are presented in Appendix~\ref{sec:vipers_bins}. From Fig.~\ref{fig:corr_vipers}, we observe that the clustering measurements are all in agreement within $1\sigma$ uncertainties. On the other hand, FS2-Wide (\citetalias{Castander2024}) tends to overestimate the clustering compared to VIPERS measurements, especially at the one-to-two-halo transition. This might be related to the lower optimised value of $f^\mathrm{cut}_\mathrm{r} \approx 2$ compared to the value adopted for FS2-Wide (\citetalias{Castander2024}; $f^\mathrm{cut}_\mathrm{r} = 3$), which suppresses the excess clustering in this regime. As for the DES results, we quantified the general agreement by using the $\chi^2_\mathrm{red}$. In this case, we obtain a value of 2.0 and 1.4 for the FS2-Wide and FS2-Deep calibrated mocks, respectively, while for the \citetalias{Castander2024} mock we obtain a value of 5.4, representing a substantial improvement in matching these clustering measurements for the calibrated versions.

\subsubsection{Horizon-AGN}

To further complement our comparison, we considered the measured clustering from the Horizon-AGN cosmological hydrodynamical simulation \citep{Dubois2014}. This simulation was generated using \texttt{RAMSES} \citep{Teyssier2002}, an adaptive mesh refinement Eulerian hydrodynamics code, with an initial gas mass resolution of $10^7 \, M_{\odot}$. The size of the simulation box has a comoving width of $100\,h^{-1}\,\mathrm{Mpc}$ with a dark matter particle mass resolution of $8\times10^7 \, M_{\odot}$. It was run by adopting WMAP7 $\Lambda$CDM cosmology \citep[$\Omega_\mathrm{m} = 0.272$, $\Omega_\Lambda = 0.728$, $\sigma_8 = 0.81$, $\Omega_\mathrm{b} = 0.045$, $H_0 = 70.4\,\mathrm{km\,s^{-1}\,Mpc^{-1}}$, and $n_\mathrm{s} = 0.967$;][]{Komatsu2011}. It takes into account several key processes in the formation of galaxies described in \citet{Dubois2014}. We performed the comparison by considering two snapshots of Horizon-AGN at $z=0$ and $z=1$, the previous release of FS2-Wide (\citetalias{Castander2024}), and the full set of mock catalogues generated in this work: the calibrated FS2-Wide and FS2-Deep light cones, as well as the FS2-Deep snapshot at $z=0$. The comparison at $z=1$ is motivated by the fact that the expected redshift distribution of galaxies observed by the EWS peaks around this redshift. The comparison at $z=0$ was performed by selecting the galaxies in the FS2-Wide light cones within $0.05 < z < 0.09$ in the whole octant. Since the ratio of the linear growth factor, $D(z)$, between $z=0$ and $z=0.1$ is approximately $1.05$, we expect an underestimation of the clustering of at most $10\%$ at large scales. For FS2-Deep we used the mock snapshot at $z=0$ within a comoving box of about $150\,h^{-1}\,\mathrm{Mpc}$ width. For the comparison at $z=1$, we selected the galaxies in the light cones within a redshift range $0.965 < z < 1.035$ and in a sky patch of $4\times4\,\mathrm{deg}^2$ to have a comoving volume equivalent to a box with width of about $140\,h^{-1}\,\mathrm{Mpc}$. 

\begin{figure}
    \captionsetup{font={small},labelfont=bf}
    \centering
    \includegraphics[scale=0.66]{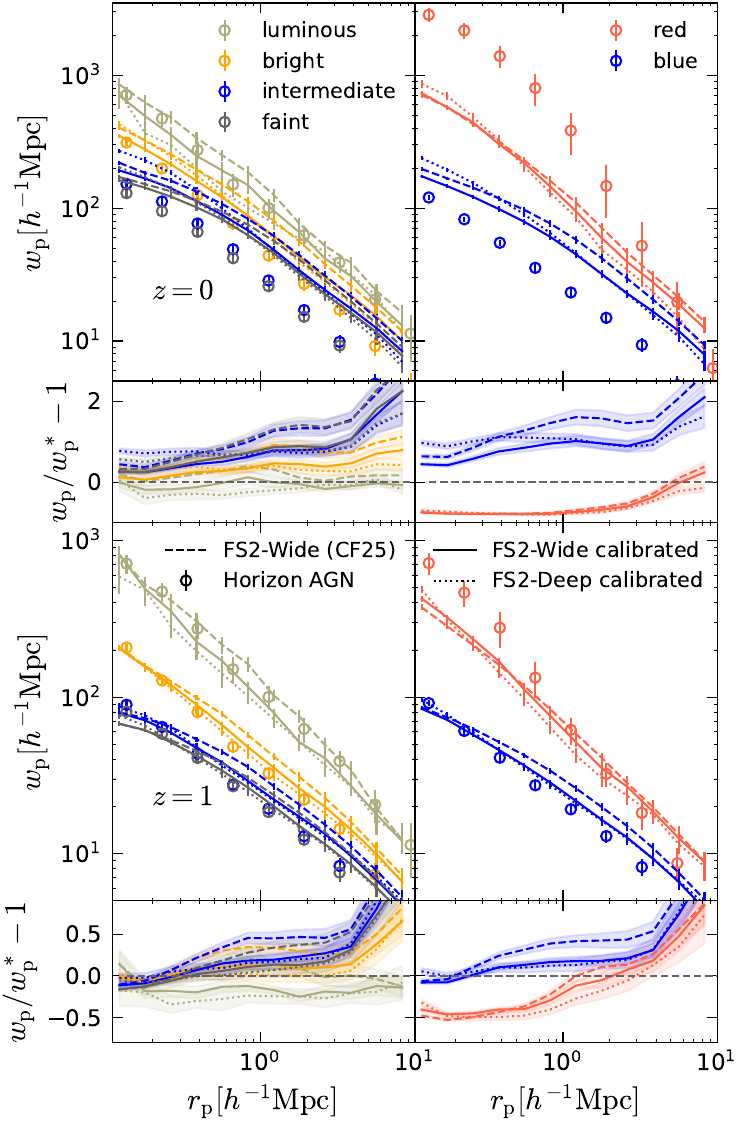}
    \caption{Clustering comparison between the Horizon-AGN galaxies and FS2 mock catalogues for luminosity- and colour-selected samples. \textit{Upper} and \textit{lower panels}: Comparison at $z=0$ and $z=1$, respectively. The panels below the correlations show the residuals between the mock clustering and the one computed for the Horizon-AGN galaxies, where the shaded region indicates the uncertainties computed for the mock clustering.}
    \label{fig:corr_Horizon}
\end{figure}

Horizon-AGN clustering has been previously compared to observational data in \citet{Hatfield2019} using VIDEO observations \citep{Jarvis2013}. In general, they find good agreement between the clustering measurements for galaxies with a stellar mass $\logten(M_{\star}/M_{\odot}) \gtrsim 10.5$. However, they find that Horizon-AGN tends to underestimate the clustering at low stellar masses and redshifts, with particularly high statistical significance ($>5\sigma$). On the other hand, at higher stellar masses and redshifts, it tends to overestimate the clustering but with a lower significance ($1\sigma$--$2\sigma$). Based on these results, we performed a clustering comparison for six galaxy subsamples. The first is a `luminous' sample selected using a cut in absolute magnitude, $M_r$, to ensure $\logten(M_{\star}/M_{\odot}) \gtrsim 10.5$. We also included a dimmer galaxy sample ($M_r < -18$), further divided into bins based on luminosity and colour. To enable a fair comparison, we adjusted the magnitude cuts in the mock galaxy catalogues to match the number densities found in Horizon-AGN. The luminosity bins (labelled as `bright', `intermediate', and `faint') were defined to contain 25\%, 25\%, and 50\% of the galaxies, respectively. Since we adopted a colour cut for the calibration, we also considered a colour-based selection. To do this, we used the $M_g - M_r$ index, classifying the top 10\% reddest galaxies as red, with the remaining 90\% designated as blue. This cut was selected to obtain a clean sample of red (elliptical) galaxies, as hydrodynamical simulations typically produce fewer large ellipticals than in observations \citep{Haslbauer2022}.

Figure~\ref{fig:corr_Horizon} presents the clustering comparison for all the galaxy samples at the two redshifts considered. In general, the clustering predictions from the FS2 mocks are consistent with each other. However, we obtain considerable discrepancies at $z=0$, with Horizon-AGN having a lower signal for the lower luminosity samples. This is in agreement with the results presented in \citet{Hatfield2019} when comparing Horizon-AGN with VIDEO clustering. In contrast, the clustering amplitude for red galaxies in Horizon-AGN is noticeably higher than that obtained from the mock catalogues, while that of blue galaxies appears to be considerably underestimated. Since blue galaxies constitute the dominant population, applying luminosity cuts results in a lower overall clustering amplitude. We evaluated the agreement between the clustering from Horizon-AGN and the galaxy mocks at $z=0$ only for the `luminous' sample, as this is the sample that most closely matches observed clustering. For this sample, we obtain $\chi^2_\mathrm{red}$ values of 1.5, 0.4, and 1.2 for the \citetalias{Castander2024} mock, the calibrated FS2-Wide, and the calibrated FS2-Deep, respectively. According to this result, all mocks are in agreement with the prediction of this hydrodynamical simulation for this particular galaxy sample. At $z=1$, we find good agreement when dividing galaxies into luminosity bins, with $\chi^2_\mathrm{red}$ values of 4.2, 2.1, and 1.9 for the \citetalias{Castander2024}, FS2-Wide, and FS2-Deep mocks, respectively. But when we split galaxies by colour, we find a substantial mismatch for the red galaxy sample. For the colour-selected samples, we obtain $\chi^2_\mathrm{red}$ values of 5.5 for the \citetalias{Castander2024} mock, while the calibrated versions yield 2.4 for FS2-Wide and 1.7 for FS2-Deep. We therefore conclude that the calibrated versions yield a substantial improvement in clustering predictions, although colour-selected galaxy samples reveal limitations in the underlying model used to assign galaxy properties.

\section{Summary and conclusions}
\label{sec:conclusion}
The unprecedented depth and volume of stage~IV galaxy surveys demand highly accurate simulations to validate observational and analysis strategies. This requires simulating volumes with box sides larger than $1 \, h^{-1}\,\mathrm{Gpc}$ and achieving sufficient mass resolution to resolve the host haloes of the faintest detectable galaxies. Generating these mocks therefore presents a massive computational challenge, as it requires populating hundreds of millions of haloes with billions of galaxies, each characterised by hundreds of distinct physical properties.
The \texttt{SciPIC} pipeline is designed to generate galaxy mock catalogues that simultaneously reproduce the observed number density and clustering across multiple galaxy populations. The produced galaxy mocks enables diverse analyses, forecasting studies, and fully end-to-end processing analogous to that applied to real observational data. Aiming to improve the clustering predictions of \texttt{SciPIC} galaxy mocks, we present in this work a calibration pipeline, \texttt{SciPICal}, to derive the parameters required for populating halo catalogues and generating FS2-like mocks. This pipeline constitutes a crucial component of the mock generation process, as the parameters optimised during the calibration are those that most substantially determine the characteristics of the resulting galaxy clustering. The calibration pipeline is also inherently challenging, owing to both the complexity of the procedure in which the galaxy properties are assigned and the substantial data volume that needs to be handled. The parameters optimised are those that control the number of assigned satellites, the distribution of galaxy colours, and the maximum virial radius used to populate haloes. One of the main advantages of this pipeline is that it produces the posterior distributions of the optimised parameters that enable the inspection of the underlying physical degeneracies and correlations between the model parameters. We tested the performance of the pipeline by applying it to a test mock catalogue, generated considering a given set of parameters. Then, we used the clustering constraints derived from this catalogue to perform the calibration. We show that the pipeline can successfully recover the set of parameters adopted within roughly $1\sigma$. This demonstrates the robustness of our pipeline and that it is not substantially affected by the stochastic effects introduced in the galaxy properties, which can impact the clustering constraints.

The presented pipeline enabled the calibration of parameters for both the FS2-Wide and FS2-Deep halo catalogues. We presented the calibration results and validated them by comparing the resulting projected clustering with SDSS measurements at low-redshifts ($z<0.1$) used as constraints in the optimisation process. Compared to the previous FS2-Wide released, our calibrated version improves the $\chi^2$ values by approximately 50\%. After obtaining the optimised parameters that jointly reproduce the observed galaxy clustering and number density at low redshifts, we applied \texttt{SciPIC} to generate three sets of galaxy mocks: one based on the FS2-Wide light cone, another on the FS2-Deep light cone, and a third using the FS2-Deep snapshot at $z=0$. The produced FS2-Wide and FS2-Deep galaxy mocks are complete up to $r$-band apparent magnitudes of approximately 26.0 and 30.0, respectively, at $z=0.1$. These catalogues are broadly useful, as they include expected fluxes for various surveys, such as LSST.
To evaluate the effectiveness of our calibration strategy, we compared the clustering predicted by these mocks with measurements from spectroscopic and photometric galaxy surveys, as well as with results from a hydrodynamical simulation. Although the FS2-Wide and FS2-Deep calibrations yield different parameter values, the clustering predictions of both calibrated mocks are consistent within the uncertainties across the different galaxy samples considered. The calibrated mocks demonstrate strong predictive power, reproducing clustering measurements within $15\%$ across a wide redshift range and for various galaxy samples. They also show improved consistency with both observational and simulated data compared to the previous FS2-Wide release \citepalias{Castander2024}. Nevertheless, the results presented indicate limitations for the current adopted methodology mainly in its ability to reproduce the clustering of colour-selected samples.

\texttt{SciPICal} is a versatile tool for generating galaxy mocks from a variety of input halo catalogues. It can be used to calibrate both a snapshot-based halo catalogue and a light-cone halo catalogue. This adaptability makes the pipeline suitable for generating galaxy mocks from a wide range of halo catalogues produced under different conditions, including variations in cosmology or halo identification methods, which can alter the halo properties, clustering, and ultimately the parameters required for mock generation. The pipeline can be also applied to explore the strategies adopted for enhancing the realism of the assigned galaxy properties. Furthermore, the calibration can be extended to incorporate additional observational constraints beyond the two-point correlation, such as galaxy-galaxy lensing measurements, which are especially sensitive to the galaxy-halo connection and therefore crucial for more accurate modelling. In future work, we plan to provide additional calibration pipelines that can be integrated into key \texttt{SciPIC} modules to ensure that all assigned properties remain accurate and up-to-date. This flexible and efficient schema will enable the incorporation of improved modelling strategies and the latest constraints from upcoming wide-field surveys. As we enter a new era of large-scale cosmological data, the development of flexible tools that enable the generation of fast and more realistic simulations is essential to fully exploit this unprecedented data quality. In this context, the pipeline presented here is well aligned with these requirements and will play a key role in shaping analysis strategies for the forthcoming survey releases, particularly for future \Euclid releases.

\begin{acknowledgements}
The authors want to thank the Horizon-AGN simulation team, esp. Yohan Dubois, Julien Devriendt and Christophe Pichon, for sharing the data and allowing us to use it in this project.

EG acknowledges support from the grant PID2021-123012NA-C44 funded by MCIN/AEI/ 10.13039/501100011033 and by “ERDF A way of making Europe”, and also from the European Union NextGenerationEU(PRTR-C17.I1) and by Generalitat de Catalunya.
EG and MM acknowledge support from the MINCINN
grant CNS2023-144328 (CARTEU), support from the CNS2023-144328 grant from MICIU/AEI /10.13039/501100011033 and from the European Union NextGenerationEU/PRTR

JC acknowledges support from the grant PID2021-123012NA-C44 funded by MCIN/AEI/ 10.13039/501100011033 and by “ERDF A way of making Europe”.

NEC and MvH acknowledge support from the project ``A rising tide: Galaxy intrinsic alignments as a new probe of cosmology and galaxy evolution'' (with project number VI.Vidi.203.011) of the Talent programme Vidi which is (partly) financed by the Dutch Research Council (NWO).

This work has made use of CosmoHub, developed by PIC (maintained by IFAE and CIEMAT) in collaboration with ICE-CSIC. It received funding from the Spanish government (grant EQC2021-007479-P funded by MCIN/AEI/10.13039/501100011033), the EU NextGeneration/PRTR (PRTR-C17.I1), and the Generalitat de Catalunya.

F.A. was supported by the European Research Council grant ERC4235. 

SR acknowledges support from project “Advanced Technologies for the exploration
of the Universe”, part of Complementary Plan ASTRO-1390 HEP, funded by
the European Union - Next Generation 
(MCIU/PRTR-C17.I1).

  \AckEC \AckCosmoHub
\end{acknowledgements}

\bibliography{bibliography}

\begin{appendix}
  
\section{Correlation function definitions}
\label{app:2pt}

To characterise the clustering in the mock catalogues, we use the projected two-point correlation function defined as
\begin{equation}
    w_\mathrm{p}(r_\mathrm{p}) = \int^\infty_{-\infty} \xi(r_\mathrm{p},\pi) \, \diff\pi \, ,
\end{equation}
where $r_\mathrm{p}$ and $\pi$ are the perpendicular and parallel separations between the galaxies, respectively, and $\xi(r_\mathrm{p},\pi)$ is the two-point correlation function. In practice, we integrate up to $\pi = 60 \, h^{-1} \, \mathrm{Mpc}$. For the calibration, we use the estimator defined as \citep{Peebles1974}
\begin{equation} \label{eq:peebles}
    \xi(r_\mathrm{p},\pi) = \frac{\mathrm{DD}(r_\mathrm{p},\pi)}{\mathrm{RR}(r_\mathrm{p},\pi)} - 1 \, ,
\end{equation}
where $\mathrm{DD}$ and $\mathrm{RR}$ are the normalised numbers of data-data and random-random pairs. We pre-compute the random number of pairs once using the observed $n(z)$ based on the LF used in \citetalias{Castander2024}. Errors are computed considering the diagonal of the covariance matrix obtained from 16 jackknife patches. This was chosen after evaluating the computational performance of the calibration and the stability of the errors. In a future work, we aim to assess the impact of incorporating the full covariance information into the optimisation.
It is well known that the usual \citet{Landy1993} estimator defined as
\begin{equation} \label{eq:landy}
    \xi(r_\mathrm{p},\pi) = \frac{\mathrm{DD}(r_\mathrm{p},\pi) - 2\,\mathrm{DR}(r_\mathrm{p},\pi) + \mathrm{RR}(r_\mathrm{p},\pi)}{\mathrm{RR}(r_\mathrm{p},\pi)}
\end{equation}
shows a better performance given that it optimally cancels shot noise and survey edge effects \citep{Kerscher2000}. We use the \citet{Peebles1974} estimator in the calibration process to speed up the computation time, while to verify the results obtained we implement the \citet{Landy1993} estimator.

\section{MCMC optimisation}
\label{sec:mcmc}
In this appendix we describe the tests performed to evaluate the performance of \texttt{SciPICal} and the results of the MCMC optimisation applied to the FS2-Wide and FS2-Deep halo catalogues. As described in Sect.~\ref{subsec:cal}, we implement a minimisation method and use its results to initialise the walkers. Details of the distributions adopted to generate the random values for the walkers are provided in Table~\ref{tab:initial_mcmc}.

\begin{table}
\captionsetup{font={small},labelfont=bf} 
\caption{Optimised parameters along with the adopted minimiser boundaries and initial values for the MCMC optimisation. The values $\mathcal{N}$ and $\mathcal{U}$ refer to normal and uniform distributions.}
    \centering
    \begin{tabular}{c c c }
    \hline
    \hline
    
Parameters & Boundaries & Initial value \\
           & minimiser  & distributions for MCMC \\[0.1cm]
\hline
\multicolumn{3}{c}{\textbf{HOD (Sect.~\ref{subsec:hod})}}\\[0.1cm]
$\alpha$ & $[0.8, 1.2]$ & $\mathcal{N}(\alpha_0; 0.05)$ \\
$f$      & $[13, 20]$   & $\mathcal{U}(f_0 \pm 2)$ \\
\multicolumn{3}{c}{\textbf{AM (Sect.~\ref{subsubsec:AM})}}\\[0.1cm]
$\sigma_L$ & $[0.05, 0.3]$ & $\mathcal{U}(\sigma_{L,0} \pm 0.05)$ \\
\multicolumn{3}{c}{\textbf{Colour class assignment (Sect.~\ref{subsubsec:colors})}}\\[0.1cm]
$M_r^\mathrm{red}$   & $[-23, -15]$ & $\mathcal{U}(M_{r,0}^\mathrm{red} \pm 1)$ \\
$M_r^\mathrm{green}$ & $[-23, -15]$ & $\mathcal{U}(M_{r,0}^\mathrm{green} \pm 1)$ \\
\multicolumn{3}{c}{\textbf{Satellite positions (Sect.~\ref{subsubsec:positions})}} \\[0.1cm]
$f^\mathrm{cut}_\mathrm{r}$ & $[0.5, 3.5]$ & $\mathcal{U}(f^\mathrm{cut}_{\mathrm{r},0} \pm 0.5)$ \\
\hline
\end{tabular}
\label{tab:initial_mcmc}
\end{table}

\subsection{Testing the calibration procedure}
\label{sec:test}
To validate the calibration procedure, we generate a mock catalogue with a set of parameters $\Theta_\mathrm{test} = \{ \alpha_\mathrm{test} = 0.85, f_\mathrm{test} = 16.5, \sigma_{\mathrm{L,test}} = 0.15, M_\mathrm{r,test}^\mathrm{red} = -17.5, M_\mathrm{r,test}^\mathrm{green} = -20.0 \}$. For this test, we do not consider $f^\mathrm{cut}_\mathrm{r}$ as a free parameter and we fix it at 3. The purpose of this exercise is to test if we can properly recover the set of parameters when running the calibration pipeline.

\begin{figure*}
    \centering
    \refstepcounter{figure}
    \label{fig:test_mcmc_pdf_hod_color}
    \begin{minipage}{0.62\textwidth}
        \includegraphics[width=\linewidth]{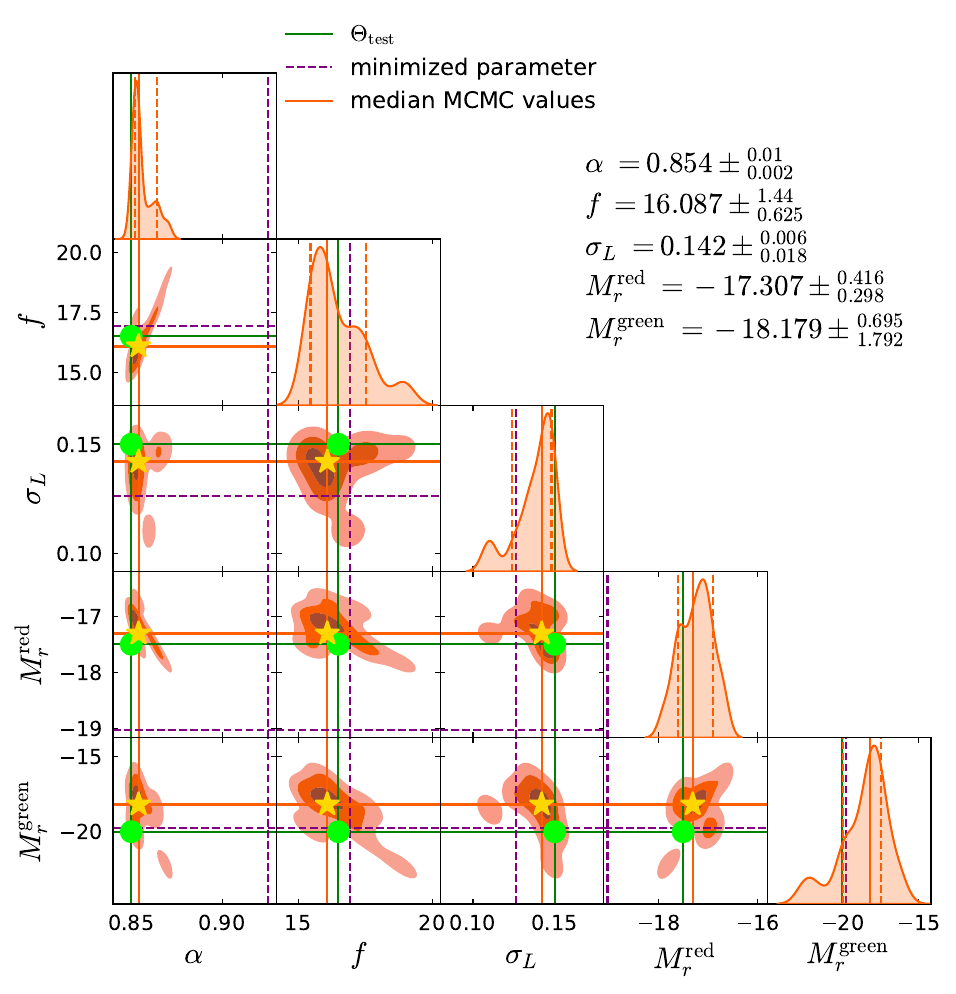}
    \end{minipage}
    \hfill
    \begin{minipage}{0.35\textwidth}
        \small\textbf{Fig. \thefigure.}Posterior distributions for the mock test fitting. The yellow stars and solid light-red lines indicate the median adopted parameters. The dashed red lines enclose 68\% of the distributions and define the adopted uncertainties. The dashed purple and solid green lines indicate the optimised parameters using the minimiser methodology and the parameters, $\Theta_\mathrm{test}$, adopted to generate the test mock catalogue from which the two-point clustering constraints are obtained. These last parameters are also indicated with a green dot.
    \end{minipage}
\end{figure*}

\begin{figure*}
    \captionsetup{font={small},labelfont=bf} 
    \centering
    \includegraphics[scale=0.7]{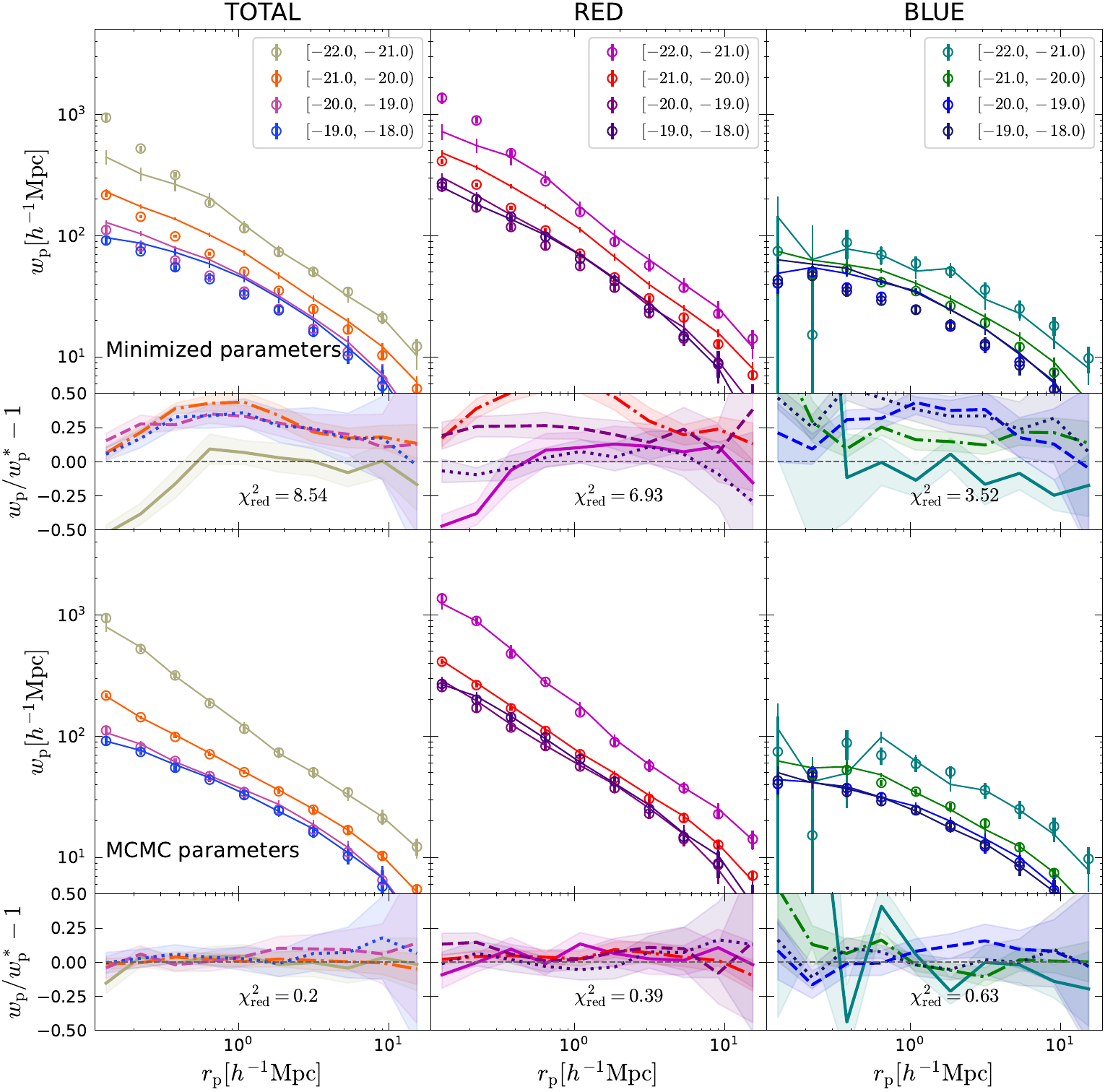}
    \caption{Two-point correlations obtained from mock test fitting for the galaxies in the mock split by luminosity, using the whole sample of galaxies (\textit{left panels}) and split according to colour (Eq.~\ref{eq:colour_cut}): red (\textit{middle panel}) and blue (\textit{right panel}). The dots indicate the measurements taken as reference for the optimisation, $w^*_\mathrm{p}$, obtained from a mock using the parameters $\Theta_\mathrm{test}$. The solid lines indicate measured clustering obtained for the mock generated with the parameters obtained using the minimiser method (\textit{upper panels}) and the optimised parameters using MCMC (\textit{bottom panel}). Below the correlations, the residuals are shown relative to the reference clustering together with the reduced chi-square values ($\chi^2_\mathrm{red}$). For clarity, the following line styles indicate luminosity bins: solid, dot-dashed, dashed, and dotted, from brightest to faintest.}
    \label{fig:test_corr}
\end{figure*}

With this aim, we apply the same calibration strategy as detailed in Sect.~\ref{subsec:cal}, by first running a minimiser method and then the MCMC optimisation. In this case, instead of the SDSS clustering measurements, we take as constraints the clustering measurements obtained for a mock generated adopting $\Theta_\mathrm{test}$. In Fig.~\ref{fig:test_mcmc_pdf_hod_color}, we show the posterior distributions of the optimised parameters. We obtain a substantial difference between the constrained parameters with the minimisation method and the parameters adopted to generate the constraints. This difference is mainly driven by the large mismatch between the initial value adopted for $\alpha$ and the centre of its allowed range, which is set around 1 (see Table~\ref{tab:initial_mcmc}). When the parameter space is relatively flat or contains multiple shallow minima, the minimiser can become trapped near the initial guess, preventing convergence to the true best-fit solution. We therefore conclude that a more extensive exploration of the parameter space is required to improve the robustness of the minimiser-based optimisation. On the other hand, the results provided by the MCMC optimisation guarantee that the optimised parameters are in agreement with the adopted values to generate the constraints and can properly recover the clustering measurements (see Fig.~\ref{fig:test_corr}). In this case, the set of parameters optimised is all in agreement within $1\sigma$, with a precision better than 10\%.

By applying this test, we are also checking that the calibration strategy is not substantially sensitive to the random process included in the mock generation. When generating a galaxy mock, stochastic effects introduced when assigning the luminosity, colour, and positions can influence the resulting clustering. Thus, the calibration can be affected by this, resulting in degeneracies in the parameters; that is, different sets of parameters can provide galaxy mocks that match the clustering constraints for particular realisations. This can affect the calibration process, misleading the optimisation. According to the results presented, the methodology applied is robust enough and provides posterior distributions that capture the different sets of parameters that provide mocks with clustering properties in agreement with the used constraints.

\subsection{MCMC posterior distributions}
\label{sec:mcmc_output}

\begin{figure*}
    \centering
    \refstepcounter{figure}
    \label{fig:fs2_mcmc_wide}
    \begin{minipage}{0.7\textwidth}
        \includegraphics[width=\linewidth]{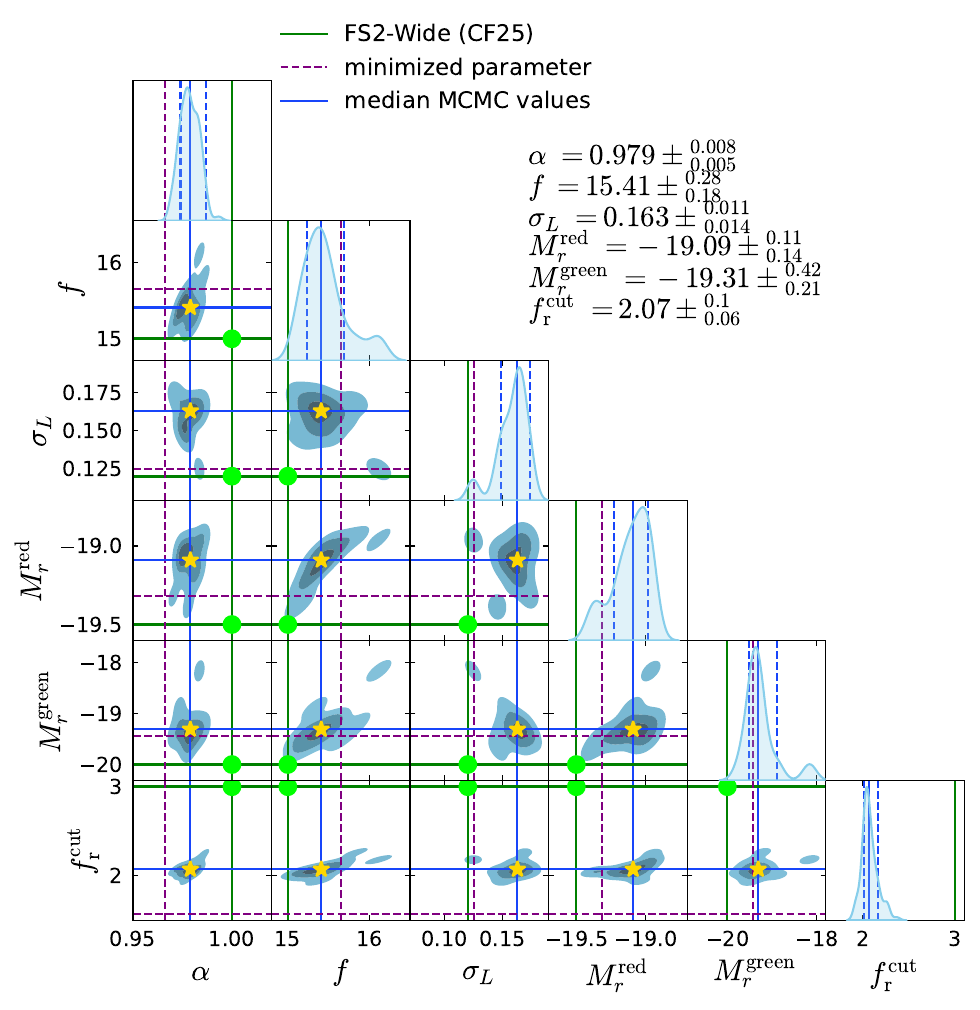}
    \end{minipage}
    \hfill
    \begin{minipage}{0.28\textwidth}
        \small\textbf{Fig. \thefigure.} Posterior distributions for the FS2-Wide fitting. The solid blue lines indicate the optimised median adopted parameters, which are also marked by the yellow star. The dashed blue lines enclose 68\% of the distributions and define the adopted uncertainties. The values obtained are included in the plot. The dashed purple and solid green lines respectively indicate the optimised parameters using the minimiser methodology and the adopted parameters by \citetalias{Castander2024}. These last parameters are also indicated by the green dot.
    \end{minipage}
\end{figure*}

\begin{figure*}
    \centering
    \refstepcounter{figure}
    \label{fig:fs2_mcmc_deep}
    \begin{minipage}{0.7\textwidth}
        \includegraphics[width=\linewidth]{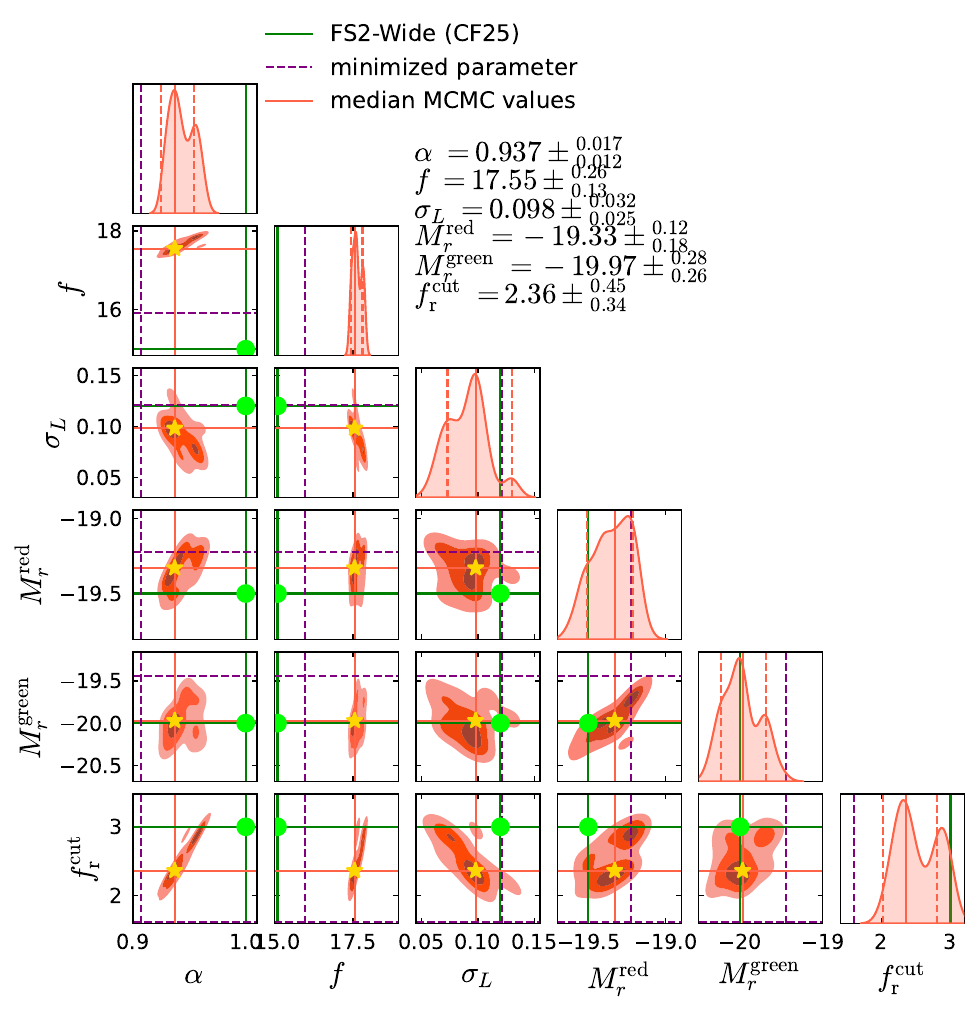}
    \end{minipage}
    \hfill
    \begin{minipage}{0.28\textwidth}
        \small\textbf{Fig. \thefigure.} Posterior distributions for the FS2-Deep fitting. The solid red lines indicate the optimised median adopted parameters, which are also marked by the yellow star. The dashed red lines enclose 68\% of the distributions and define the adopted uncertainties. The values obtained are included in the plot. The dashed purple and solid green lines respectively indicate the optimised parameters using the minimiser methodology and the adopted parameters by \citetalias{Castander2024}. These last parameters are also indicated by the green dot.
    \end{minipage}
\end{figure*}

In this section we show the results of the calibration procedure. Convergence was assessed via the stability of the $\chi^2$ values and parameter trace plots, which indicate that the walkers reached a stationary state after the 300-step burn-in period. This suggests that the walkers have successfully identified and constrained the global minimum. Although the resulting PDFs are likely under-sampled and should be interpreted with caution for formal error estimation, they serve as reliable indicators of the underlying physical degeneracies and correlations between the model parameters. We show in Figs.~\ref{fig:fs2_mcmc_wide} and ~\ref{fig:fs2_mcmc_deep} the obtained posterior distributions for the FS2-Wide and FS2-Deep halo catalogues, respectively. There are no considerable correlations among the calibrated parameters for FS2-Wide. On the other hand, we obtain a significant correlation between the HOD parameters, $\alpha$ and $f$, and $f^\mathrm{vir}_\mathrm{r}$. We also obtain wider posterior distributions for $\alpha$ and $f^\mathrm{vir}_\mathrm{r}$. The differences observed can be related to the lower uncertainties from the clustering measurements for the FS2-Deep mock, given that for this halo catalogue, we perform the calibration using the galaxies in a snapshot. This results in a larger number of galaxies and thus, in a higher constraining power. This reduction in the uncertainties is then translated to larger uncertainties in the optimised parameters.

\section{FS2-Deep halo catalogue characterisation}
\label{sec:deep_halo}

To generate FS2-Deep mock catalogues, we first characterise the HMFs in the light cone for the whole redshift range. We also correct the halo masses and concentrations for lower-mass haloes due to resolution effects. In this section we describe this procedure in order to obtain the characterised and corrected halo catalogue that will be taken as input by the \texttt{SciPIC} pipeline.

\subsection{Characterising the HMF}

\begin{figure*}
    \captionsetup{font={small},labelfont=bf}
    \centering
    \includegraphics[scale=0.8]{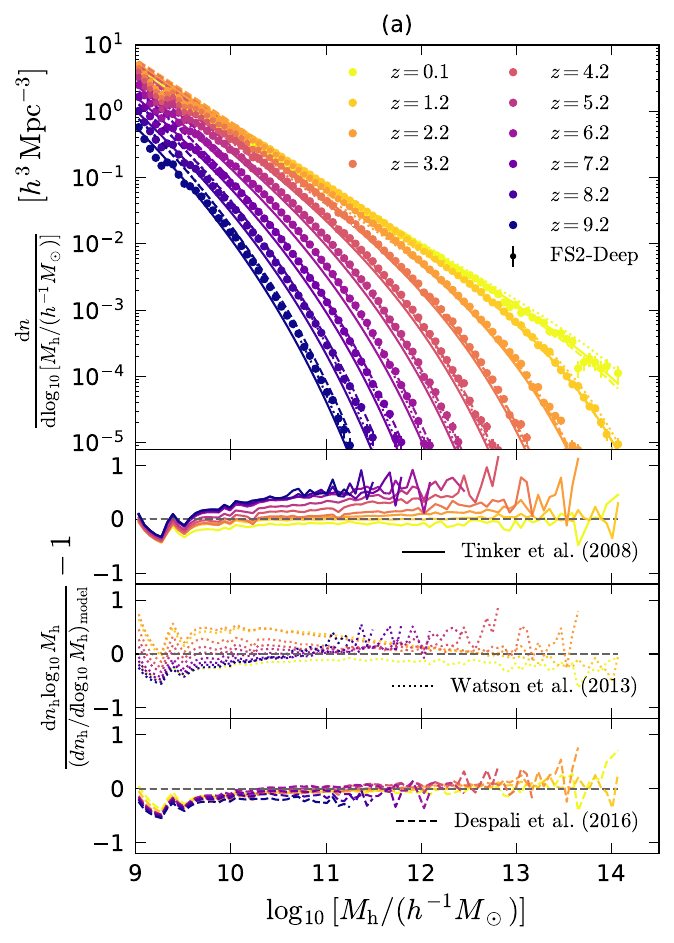}
    \includegraphics[scale=0.8]{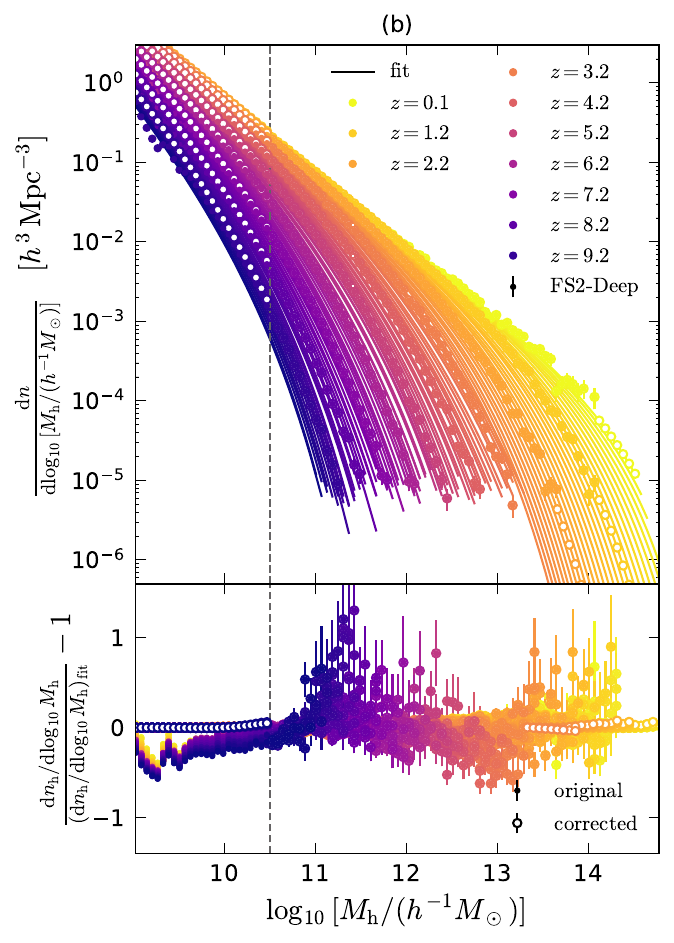}
    \caption{HMF characterisation of the FS2-Deep light cone. The colour-coding indicates the redshift bin. (a) Measured halo mass distributions compared with the \citet{Tinker2008}, \citet{Watson2013}, and \citet{Despali2016} measurements. (b) Fitted HMF for the FS2-Deep light cone in redshift bins. The open circles represent the corrected measurements used for the fit, considering D16 for the low-mass end (delimited by the vertical dashed line) and the measurements from FS2-Wide for the high-mass end.}
    \label{fig:HMF_comparison}
\end{figure*}

To constrain the HMF for the whole FS2-Deep light cone, we start by computing the number of haloes in 100 bins of halo mass and in comoving shells using equispaced redshift bins of size $\Delta z = 0.1$. This computation was performed interactively using the \texttt{CosmoHub} platform \citep{Tallada2024,Carretero:17}. Then, we compute the number density only considering the bins with more than ten haloes and compare it with different HMFs provided in the literature. The halo mass considered in this analysis is the virial mass computed by \texttt{ROCKSTAR} using the particles classified as bound. In the left panel of Fig.~\ref{fig:HMF_comparison}, we show the measurements for ten redshift bins and the comparison with three HMFs available in the \texttt{COLOSSUS} astrophysics toolkit \citep{Diemer2018}.\footnote{\href{https://bitbucket.org/bdiemer/colossus/src/master/}{https://bitbucket.org/bdiemer/colossus/src/master/}} In particular, we find a tighter agreement between our measurements and the \citet[][hereafter D16]{Despali2016} HMF.

Taking this into account, we correct the measurements for the low-mass end, $\logten(M_\mathrm{h}/h^{-1}\,M_{\odot}) < 10.5$, by considering the D16 HMF, after shifting this prediction according to the mean difference between the D16 and FS2-Deep HMF within a mass range of $10.5 < \logten(M_\mathrm{h}/h^{-1}\,M_{\odot}) < 12.0$. We also include FS2-Wide measurements for the redshift range $z < 3$ to complete the higher-mass end, given that this simulation is performed in a larger box; hence, it provides better statistical power for higher-mass haloes. Then, we model the measured distributions using a Schechter function. In the right panel of Fig.~\ref{fig:HMF_comparison}, we show the fit together with the residuals for the whole redshift range considered. After this, we model the Schechter function parameters as a function of redshift to obtain the $n_\text{h}(M_\mathrm{h}, z)$.

\subsection{Mass corrections}

Halo mass measurements at the low-mass end are affected by the particle mass resolution. Following the \citetalias{Castander2024} procedure, we perform the corrections to the halo masses for incompleteness and discreteness. To perform these corrections, we take advantage of the fact that the two-point correlation of haloes is approximately independent of mass at low halo masses. Then, the correlation properties of all haloes detected will not differ from those obtained if the catalogue had been complete. The procedure applied is the same as the one detailed in Sect.~4.2 in \citetalias{Castander2024}. Here, we provide a brief description of the application.

\begin{figure*}
    \captionsetup{font={small},labelfont=bf}
    \centering
    \begin{minipage}{.5\textwidth}
        \centering
        \includegraphics[scale=0.7]{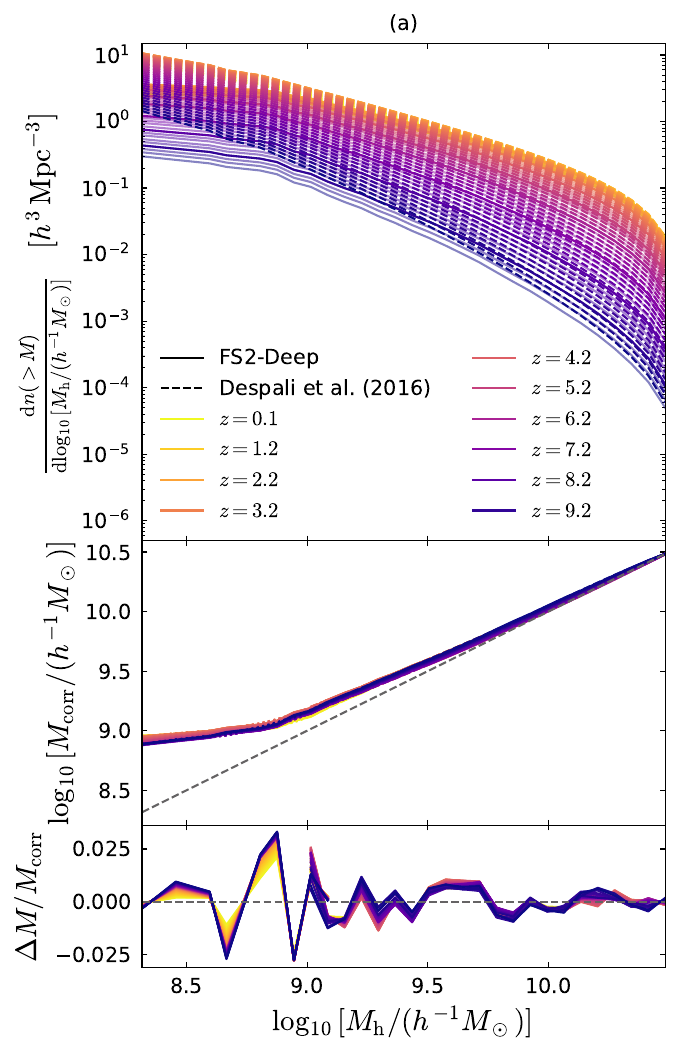}
    \end{minipage}%
    \begin{minipage}{0.5\textwidth}
        \centering
        \includegraphics[scale=0.75]{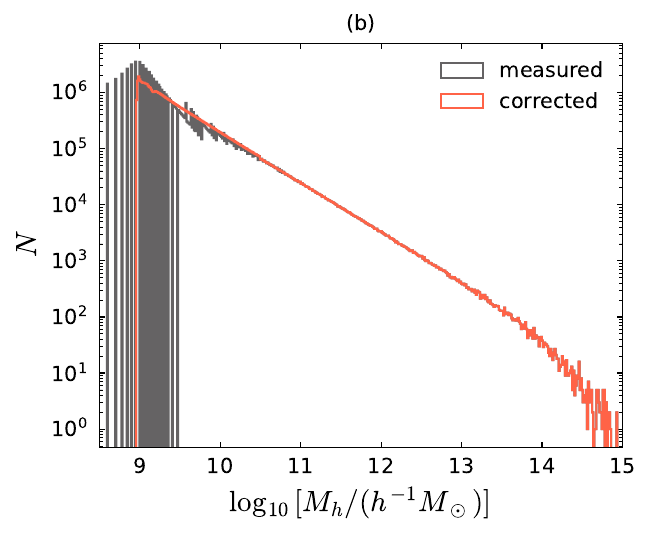}
        \includegraphics[scale=0.75]{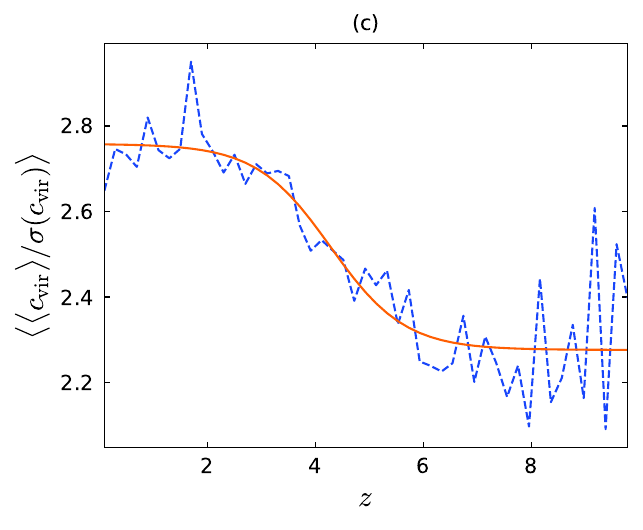}
    \end{minipage}
    \caption{(a) \textit{Upper panel}: Cumulative HMF for FS2-Deep and D16. \textit{Middle panel}: Fitted relation between the halo mass and the corrected mass computed using an AM performed between both cumulative relations, together with the fit. \textit{Bottom panel}: Residuals between the predicted corrected mass according to the fit and the corrected mass considering the AM relation. (b) Distribution of the virial halo masses computed using \texttt{ROCKSTAR} for the haloes considered in the calibration in a snapshot at $z=0$ (grey line), compared with the distribution after assigning a corrected mass, $M_\mathrm{h}$, which accounts for incompleteness and discreteness effects. (c) Mean of the ratio between the mean virial halo concentration and its standard deviation as a function of redshift (dashed blue line) together with the obtained fit (solid orange line).}
    \label{fig:HMF_corrections}
\end{figure*}

To correct the masses for incompleteness, we performed AM by comparing the cumulative HMF measured in the FS2-Deep to that computed from D16, after shifting it by the mean difference between both relations in the mass bin $10.5 < \logten(M_\mathrm{h}/h^{-1}\,M_{\odot}) < 12.0$. From this comparison, we reassigned the halo masses to match the slope expected by the D16 relation. To implement this correction in an efficient way, we fitted the relation as a function of halo mass and redshift to reassign the corrected mass for the haloes with masses $\logten(M_\mathrm{h}/h^{-1}\,M_{\odot}) < 10.5$. In the left panel of Fig.~\ref{fig:HMF_corrections}, we show the comparison between the cumulative HMF together with the fitted relations for the whole redshift range considered.

Finally, we corrected the masses for the discreteness effect by interpolating the halo abundance between $\logten(M_\mathrm{h}/h^{-1}\,M_{\odot})$ and $\logten[(M_\mathrm{h}+m_\mathrm{p})/h^{-1}\,M_{\odot}]$, where $m_\mathrm{p}$ is the particle mass resolution, for masses within that range. To perform this interpolation, we accounted for the D16 HMF slope in the considered mass bin. We show in the right panel of Fig.~\ref{fig:HMF_corrections} the result of applying both corrections to the distribution of the halo masses at $z=0$.

\begin{figure*}
    \captionsetup{font={small},labelfont=bf}
    \centering
    \includegraphics[scale=0.8]{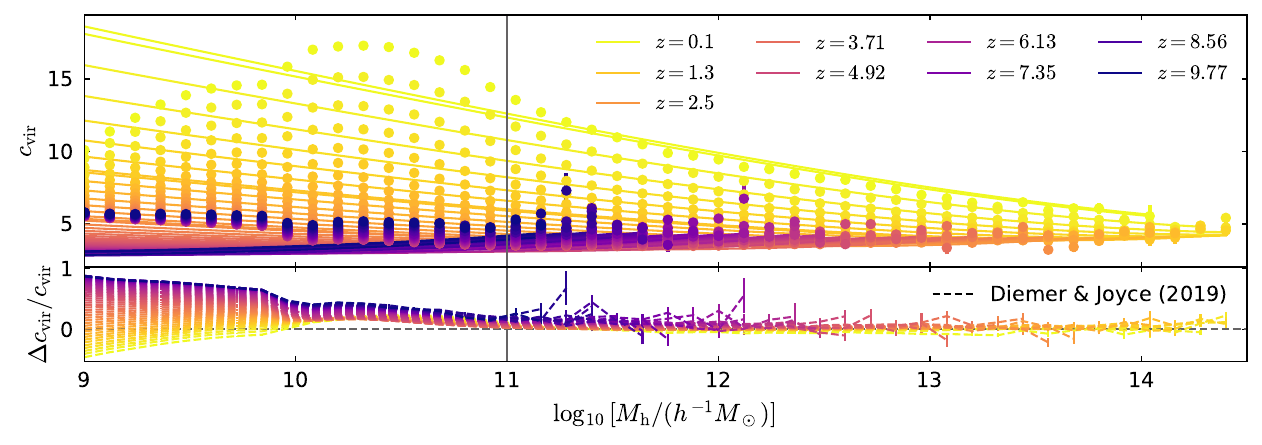}
    \caption{FS2-Deep mean halo concentrations measured in mass and redshift bins. The error bars indicate the error on the mean, computed according to the standard deviation and the number of haloes in each bin. The colour-coding indicates the redshift bin. The solid lines represent the concentration--redshift--mass relations given by \citet{Diemer2019}. \textit{Bottom panels}: Scaled differences between the \citet{Diemer2019} relation and the FS2-Deep measurements.}
    \label{fig:concentrations}
\end{figure*}

\subsection{Concentration correction}

As well as for the mass measurements, concentration estimates are affected by shot noise due to the low number of particles considered to derive this parameter \citep{Ramakrishnan2025}. Therefore, we correct the concentrations computed by \texttt{ROCKSTAR} applying a similar approach as the one described in \citetalias{Castander2024}. We first measure the mean and standard deviation of the concentrations in bins of halo mass and redshift. In Fig.~\ref{fig:concentrations} we show the comparison between our mean values and the relation provided by \citet{Diemer2019}. We obtain a good correspondence between the two for haloes with $\logten(M_\mathrm{h}/h^{-1}\,\mathrm{M}_{\odot}) \gtrsim 11$, which corresponds to haloes with more than 1000 particles.

In the bottom right panel of Fig.~\ref{fig:HMF_corrections}, we show the mean of the ratio between the mean concentration values and the standard deviation in redshift bins. We fitted a sigmoid function to this relation to predict the standard deviation at the whole redshift range. Then, we assign a new concentration value for the haloes with $\logten(M_\mathrm{h}/h^{-1}\,M_{\odot}) < 11$ considering a Gaussian distribution with mean given by the \citet{Diemer2019} relation and with a standard deviation computed according to the sigmoid fit.

\section{Abundance matching fit for Flagship mocks}
\label{sec:AM_deep}

\begin{figure*}
    \captionsetup{font={small},labelfont=bf}
    \centering
    \includegraphics[scale=0.8]{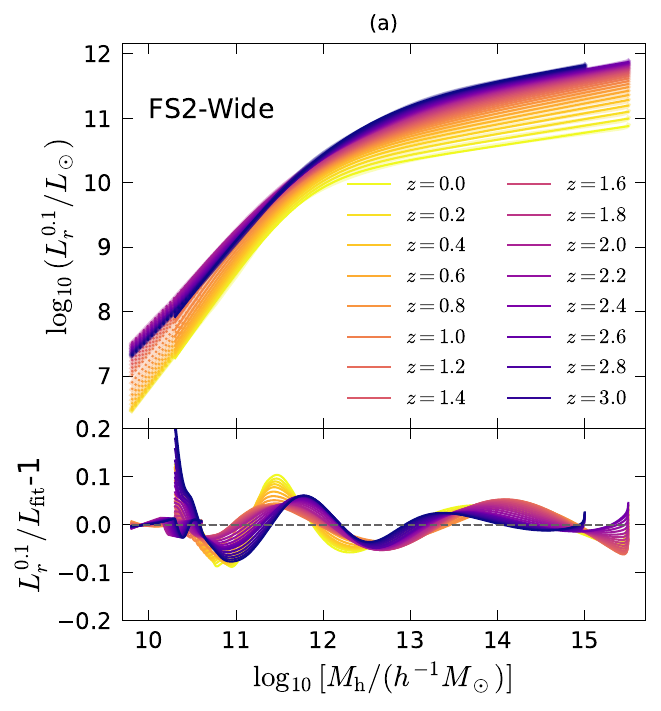}
    \includegraphics[scale=0.8]{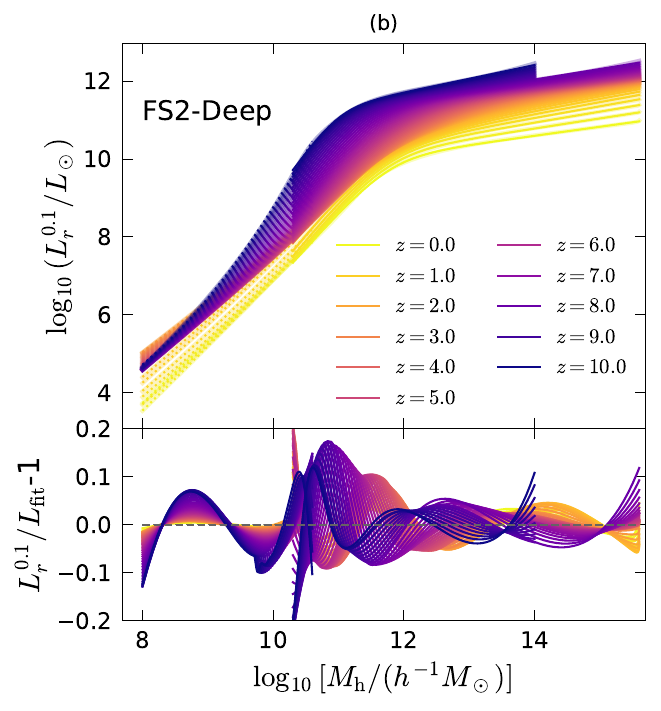}
    \caption{Relations between halo mass and the unscattered luminosity of central galaxies for the FS2-Wide (a) and the FS2-Deep (b). The solid lines correspond to the fitted relation (Eq.~\ref{eq:AMfit}), and the dotted lines correspond to the second-order polynomial fit. The residuals are shown in the \textit{bottom panels}.}
    \label{fig:AM}
\end{figure*}

To assign luminosities to central galaxies, we apply an AM procedure described in Sect.~\ref{subsubsec:AM}, comparing the cumulative galaxy number density function, $n_\text{h}(> M_\mathrm{h})$ defined in Eq.~(\ref{eq:cum_gal}), with the unscattered cumulative LF, $\Phi_\mathrm{unscat}(>L)$. By matching these functions, we can compute a relation between the halo mass and the luminosity that will be assigned to the central galaxy without any scatter. This can be performed for the whole redshift range considered for the mock production. We compute $n_\text{h}(> M_\mathrm{h})$ using the optimised parameters, $\alpha$ and $f$, and $\Phi_\mathrm{unscat}(>L)$ taking into account $\sigma_L$. We fitted these relations considering two different functional forms:
\begin{equation}
\label{eq:AMfit}
    \left(\frac{L^{0.1}_\mathrm{r}}{L_0}\right)^{1/\gamma} = \left(\frac{M_\mathrm{h}}{M_0}\right)^{\beta/\gamma} + \left(\frac{M_\mathrm{h}}{M_0}\right)^{\delta/\gamma},\text{for}\,\, M_\mathrm{h}/(h^{-1}\,M_\odot) > 10^{10},
\end{equation}
and a second-order polynomial function in logarithmic space for $M_\mathrm{h}/(h^{-1}\,M_\odot) < 10^{10}$. LFs up to redshift 10 were obtained by mimicking the number density distribution observed in COSMOS2020~\citep{Ramakrishnan2025b}. By fitting these relations for the whole redshift range considered, we build a grid of parameters to be ingested in \texttt{SciPIC}. In Fig.~\ref{fig:AM}, we show the relation obtained between the halo mass and the central galaxy luminosity for FS2-Wide and FS2-Deep, together with the fits and the residuals obtained from the fit function.

\section{VIPERS comparison using stellar mass bins}
\label{sec:vipers_bins}

\begin{figure*}
    
    \captionsetup{font={small},labelfont=bf} 
    \centering
    \includegraphics[scale=0.7]{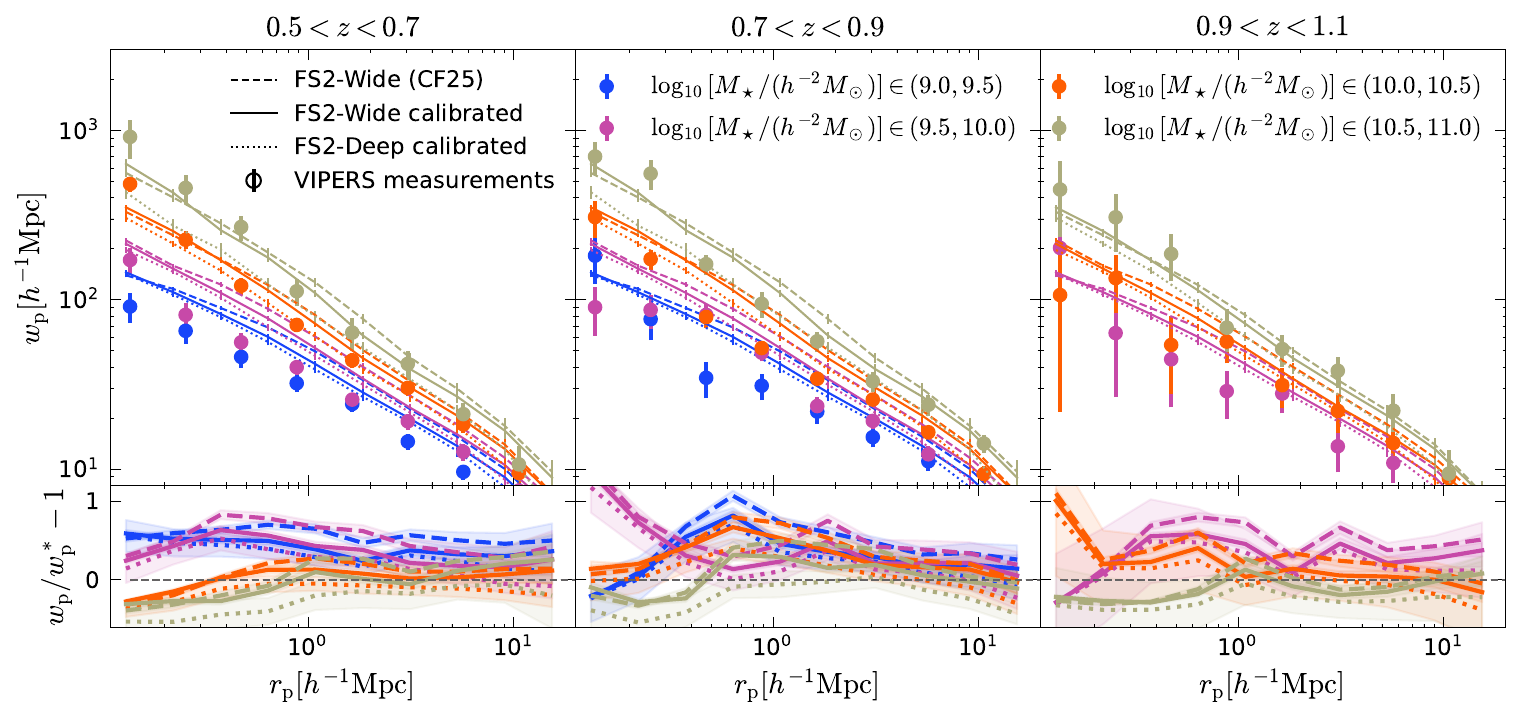}
    \caption{Comparison of the projected two-point correlations derived from the VIPERS galaxy samples in redshift bins, considering stellar mass cuts. Below the correlations, the residuals are shown relative to the VIPERS clustering, with the shaded region indicating the uncertainties of the VIPERS measurements.}
    \label{fig:corr_vipers_lM_bins}
\end{figure*}

In Fig.~\ref{fig:corr_vipers_lM_bins} we show the comparison between VIPERS clustering measurements, FS2-Wide (\citetalias{Castander2024}), and the mocks produced with calibrated parameters in this work. Overall, the clustering measurements are consistent within the uncertainties. However, we obtain significant differences for the lowest mass bins. These differences can be related to selection effects introduced in the galaxy selection of the VIPERS samples. In particular, the flux cut applied ($i_\mathrm{AB} < 22.5$) results in partial incompleteness in the mass-selected galaxy samples, preferentially excluding high mass-to-light-ratio galaxies. According to semi-analytic models, these discarded galaxies are mainly faint and red galaxies located in overdense regions. Hence, their removal suppresses the observed clustering signal, especially on small scales. We also observe that the discrepancies are reduced when using the calibrated versions. Therefore, we do not discard the possibility that these differences could be driven by our strategy adopted to populate the haloes.

\end{appendix}

\end{document}

%% file: authors.tex
\newcommand{\orcid}[1]{} 
\author{Euclid Collaboration: E.~J.~Gonzalez\orcid{0000-0002-0226-9893}\thanks{\email{egonzalez@pic.es}}\inst{\ref{aff1},\ref{aff2},\ref{aff3}}
\and J.~Carretero\orcid{0000-0002-3130-0204}\inst{\ref{aff4},\ref{aff5}}
\and Z.~Baghkhani\orcid{0000-0002-6632-2614}\inst{\ref{aff1},\ref{aff2}}
\and F.~J.~Castander\orcid{0000-0001-7316-4573}\inst{\ref{aff1},\ref{aff2}}
\and P.~Fosalba\orcid{0000-0002-1510-5214}\inst{\ref{aff2},\ref{aff1}}
\and P.~Tallada-Cresp\'{i}\orcid{0000-0002-1336-8328}\inst{\ref{aff4},\ref{aff5}}
\and J.~Stadel\orcid{0000-0001-7565-8622}\inst{\ref{aff6}}
\and D.~Potter\orcid{0000-0002-0757-5195}\inst{\ref{aff6}}
\and I.~Tutusaus\orcid{0000-0002-3199-0399}\inst{\ref{aff1},\ref{aff2},\ref{aff7}}
\and S.~Ramakrishnan\orcid{0000-0003-1113-1834}\inst{\ref{aff1}}
\and M.~L.~van~Heukelum\orcid{0009-0008-3780-1617}\inst{\ref{aff8}}
\and N.~E.~Chisari\orcid{0000-0003-4221-6718}\inst{\ref{aff8},\ref{aff9}}
\and F.~Marulli\orcid{0000-0002-8850-0303}\inst{\ref{aff10},\ref{aff11},\ref{aff12}}
\and M.~Bolzonella\orcid{0000-0003-3278-4607}\inst{\ref{aff11}}
\and L.~Pozzetti\orcid{0000-0001-7085-0412}\inst{\ref{aff11}}
\and D.~Navarro-Giron\'{e}s\orcid{0000-0003-0507-372X}\inst{\ref{aff9}}
\and J.~Chaves-Montero\orcid{0000-0002-9553-4261}\inst{\ref{aff13}}
\and G.~Parimbelli\orcid{0000-0002-2539-2472}\inst{\ref{aff1},\ref{aff14}}
\and M.~Manera\orcid{0000-0003-4962-8934}\inst{\ref{aff15},\ref{aff13}}
\and L.~Blot\orcid{0000-0002-9622-7167}\inst{\ref{aff16},\ref{aff17}}
\and K.~Hoffmann\orcid{0000-0002-7885-5274}\inst{\ref{aff1}}
\and M.~Huertas-Company\orcid{0000-0002-1416-8483}\inst{\ref{aff18},\ref{aff19},\ref{aff20}}
\and P.~Monaco\orcid{0000-0003-2083-7564}\inst{\ref{aff21},\ref{aff22},\ref{aff23},\ref{aff24}}
\and C.~Scarlata\orcid{0000-0002-9136-8876}\inst{\ref{aff25}}
\and M.-A.~Breton\inst{\ref{aff26}}
\and S.-S.~Li\orcid{0000-0001-9952-7408}\inst{\ref{aff27},\ref{aff28}}
\and R.~Teyssier\orcid{0000-0001-7689-0933}\inst{\ref{aff29}}
\and M.~Crocce\orcid{0000-0002-9745-6228}\inst{\ref{aff1},\ref{aff2}}
\and G.~Congedo\orcid{0000-0003-2508-0046}\inst{\ref{aff30}}
\and A.~Biviano\orcid{0000-0002-0857-0732}\inst{\ref{aff22},\ref{aff24}}
\and M.~Hirschmann\orcid{0000-0002-3301-3321}\inst{\ref{aff31}}
\and A.~Pezzotta\orcid{0000-0003-0726-2268}\inst{\ref{aff32}}
\and H.~Hoekstra\orcid{0000-0002-0641-3231}\inst{\ref{aff9}}
\and W.~J.~Percival\orcid{0000-0002-0644-5727}\inst{\ref{aff33},\ref{aff34},\ref{aff35}}
\and P.~A.~Oesch\orcid{0000-0001-5851-6649}\inst{\ref{aff36},\ref{aff37},\ref{aff38}}
\and R.~A.~A.~Bowler\orcid{0000-0003-3917-1678}\inst{\ref{aff39}}
\and V.~Gonzalez-Perez\orcid{0000-0001-9938-2755}\inst{\ref{aff40},\ref{aff41}}
\and S.~Avila\orcid{0000-0001-5043-3662}\inst{\ref{aff4}}
\and A.~Kov\'acs\orcid{0000-0002-5825-579X}\inst{\ref{aff42},\ref{aff43}}
\and B.~Altieri\orcid{0000-0003-3936-0284}\inst{\ref{aff44}}
\and S.~Andreon\orcid{0000-0002-2041-8784}\inst{\ref{aff32}}
\and N.~Auricchio\orcid{0000-0003-4444-8651}\inst{\ref{aff11}}
\and C.~Baccigalupi\orcid{0000-0002-8211-1630}\inst{\ref{aff24},\ref{aff22},\ref{aff23},\ref{aff14}}
\and M.~Baldi\orcid{0000-0003-4145-1943}\inst{\ref{aff45},\ref{aff11},\ref{aff12}}
\and S.~Bardelli\orcid{0000-0002-8900-0298}\inst{\ref{aff11}}
\and P.~Battaglia\orcid{0000-0002-7337-5909}\inst{\ref{aff11}}
\and E.~Branchini\orcid{0000-0002-0808-6908}\inst{\ref{aff46},\ref{aff47},\ref{aff32}}
\and M.~Brescia\orcid{0000-0001-9506-5680}\inst{\ref{aff48},\ref{aff49}}
\and S.~Camera\orcid{0000-0003-3399-3574}\inst{\ref{aff50},\ref{aff51},\ref{aff52}}
\and V.~Capobianco\orcid{0000-0002-3309-7692}\inst{\ref{aff52}}
\and C.~Carbone\orcid{0000-0003-0125-3563}\inst{\ref{aff53}}
\and S.~Casas\orcid{0000-0002-4751-5138}\inst{\ref{aff54},\ref{aff55}}
\and M.~Castellano\orcid{0000-0001-9875-8263}\inst{\ref{aff56}}
\and G.~Castignani\orcid{0000-0001-6831-0687}\inst{\ref{aff11}}
\and S.~Cavuoti\orcid{0000-0002-3787-4196}\inst{\ref{aff49},\ref{aff57}}
\and A.~Cimatti\inst{\ref{aff58}}
\and C.~Colodro-Conde\inst{\ref{aff18}}
\and L.~Conversi\orcid{0000-0002-6710-8476}\inst{\ref{aff59},\ref{aff44}}
\and Y.~Copin\orcid{0000-0002-5317-7518}\inst{\ref{aff60}}
\and F.~Courbin\orcid{0000-0003-0758-6510}\inst{\ref{aff61},\ref{aff62},\ref{aff63}}
\and H.~M.~Courtois\orcid{0000-0003-0509-1776}\inst{\ref{aff64}}
\and H.~Degaudenzi\orcid{0000-0002-5887-6799}\inst{\ref{aff36}}
\and S.~de~la~Torre\inst{\ref{aff65}}
\and G.~De~Lucia\orcid{0000-0002-6220-9104}\inst{\ref{aff22}}
\and H.~Dole\orcid{0000-0002-9767-3839}\inst{\ref{aff66}}
\and F.~Dubath\orcid{0000-0002-6533-2810}\inst{\ref{aff36}}
\and X.~Dupac\inst{\ref{aff44}}
\and S.~Escoffier\orcid{0000-0002-2847-7498}\inst{\ref{aff67}}
\and M.~Farina\orcid{0000-0002-3089-7846}\inst{\ref{aff68}}
\and R.~Farinelli\inst{\ref{aff11}}
\and S.~Farrens\orcid{0000-0002-9594-9387}\inst{\ref{aff26}}
\and F.~Faustini\orcid{0000-0001-6274-5145}\inst{\ref{aff56},\ref{aff69}}
\and S.~Ferriol\inst{\ref{aff60}}
\and F.~Finelli\orcid{0000-0002-6694-3269}\inst{\ref{aff11},\ref{aff70}}
\and S.~Fotopoulou\orcid{0000-0002-9686-254X}\inst{\ref{aff71}}
\and N.~Fourmanoit\orcid{0009-0005-6816-6925}\inst{\ref{aff67}}
\and M.~Frailis\orcid{0000-0002-7400-2135}\inst{\ref{aff22}}
\and E.~Franceschi\orcid{0000-0002-0585-6591}\inst{\ref{aff11}}
\and M.~Fumana\orcid{0000-0001-6787-5950}\inst{\ref{aff53}}
\and L.~Gabarra\orcid{0000-0002-8486-8856}\inst{\ref{aff72}}
\and S.~Galeotta\orcid{0000-0002-3748-5115}\inst{\ref{aff22}}
\and K.~George\orcid{0000-0002-1734-8455}\inst{\ref{aff73}}
\and B.~Gillis\orcid{0000-0002-4478-1270}\inst{\ref{aff30}}
\and C.~Giocoli\orcid{0000-0002-9590-7961}\inst{\ref{aff11},\ref{aff12}}
\and J.~Gracia-Carpio\orcid{0000-0003-4689-3134}\inst{\ref{aff74}}
\and A.~Grazian\orcid{0000-0002-5688-0663}\inst{\ref{aff75}}
\and F.~Grupp\inst{\ref{aff74},\ref{aff76}}
\and L.~Guzzo\orcid{0000-0001-8264-5192}\inst{\ref{aff77},\ref{aff32},\ref{aff78}}
\and S.~V.~H.~Haugan\orcid{0000-0001-9648-7260}\inst{\ref{aff79}}
\and W.~Holmes\orcid{0009-0007-8554-4646}\inst{\ref{aff80}}
\and F.~Hormuth\inst{\ref{aff81}}
\and A.~Hornstrup\orcid{0000-0002-3363-0936}\inst{\ref{aff82},\ref{aff83}}
\and K.~Jahnke\orcid{0000-0003-3804-2137}\inst{\ref{aff84}}
\and M.~Jhabvala\inst{\ref{aff85}}
\and B.~Joachimi\orcid{0000-0001-7494-1303}\inst{\ref{aff86}}
\and S.~Kermiche\orcid{0000-0002-0302-5735}\inst{\ref{aff67}}
\and A.~Kiessling\orcid{0000-0002-2590-1273}\inst{\ref{aff80}}
\and B.~Kubik\orcid{0009-0006-5823-4880}\inst{\ref{aff60}}
\and M.~K\"ummel\orcid{0000-0003-2791-2117}\inst{\ref{aff76}}
\and M.~Kunz\orcid{0000-0002-3052-7394}\inst{\ref{aff87}}
\and H.~Kurki-Suonio\orcid{0000-0002-4618-3063}\inst{\ref{aff88},\ref{aff89}}
\and A.~M.~C.~Le~Brun\orcid{0000-0002-0936-4594}\inst{\ref{aff17}}
\and S.~Ligori\orcid{0000-0003-4172-4606}\inst{\ref{aff52}}
\and P.~B.~Lilje\orcid{0000-0003-4324-7794}\inst{\ref{aff79}}
\and V.~Lindholm\orcid{0000-0003-2317-5471}\inst{\ref{aff88},\ref{aff89}}
\and I.~Lloro\orcid{0000-0001-5966-1434}\inst{\ref{aff90}}
\and M.~Magliocchetti\orcid{0000-0001-9158-4838}\inst{\ref{aff68}}
\and G.~Mainetti\orcid{0000-0003-2384-2377}\inst{\ref{aff91}}
\and O.~Mansutti\orcid{0000-0001-5758-4658}\inst{\ref{aff22}}
\and O.~Marggraf\orcid{0000-0001-7242-3852}\inst{\ref{aff92}}
\and M.~Martinelli\orcid{0000-0002-6943-7732}\inst{\ref{aff56},\ref{aff93}}
\and N.~Martinet\orcid{0000-0003-2786-7790}\inst{\ref{aff65}}
\and R.~J.~Massey\orcid{0000-0002-6085-3780}\inst{\ref{aff94}}
\and N.~Mauri\orcid{0000-0001-8196-1548}\inst{\ref{aff58},\ref{aff12}}
\and E.~Medinaceli\orcid{0000-0002-4040-7783}\inst{\ref{aff11}}
\and S.~Mei\orcid{0000-0002-2849-559X}\inst{\ref{aff95},\ref{aff96}}
\and M.~Meneghetti\orcid{0000-0003-1225-7084}\inst{\ref{aff11},\ref{aff12}}
\and E.~Merlin\orcid{0000-0001-6870-8900}\inst{\ref{aff56}}
\and G.~Meylan\inst{\ref{aff97}}
\and A.~Mora\orcid{0000-0002-1922-8529}\inst{\ref{aff98}}
\and M.~Moresco\orcid{0000-0002-7616-7136}\inst{\ref{aff10},\ref{aff11}}
\and C.~Moretti\orcid{0000-0003-3314-8936}\inst{\ref{aff22},\ref{aff24},\ref{aff23}}
\and L.~Moscardini\orcid{0000-0002-3473-6716}\inst{\ref{aff10},\ref{aff11},\ref{aff12}}
\and R.~Nakajima\orcid{0009-0009-1213-7040}\inst{\ref{aff92}}
\and C.~Neissner\orcid{0000-0001-8524-4968}\inst{\ref{aff13},\ref{aff5}}
\and S.-M.~Niemi\orcid{0009-0005-0247-0086}\inst{\ref{aff99}}
\and J.~W.~Nightingale\orcid{0000-0002-8987-7401}\inst{\ref{aff100}}
\and C.~Padilla\orcid{0000-0001-7951-0166}\inst{\ref{aff13}}
\and S.~Paltani\orcid{0000-0002-8108-9179}\inst{\ref{aff36}}
\and F.~Pasian\orcid{0000-0002-4869-3227}\inst{\ref{aff22}}
\and K.~Pedersen\inst{\ref{aff101}}
\and V.~Pettorino\orcid{0000-0002-4203-9320}\inst{\ref{aff99}}
\and S.~Pires\orcid{0000-0002-0249-2104}\inst{\ref{aff26}}
\and G.~Polenta\orcid{0000-0003-4067-9196}\inst{\ref{aff69}}
\and L.~A.~Popa\inst{\ref{aff102}}
\and F.~Raison\orcid{0000-0002-7819-6918}\inst{\ref{aff74}}
\and A.~Renzi\orcid{0000-0001-9856-1970}\inst{\ref{aff103},\ref{aff104},\ref{aff11}}
\and J.~Rhodes\orcid{0000-0002-4485-8549}\inst{\ref{aff80}}
\and G.~Riccio\inst{\ref{aff49}}
\and E.~Romelli\orcid{0000-0003-3069-9222}\inst{\ref{aff22}}
\and M.~Roncarelli\orcid{0000-0001-9587-7822}\inst{\ref{aff11}}
\and R.~Saglia\orcid{0000-0003-0378-7032}\inst{\ref{aff76},\ref{aff74}}
\and Z.~Sakr\orcid{0000-0002-4823-3757}\inst{\ref{aff105},\ref{aff7},\ref{aff106}}
\and A.~G.~S\'anchez\orcid{0000-0003-1198-831X}\inst{\ref{aff74}}
\and D.~Sapone\orcid{0000-0001-7089-4503}\inst{\ref{aff107}}
\and B.~Sartoris\orcid{0000-0003-1337-5269}\inst{\ref{aff76},\ref{aff22}}
\and P.~Schneider\orcid{0000-0001-8561-2679}\inst{\ref{aff92}}
\and T.~Schrabback\orcid{0000-0002-6987-7834}\inst{\ref{aff108}}
\and A.~Secroun\orcid{0000-0003-0505-3710}\inst{\ref{aff67}}
\and S.~Serrano\orcid{0000-0002-0211-2861}\inst{\ref{aff2},\ref{aff109},\ref{aff1}}
\and E.~Sihvola\orcid{0000-0003-1804-7715}\inst{\ref{aff110}}
\and P.~Simon\inst{\ref{aff92}}
\and C.~Sirignano\orcid{0000-0002-0995-7146}\inst{\ref{aff103},\ref{aff104}}
\and G.~Sirri\orcid{0000-0003-2626-2853}\inst{\ref{aff12}}
\and A.~Spurio~Mancini\orcid{0000-0001-5698-0990}\inst{\ref{aff111}}
\and L.~Stanco\orcid{0000-0002-9706-5104}\inst{\ref{aff104}}
\and A.~N.~Taylor\inst{\ref{aff30}}
\and I.~Tereno\orcid{0000-0002-4537-6218}\inst{\ref{aff112},\ref{aff113}}
\and S.~Toft\orcid{0000-0003-3631-7176}\inst{\ref{aff37},\ref{aff38}}
\and R.~Toledo-Moreo\orcid{0000-0002-2997-4859}\inst{\ref{aff114}}
\and F.~Torradeflot\orcid{0000-0003-1160-1517}\inst{\ref{aff5},\ref{aff4}}
\and J.~Valiviita\orcid{0000-0001-6225-3693}\inst{\ref{aff88},\ref{aff89}}
\and T.~Vassallo\orcid{0000-0001-6512-6358}\inst{\ref{aff22},\ref{aff73}}
\and G.~Verdoes~Kleijn\orcid{0000-0001-5803-2580}\inst{\ref{aff115}}
\and Y.~Wang\orcid{0000-0002-4749-2984}\inst{\ref{aff116}}
\and J.~Weller\orcid{0000-0002-8282-2010}\inst{\ref{aff76},\ref{aff74}}
\and G.~Zamorani\orcid{0000-0002-2318-301X}\inst{\ref{aff11}}
\and F.~M.~Zerbi\orcid{0000-0002-9996-973X}\inst{\ref{aff32}}
\and E.~Zucca\orcid{0000-0002-5845-8132}\inst{\ref{aff11}}
\and V.~Allevato\orcid{0000-0001-7232-5152}\inst{\ref{aff49}}
\and M.~Ballardini\orcid{0000-0003-4481-3559}\inst{\ref{aff117},\ref{aff118},\ref{aff11}}
\and C.~Burigana\orcid{0000-0002-3005-5796}\inst{\ref{aff119},\ref{aff70}}
\and R.~Cabanac\orcid{0000-0001-6679-2600}\inst{\ref{aff7}}
\and M.~Calabrese\orcid{0000-0002-2637-2422}\inst{\ref{aff120},\ref{aff53}}
\and A.~Cappi\inst{\ref{aff121},\ref{aff11}}
\and T.~Castro\orcid{0000-0002-6292-3228}\inst{\ref{aff22},\ref{aff23},\ref{aff24},\ref{aff122}}
\and J.~A.~Escartin~Vigo\inst{\ref{aff74}}
\and J.~Garc\'ia-Bellido\orcid{0000-0002-9370-8360}\inst{\ref{aff105}}
\and J.~Macias-Perez\orcid{0000-0002-5385-2763}\inst{\ref{aff123}}
\and R.~Maoli\orcid{0000-0002-6065-3025}\inst{\ref{aff124},\ref{aff56}}
\and J.~Mart\'{i}n-Fleitas\orcid{0000-0002-8594-569X}\inst{\ref{aff125}}
\and R.~B.~Metcalf\orcid{0000-0003-3167-2574}\inst{\ref{aff10},\ref{aff11}}
\and A.~Montoro\orcid{0000-0003-4730-8590}\inst{\ref{aff1},\ref{aff2}}
\and A.~A.~Nucita\inst{\ref{aff126},\ref{aff127},\ref{aff128}}
\and M.~P\"ontinen\orcid{0000-0001-5442-2530}\inst{\ref{aff88}}
\and V.~Scottez\orcid{0009-0008-3864-940X}\inst{\ref{aff129},\ref{aff130}}
\and M.~Sereno\orcid{0000-0003-0302-0325}\inst{\ref{aff11},\ref{aff12}}
\and M.~Tenti\orcid{0000-0002-4254-5901}\inst{\ref{aff12}}
\and M.~Tucci\inst{\ref{aff36}}
\and M.~Viel\orcid{0000-0002-2642-5707}\inst{\ref{aff24},\ref{aff22},\ref{aff14},\ref{aff23},\ref{aff122}}
\and M.~Wiesmann\orcid{0009-0000-8199-5860}\inst{\ref{aff79}}
\and Y.~Akrami\orcid{0000-0002-2407-7956}\inst{\ref{aff105},\ref{aff131}}
\and I.~T.~Andika\orcid{0000-0001-6102-9526}\inst{\ref{aff76}}
\and G.~Angora\orcid{0000-0002-0316-6562}\inst{\ref{aff49},\ref{aff117}}
\and S.~Anselmi\orcid{0000-0002-3579-9583}\inst{\ref{aff104},\ref{aff103},\ref{aff132}}
\and M.~Archidiacono\orcid{0000-0003-4952-9012}\inst{\ref{aff77},\ref{aff78}}
\and F.~Atrio-Barandela\orcid{0000-0002-2130-2513}\inst{\ref{aff133}}
\and L.~Bazzanini\orcid{0000-0003-0727-0137}\inst{\ref{aff117},\ref{aff11}}
\and J.~Bel\inst{\ref{aff134}}
\and D.~Bertacca\orcid{0000-0002-2490-7139}\inst{\ref{aff103},\ref{aff75},\ref{aff104}}
\and M.~Bethermin\orcid{0000-0002-3915-2015}\inst{\ref{aff135}}
\and F.~Beutler\orcid{0000-0003-0467-5438}\inst{\ref{aff30}}
\and A.~Blanchard\orcid{0000-0001-8555-9003}\inst{\ref{aff7}}
\and M.~Bonici\orcid{0000-0002-8430-126X}\inst{\ref{aff33},\ref{aff53}}
\and M.~L.~Brown\orcid{0000-0002-0370-8077}\inst{\ref{aff39}}
\and S.~Bruton\orcid{0000-0002-6503-5218}\inst{\ref{aff136}}
\and A.~Calabro\orcid{0000-0003-2536-1614}\inst{\ref{aff56}}
\and B.~Camacho~Quevedo\orcid{0000-0002-8789-4232}\inst{\ref{aff24},\ref{aff14},\ref{aff22}}
\and F.~Caro\orcid{0009-0003-1053-0507}\inst{\ref{aff56}}
\and C.~S.~Carvalho\inst{\ref{aff113}}
\and F.~Cogato\orcid{0000-0003-4632-6113}\inst{\ref{aff10},\ref{aff11}}
\and A.~R.~Cooray\orcid{0000-0002-3892-0190}\inst{\ref{aff137}}
\and O.~Cucciati\orcid{0000-0002-9336-7551}\inst{\ref{aff11}}
\and T.~de~Boer\orcid{0000-0001-5486-2747}\inst{\ref{aff138}}
\and G.~Desprez\orcid{0000-0001-8325-1742}\inst{\ref{aff115}}
\and A.~D\'iaz-S\'anchez\orcid{0000-0003-0748-4768}\inst{\ref{aff139}}
\and S.~Di~Domizio\orcid{0000-0003-2863-5895}\inst{\ref{aff46},\ref{aff47}}
\and J.~M.~Diego\orcid{0000-0001-9065-3926}\inst{\ref{aff140}}
\and V.~Duret\orcid{0009-0009-0383-4960}\inst{\ref{aff67}}
\and M.~Y.~Elkhashab\orcid{0000-0001-9306-2603}\inst{\ref{aff22},\ref{aff23},\ref{aff21},\ref{aff24}}
\and A.~Enia\orcid{0000-0002-0200-2857}\inst{\ref{aff11}}
\and Y.~Fang\orcid{0000-0002-0334-6950}\inst{\ref{aff76}}
\and A.~Finoguenov\orcid{0000-0002-4606-5403}\inst{\ref{aff88}}
\and A.~Franco\orcid{0000-0002-4761-366X}\inst{\ref{aff127},\ref{aff126},\ref{aff128}}
\and K.~Ganga\orcid{0000-0001-8159-8208}\inst{\ref{aff95}}
\and T.~Gasparetto\orcid{0000-0002-7913-4866}\inst{\ref{aff56}}
\and R.~Gavazzi\orcid{0000-0002-5540-6935}\inst{\ref{aff65},\ref{aff141}}
\and E.~Gaztanaga\orcid{0000-0001-9632-0815}\inst{\ref{aff1},\ref{aff2},\ref{aff142}}
\and F.~Giacomini\orcid{0000-0002-3129-2814}\inst{\ref{aff12}}
\and F.~Gianotti\orcid{0000-0003-4666-119X}\inst{\ref{aff11}}
\and G.~Gozaliasl\orcid{0000-0002-0236-919X}\inst{\ref{aff143},\ref{aff88}}
\and A.~Gruppuso\orcid{0000-0001-9272-5292}\inst{\ref{aff11},\ref{aff12}}
\and M.~Guidi\orcid{0000-0001-9408-1101}\inst{\ref{aff45},\ref{aff11}}
\and C.~M.~Gutierrez\orcid{0000-0001-7854-783X}\inst{\ref{aff18},\ref{aff144}}
\and A.~Hall\orcid{0000-0002-3139-8651}\inst{\ref{aff30}}
\and C.~Hern\'andez-Monteagudo\orcid{0000-0001-5471-9166}\inst{\ref{aff144},\ref{aff18}}
\and H.~Hildebrandt\orcid{0000-0002-9814-3338}\inst{\ref{aff145}}
\and J.~Hjorth\orcid{0000-0002-4571-2306}\inst{\ref{aff101}}
\and J.~J.~E.~Kajava\orcid{0000-0002-3010-8333}\inst{\ref{aff146},\ref{aff147},\ref{aff148}}
\and Y.~Kang\orcid{0009-0000-8588-7250}\inst{\ref{aff36}}
\and V.~Kansal\orcid{0000-0002-4008-6078}\inst{\ref{aff149},\ref{aff150}}
\and D.~Karagiannis\orcid{0000-0002-4927-0816}\inst{\ref{aff117},\ref{aff151}}
\and K.~Kiiveri\inst{\ref{aff110}}
\and J.~Kim\orcid{0000-0003-2776-2761}\inst{\ref{aff72}}
\and C.~C.~Kirkpatrick\inst{\ref{aff110}}
\and K.~Koyama\orcid{0000-0001-6727-6915}\inst{\ref{aff142}}
\and S.~Kruk\orcid{0000-0001-8010-8879}\inst{\ref{aff44}}
\and M.~C.~Lam\orcid{0000-0002-9347-2298}\inst{\ref{aff30}}
\and M.~Lattanzi\orcid{0000-0003-1059-2532}\inst{\ref{aff118}}
\and L.~Legrand\orcid{0000-0003-0610-5252}\inst{\ref{aff152},\ref{aff153}}
\and M.~Lembo\orcid{0000-0002-5271-5070}\inst{\ref{aff141}}
\and G.~Leroy\orcid{0009-0004-2523-4425}\inst{\ref{aff154},\ref{aff94}}
\and G.~F.~Lesci\orcid{0000-0002-4607-2830}\inst{\ref{aff10},\ref{aff11}}
\and J.~Lesgourgues\orcid{0000-0001-7627-353X}\inst{\ref{aff54}}
\and T.~I.~Liaudat\orcid{0000-0002-9104-314X}\inst{\ref{aff155}}
\and S.~J.~Liu\orcid{0000-0001-7680-2139}\inst{\ref{aff68}}
\and X.~Lopez~Lopez\orcid{0009-0008-5194-5908}\inst{\ref{aff11}}
\and A.~Manj\'on-Garc\'ia\orcid{0000-0002-7413-8825}\inst{\ref{aff139}}
\and F.~Mannucci\orcid{0000-0002-4803-2381}\inst{\ref{aff156}}
\and C.~J.~A.~P.~Martins\orcid{0000-0002-4886-9261}\inst{\ref{aff157},\ref{aff158}}
\and L.~Maurin\orcid{0000-0002-8406-0857}\inst{\ref{aff66}}
\and M.~Miluzio\inst{\ref{aff44},\ref{aff159}}
\and G.~Morgante\inst{\ref{aff11}}
\and S.~Nadathur\orcid{0000-0001-9070-3102}\inst{\ref{aff142}}
\and K.~Naidoo\orcid{0000-0002-9182-1802}\inst{\ref{aff142},\ref{aff84}}
\and A.~Navarro-Alsina\orcid{0000-0002-3173-2592}\inst{\ref{aff92}}
\and S.~Nesseris\orcid{0000-0002-0567-0324}\inst{\ref{aff105}}
\and F.~Pace\orcid{0000-0001-8039-0480}\inst{\ref{aff50},\ref{aff51},\ref{aff52}}
\and D.~Paoletti\orcid{0000-0003-4761-6147}\inst{\ref{aff11},\ref{aff70}}
\and F.~Passalacqua\orcid{0000-0002-8606-4093}\inst{\ref{aff103},\ref{aff104}}
\and K.~Paterson\orcid{0000-0001-8340-3486}\inst{\ref{aff84}}
\and L.~Patrizii\inst{\ref{aff12}}
\and C.~Pattison\orcid{0000-0003-3272-2617}\inst{\ref{aff142}}
\and R.~Paviot\orcid{0009-0002-8108-3460}\inst{\ref{aff26},\ref{aff160}}
\and A.~Pisani\orcid{0000-0002-6146-4437}\inst{\ref{aff67}}
\and G.~W.~Pratt\inst{\ref{aff26}}
\and S.~Quai\orcid{0000-0002-0449-8163}\inst{\ref{aff10},\ref{aff11}}
\and M.~Radovich\orcid{0000-0002-3585-866X}\inst{\ref{aff75}}
\and K.~Rojas\orcid{0000-0003-1391-6854}\inst{\ref{aff161}}
\and W.~Roster\orcid{0000-0002-9149-6528}\inst{\ref{aff74}}
\and S.~Sacquegna\orcid{0000-0002-8433-6630}\inst{\ref{aff162}}
\and M.~Sahl\'en\orcid{0000-0003-0973-4804}\inst{\ref{aff163}}
\and D.~B.~Sanders\orcid{0000-0002-1233-9998}\inst{\ref{aff138}}
\and E.~Sarpa\orcid{0000-0002-1256-655X}\inst{\ref{aff22}}
\and A.~Schneider\orcid{0000-0001-7055-8104}\inst{\ref{aff6}}
\and M.~Schultheis\inst{\ref{aff121}}
\and D.~Sciotti\orcid{0009-0008-4519-2620}\inst{\ref{aff56},\ref{aff93}}
\and E.~Sellentin\inst{\ref{aff164},\ref{aff9}}
\and L.~C.~Smith\orcid{0000-0002-3259-2771}\inst{\ref{aff165}}
\and J.~G.~Sorce\orcid{0000-0002-2307-2432}\inst{\ref{aff166},\ref{aff66}}
\and I.~Szapudi\orcid{0000-0003-2274-0301}\inst{\ref{aff138}}
\and K.~Tanidis\orcid{0000-0001-9843-5130}\inst{\ref{aff167}}
\and F.~Tarsitano\orcid{0000-0002-5919-0238}\inst{\ref{aff168},\ref{aff36}}
\and G.~Testera\inst{\ref{aff47}}
\and S.~Tosi\orcid{0000-0002-7275-9193}\inst{\ref{aff46},\ref{aff32},\ref{aff47}}
\and A.~Troja\orcid{0000-0003-0239-4595}\inst{\ref{aff22}}
\and C.~Uhlemann\orcid{0000-0001-7831-1579}\inst{\ref{aff169},\ref{aff100}}
\and C.~Valieri\inst{\ref{aff12}}
\and A.~Venhola\orcid{0000-0001-6071-4564}\inst{\ref{aff170}}
\and D.~Vergani\orcid{0000-0003-0898-2216}\inst{\ref{aff11}}
\and G.~Verza\orcid{0000-0002-1886-8348}\inst{\ref{aff171},\ref{aff172}}
\and P.~Vielzeuf\orcid{0000-0003-2035-9339}\inst{\ref{aff67}}
\and S.~Vinciguerra\orcid{0009-0005-4018-3184}\inst{\ref{aff65}}
\and M.~von~Wietersheim-Kramsta\orcid{0000-0003-4986-5091}\inst{\ref{aff94},\ref{aff154}}
\and N.~A.~Walton\orcid{0000-0003-3983-8778}\inst{\ref{aff165}}
\and A.~H.~Wright\orcid{0000-0001-7363-7932}\inst{\ref{aff145}}
\and H.~W.~Yeung\orcid{0000-0002-4993-9014}\inst{\ref{aff30}}}
										   
\institute{Institute of Space Sciences (ICE, CSIC), Campus UAB, Carrer de Can Magrans, s/n, 08193 Barcelona, Spain\label{aff1}
\and
Institut d'Estudis Espacials de Catalunya (IEEC),  Edifici RDIT, Campus UPC, 08860 Castelldefels, Barcelona, Spain\label{aff2}
\and
Instituto de Astronomia Teorica y Experimental (IATE-CONICET), Laprida 854, X5000BGR, C\'ordoba, Argentina\label{aff3}
\and
Centro de Investigaciones Energ\'eticas, Medioambientales y Tecnol\'ogicas (CIEMAT), Avenida Complutense 40, 28040 Madrid, Spain\label{aff4}
\and
Port d'Informaci\'{o} Cient\'{i}fica, Campus UAB, C. Albareda s/n, 08193 Bellaterra (Barcelona), Spain\label{aff5}
\and
Department of Astrophysics, University of Zurich, Winterthurerstrasse 190, 8057 Zurich, Switzerland\label{aff6}
\and
Institut de Recherche en Astrophysique et Plan\'etologie (IRAP), Universit\'e de Toulouse, CNRS, UPS, CNES, 14 Av. Edouard Belin, 31400 Toulouse, France\label{aff7}
\and
Institute for Theoretical Physics, Utrecht University, Princetonplein 5, 3584 CE Utrecht, The Netherlands\label{aff8}
\and
Leiden Observatory, Leiden University, Einsteinweg 55, 2333 CC Leiden, The Netherlands\label{aff9}
\and
Dipartimento di Fisica e Astronomia "Augusto Righi" - Alma Mater Studiorum Universit\`a di Bologna, via Piero Gobetti 93/2, 40129 Bologna, Italy\label{aff10}
\and
INAF-Osservatorio di Astrofisica e Scienza dello Spazio di Bologna, Via Piero Gobetti 93/3, 40129 Bologna, Italy\label{aff11}
\and
INFN-Sezione di Bologna, Viale Berti Pichat 6/2, 40127 Bologna, Italy\label{aff12}
\and
Institut de F\'{i}sica d'Altes Energies (IFAE), The Barcelona Institute of Science and Technology, Campus UAB, 08193 Bellaterra (Barcelona), Spain\label{aff13}
\and
SISSA, International School for Advanced Studies, Via Bonomea 265, 34136 Trieste TS, Italy\label{aff14}
\and
Serra H\'unter Fellow, Departament de F\'{\i}sica, Universitat Aut\`onoma de Barcelona, E-08193 Bellaterra, Spain\label{aff15}
\and
Center for Data-Driven Discovery, Kavli IPMU (WPI), UTIAS, The University of Tokyo, Kashiwa, Chiba 277-8583, Japan\label{aff16}
\and
Laboratoire d'etude de l'Univers et des phenomenes eXtremes, Observatoire de Paris, Universit\'e PSL, Sorbonne Universit\'e, CNRS, 92190 Meudon, France\label{aff17}
\and
Instituto de Astrof\'{\i}sica de Canarias, E-38205 La Laguna, Tenerife, Spain\label{aff18}
\and
Universit\'e PSL, Observatoire de Paris, Sorbonne Universit\'e, CNRS, LERMA, 75014, Paris, France\label{aff19}
\and
Universit\'e Paris-Cit\'e, 5 Rue Thomas Mann, 75013, Paris, France\label{aff20}
\and
Dipartimento di Fisica - Sezione di Astronomia, Universit\`a di Trieste, Via Tiepolo 11, 34131 Trieste, Italy\label{aff21}
\and
INAF-Osservatorio Astronomico di Trieste, Via G. B. Tiepolo 11, 34143 Trieste, Italy\label{aff22}
\and
INFN, Sezione di Trieste, Via Valerio 2, 34127 Trieste TS, Italy\label{aff23}
\and
IFPU, Institute for Fundamental Physics of the Universe, via Beirut 2, 34151 Trieste, Italy\label{aff24}
\and
Minnesota Institute for Astrophysics, University of Minnesota, 116 Church St SE, Minneapolis, MN 55455, USA\label{aff25}
\and
Universit\'e Paris-Saclay, Universit\'e Paris Cit\'e, CEA, CNRS, AIM, 91191, Gif-sur-Yvette, France\label{aff26}
\and
Kavli Institute for Particle Astrophysics \& Cosmology (KIPAC), Stanford University, Stanford, CA 94305, USA\label{aff27}
\and
SLAC National Accelerator Laboratory, 2575 Sand Hill Road, Menlo Park, CA 94025, USA\label{aff28}
\and
Department of Astrophysical Sciences, Peyton Hall, Princeton University, Princeton, NJ 08544, USA\label{aff29}
\and
Institute for Astronomy, University of Edinburgh, Royal Observatory, Blackford Hill, Edinburgh EH9 3HJ, UK\label{aff30}
\and
Institute of Physics, Laboratory for Galaxy Evolution, Ecole Polytechnique F\'ed\'erale de Lausanne, Observatoire de Sauverny, CH-1290 Versoix, Switzerland\label{aff31}
\and
INAF-Osservatorio Astronomico di Brera, Via Brera 28, 20122 Milano, Italy\label{aff32}
\and
Waterloo Centre for Astrophysics, University of Waterloo, Waterloo, Ontario N2L 3G1, Canada\label{aff33}
\and
Department of Physics and Astronomy, University of Waterloo, Waterloo, Ontario N2L 3G1, Canada\label{aff34}
\and
Perimeter Institute for Theoretical Physics, Waterloo, Ontario N2L 2Y5, Canada\label{aff35}
\and
Department of Astronomy, University of Geneva, ch. d'Ecogia 16, 1290 Versoix, Switzerland\label{aff36}
\and
Cosmic Dawn Center (DAWN)\label{aff37}
\and
Niels Bohr Institute, University of Copenhagen, Jagtvej 128, 2200 Copenhagen, Denmark\label{aff38}
\and
Jodrell Bank Centre for Astrophysics, Department of Physics and Astronomy, University of Manchester, Oxford Road, Manchester M13 9PL, UK\label{aff39}
\and
Departamento de F\'isica Te\'orica, Facultad de Ciencias, Universidad Aut\'onoma de Madrid, 28049 Cantoblanco, Madrid, Spain\label{aff40}
\and
Centro de Investigaci\'{o}n Avanzada en F\'isica Fundamental (CIAFF), Facultad de Ciencias, Universidad Aut\'{o}noma de Madrid, 28049 Madrid, Spain\label{aff41}
\and
MTA-CSFK Lend\"ulet Large-Scale Structure Research Group, Konkoly-Thege Mikl\'os \'ut 15-17, H-1121 Budapest, Hungary\label{aff42}
\and
Konkoly Observatory, HUN-REN CSFK, MTA Centre of Excellence, Budapest, Konkoly Thege Mikl\'os {\'u}t 15-17. H-1121, Hungary\label{aff43}
\and
ESAC/ESA, Camino Bajo del Castillo, s/n., Urb. Villafranca del Castillo, 28692 Villanueva de la Ca\~nada, Madrid, Spain\label{aff44}
\and
Dipartimento di Fisica e Astronomia, Universit\`a di Bologna, Via Gobetti 93/2, 40129 Bologna, Italy\label{aff45}
\and
Dipartimento di Fisica, Universit\`a di Genova, Via Dodecaneso 33, 16146, Genova, Italy\label{aff46}
\and
INFN-Sezione di Genova, Via Dodecaneso 33, 16146, Genova, Italy\label{aff47}
\and
Department of Physics "E. Pancini", University Federico II, Via Cinthia 6, 80126, Napoli, Italy\label{aff48}
\and
INAF-Osservatorio Astronomico di Capodimonte, Via Moiariello 16, 80131 Napoli, Italy\label{aff49}
\and
Dipartimento di Fisica, Universit\`a degli Studi di Torino, Via P. Giuria 1, 10125 Torino, Italy\label{aff50}
\and
INFN-Sezione di Torino, Via P. Giuria 1, 10125 Torino, Italy\label{aff51}
\and
INAF-Osservatorio Astrofisico di Torino, Via Osservatorio 20, 10025 Pino Torinese (TO), Italy\label{aff52}
\and
INAF-IASF Milano, Via Alfonso Corti 12, 20133 Milano, Italy\label{aff53}
\and
Institute for Theoretical Particle Physics and Cosmology (TTK), RWTH Aachen University, 52056 Aachen, Germany\label{aff54}
\and
Deutsches Zentrum f\"ur Luft- und Raumfahrt e. V. (DLR), Linder H\"ohe, 51147 K\"oln, Germany\label{aff55}
\and
INAF-Osservatorio Astronomico di Roma, Via Frascati 33, 00078 Monteporzio Catone, Italy\label{aff56}
\and
INFN section of Naples, Via Cinthia 6, 80126, Napoli, Italy\label{aff57}
\and
Dipartimento di Fisica e Astronomia "Augusto Righi" - Alma Mater Studiorum Universit\`a di Bologna, Viale Berti Pichat 6/2, 40127 Bologna, Italy\label{aff58}
\and
European Space Agency/ESRIN, Largo Galileo Galilei 1, 00044 Frascati, Roma, Italy\label{aff59}
\and
Universit\'e Claude Bernard Lyon 1, CNRS/IN2P3, IP2I Lyon, UMR 5822, Villeurbanne, F-69100, France\label{aff60}
\and
Institut de Ci\`{e}ncies del Cosmos (ICCUB), Universitat de Barcelona (IEEC-UB), Mart\'{i} i Franqu\`{e}s 1, 08028 Barcelona, Spain\label{aff61}
\and
Instituci\'o Catalana de Recerca i Estudis Avan\c{c}ats (ICREA), Passeig de Llu\'{\i}s Companys 23, 08010 Barcelona, Spain\label{aff62}
\and
Institut de Ciencies de l'Espai (IEEC-CSIC), Campus UAB, Carrer de Can Magrans, s/n Cerdanyola del Vall\'es, 08193 Barcelona, Spain\label{aff63}
\and
UCB Lyon 1, CNRS/IN2P3, IUF, IP2I Lyon, 4 rue Enrico Fermi, 69622 Villeurbanne, France\label{aff64}
\and
Aix-Marseille Universit\'e, CNRS, CNES, LAM, Marseille, France\label{aff65}
\and
Universit\'e Paris-Saclay, CNRS, Institut d'astrophysique spatiale, 91405, Orsay, France\label{aff66}
\and
Aix-Marseille Universit\'e, CNRS/IN2P3, CPPM, Marseille, France\label{aff67}
\and
INAF-Istituto di Astrofisica e Planetologia Spaziali, via del Fosso del Cavaliere, 100, 00100 Roma, Italy\label{aff68}
\and
Space Science Data Center, Italian Space Agency, via del Politecnico snc, 00133 Roma, Italy\label{aff69}
\and
INFN-Bologna, Via Irnerio 46, 40126 Bologna, Italy\label{aff70}
\and
School of Physics, HH Wills Physics Laboratory, University of Bristol, Tyndall Avenue, Bristol, BS8 1TL, UK\label{aff71}
\and
Department of Physics, Oxford University, Keble Road, Oxford OX1 3RH, UK\label{aff72}
\and
University Observatory, LMU Faculty of Physics, Scheinerstr.~1, 81679 Munich, Germany\label{aff73}
\and
Max Planck Institute for Extraterrestrial Physics, Giessenbachstr. 1, 85748 Garching, Germany\label{aff74}
\and
INAF-Osservatorio Astronomico di Padova, Via dell'Osservatorio 5, 35122 Padova, Italy\label{aff75}
\and
Universit\"ats-Sternwarte M\"unchen, Fakult\"at f\"ur Physik, Ludwig-Maximilians-Universit\"at M\"unchen, Scheinerstr.~1, 81679 M\"unchen, Germany\label{aff76}
\and
Dipartimento di Fisica "Aldo Pontremoli", Universit\`a degli Studi di Milano, Via Celoria 16, 20133 Milano, Italy\label{aff77}
\and
INFN-Sezione di Milano, Via Celoria 16, 20133 Milano, Italy\label{aff78}
\and
Institute of Theoretical Astrophysics, University of Oslo, P.O. Box 1029 Blindern, 0315 Oslo, Norway\label{aff79}
\and
Jet Propulsion Laboratory, California Institute of Technology, 4800 Oak Grove Drive, Pasadena, CA, 91109, USA\label{aff80}
\and
Felix Hormuth Engineering, Goethestr. 17, 69181 Leimen, Germany\label{aff81}
\and
Technical University of Denmark, Elektrovej 327, 2800 Kgs. Lyngby, Denmark\label{aff82}
\and
Cosmic Dawn Center (DAWN), Denmark\label{aff83}
\and
Max-Planck-Institut f\"ur Astronomie, K\"onigstuhl 17, 69117 Heidelberg, Germany\label{aff84}
\and
NASA Goddard Space Flight Center, Greenbelt, MD 20771, USA\label{aff85}
\and
Department of Physics and Astronomy, University College London, Gower Street, London WC1E 6BT, UK\label{aff86}
\and
Universit\'e de Gen\`eve, D\'epartement de Physique Th\'eorique and Centre for Astroparticle Physics, 24 quai Ernest-Ansermet, CH-1211 Gen\`eve 4, Switzerland\label{aff87}
\and
Department of Physics, P.O. Box 64, University of Helsinki, 00014 Helsinki, Finland\label{aff88}
\and
Helsinki Institute of Physics, Gustaf H{\"a}llstr{\"o}min katu 2, University of Helsinki, 00014 Helsinki, Finland\label{aff89}
\and
SKAO, Jodrell Bank, Lower Withington, Macclesfield SK11 9FT, UK\label{aff90}
\and
Centre de Calcul de l'IN2P3/CNRS, 21 avenue Pierre de Coubertin 69627 Villeurbanne Cedex, France\label{aff91}
\and
Universit\"at Bonn, Argelander-Institut f\"ur Astronomie, Auf dem H\"ugel 71, 53121 Bonn, Germany\label{aff92}
\and
INFN-Sezione di Roma, Piazzale Aldo Moro, 2 - c/o Dipartimento di Fisica, Edificio G. Marconi, 00185 Roma, Italy\label{aff93}
\and
Department of Physics, Institute for Computational Cosmology, Durham University, South Road, Durham, DH1 3LE, UK\label{aff94}
\and
Universit\'e Paris Cit\'e, CNRS, Astroparticule et Cosmologie, 75013 Paris, France\label{aff95}
\and
CNRS-UCB International Research Laboratory, Centre Pierre Bin\'etruy, IRL2007, CPB-IN2P3, Berkeley, USA\label{aff96}
\and
Institute of Physics, Laboratory of Astrophysics, Ecole Polytechnique F\'ed\'erale de Lausanne (EPFL), Observatoire de Sauverny, 1290 Versoix, Switzerland\label{aff97}
\and
Telespazio UK S.L. for European Space Agency (ESA), Camino bajo del Castillo, s/n, Urbanizacion Villafranca del Castillo, Villanueva de la Ca\~nada, 28692 Madrid, Spain\label{aff98}
\and
European Space Agency/ESTEC, Keplerlaan 1, 2201 AZ Noordwijk, The Netherlands\label{aff99}
\and
School of Mathematics, Statistics and Physics, Newcastle University, Herschel Building, Newcastle-upon-Tyne, NE1 7RU, UK\label{aff100}
\and
DARK, Niels Bohr Institute, University of Copenhagen, Jagtvej 155, 2200 Copenhagen, Denmark\label{aff101}
\and
Institute of Space Science, Str. Atomistilor, nr. 409 M\u{a}gurele, Ilfov, 077125, Romania\label{aff102}
\and
Dipartimento di Fisica e Astronomia "G. Galilei", Universit\`a di Padova, Via Marzolo 8, 35131 Padova, Italy\label{aff103}
\and
INFN-Padova, Via Marzolo 8, 35131 Padova, Italy\label{aff104}
\and
Instituto de F\'isica Te\'orica UAM-CSIC, Campus de Cantoblanco, 28049 Madrid, Spain\label{aff105}
\and
Universit\'e St Joseph; Faculty of Sciences, Beirut, Lebanon\label{aff106}
\and
Departamento de F\'isica, FCFM, Universidad de Chile, Blanco Encalada 2008, Santiago, Chile\label{aff107}
\and
Universit\"at Innsbruck, Institut f\"ur Astro- und Teilchenphysik, Technikerstr. 25/8, 6020 Innsbruck, Austria\label{aff108}
\and
Satlantis, University Science Park, Sede Bld 48940, Leioa-Bilbao, Spain\label{aff109}
\and
Department of Physics and Helsinki Institute of Physics, Gustaf H\"allstr\"omin katu 2, University of Helsinki, 00014 Helsinki, Finland\label{aff110}
\and
Department of Physics, Royal Holloway, University of London, Surrey TW20 0EX, UK\label{aff111}
\and
Departamento de F\'isica, Faculdade de Ci\^encias, Universidade de Lisboa, Edif\'icio C8, Campo Grande, PT1749-016 Lisboa, Portugal\label{aff112}
\and
Instituto de Astrof\'isica e Ci\^encias do Espa\c{c}o, Faculdade de Ci\^encias, Universidade de Lisboa, Tapada da Ajuda, 1349-018 Lisboa, Portugal\label{aff113}
\and
Universidad Polit\'ecnica de Cartagena, Departamento de Electr\'onica y Tecnolog\'ia de Computadoras,  Plaza del Hospital 1, 30202 Cartagena, Spain\label{aff114}
\and
Kapteyn Astronomical Institute, University of Groningen, PO Box 800, 9700 AV Groningen, The Netherlands\label{aff115}
\and
Caltech/IPAC, 1200 E. California Blvd., Pasadena, CA 91125, USA\label{aff116}
\and
Dipartimento di Fisica e Scienze della Terra, Universit\`a degli Studi di Ferrara, Via Giuseppe Saragat 1, 44122 Ferrara, Italy\label{aff117}
\and
Istituto Nazionale di Fisica Nucleare, Sezione di Ferrara, Via Giuseppe Saragat 1, 44122 Ferrara, Italy\label{aff118}
\and
INAF, Istituto di Radioastronomia, Via Piero Gobetti 101, 40129 Bologna, Italy\label{aff119}
\and
Astronomical Observatory of the Autonomous Region of the Aosta Valley (OAVdA), Loc. Lignan 39, I-11020, Nus (Aosta Valley), Italy\label{aff120}
\and
Universit\'e C\^{o}te d'Azur, Observatoire de la C\^{o}te d'Azur, CNRS, Laboratoire Lagrange, Bd de l'Observatoire, CS 34229, 06304 Nice cedex 4, France\label{aff121}
\and
ICSC - Centro Nazionale di Ricerca in High Performance Computing, Big Data e Quantum Computing, Via Magnanelli 2, Bologna, Italy\label{aff122}
\and
Univ. Grenoble Alpes, CNRS, Grenoble INP, LPSC-IN2P3, 53, Avenue des Martyrs, 38000, Grenoble, France\label{aff123}
\and
Dipartimento di Fisica, Sapienza Universit\`a di Roma, Piazzale Aldo Moro 2, 00185 Roma, Italy\label{aff124}
\and
Aurora Technology for European Space Agency (ESA), Camino bajo del Castillo, s/n, Urbanizacion Villafranca del Castillo, Villanueva de la Ca\~nada, 28692 Madrid, Spain\label{aff125}
\and
Department of Mathematics and Physics E. De Giorgi, University of Salento, Via per Arnesano, CP-I93, 73100, Lecce, Italy\label{aff126}
\and
INFN, Sezione di Lecce, Via per Arnesano, CP-193, 73100, Lecce, Italy\label{aff127}
\and
INAF-Sezione di Lecce, c/o Dipartimento Matematica e Fisica, Via per Arnesano, 73100, Lecce, Italy\label{aff128}
\and
Institut d'Astrophysique de Paris, 98bis Boulevard Arago, 75014, Paris, France\label{aff129}
\and
ICL, Junia, Universit\'e Catholique de Lille, LITL, 59000 Lille, France\label{aff130}
\and
CERCA/ISO, Department of Physics, Case Western Reserve University, 10900 Euclid Avenue, Cleveland, OH 44106, USA\label{aff131}
\and
Laboratoire Univers et Th\'eorie, Observatoire de Paris, Universit\'e PSL, Universit\'e Paris Cit\'e, CNRS, 92190 Meudon, France\label{aff132}
\and
Departamento de F{\'\i}sica Fundamental. Universidad de Salamanca. Plaza de la Merced s/n. 37008 Salamanca, Spain\label{aff133}
\and
Aix-Marseille Universit\'e, Universit\'e de Toulon, CNRS, CPT, Marseille, France\label{aff134}
\and
Universit\'e de Strasbourg, CNRS, Observatoire astronomique de Strasbourg, UMR 7550, 67000 Strasbourg, France\label{aff135}
\and
California Institute of Technology, 1200 E California Blvd, Pasadena, CA 91125, USA\label{aff136}
\and
Department of Physics \& Astronomy, University of California Irvine, Irvine CA 92697, USA\label{aff137}
\and
Institute for Astronomy, University of Hawaii, 2680 Woodlawn Drive, Honolulu, HI 96822, USA\label{aff138}
\and
Departamento F\'isica Aplicada, Universidad Polit\'ecnica de Cartagena, Campus Muralla del Mar, 30202 Cartagena, Murcia, Spain\label{aff139}
\and
Instituto de F\'isica de Cantabria, Edificio Juan Jord\'a, Avenida de los Castros, 39005 Santander, Spain\label{aff140}
\and
Institut d'Astrophysique de Paris, UMR 7095, CNRS, and Sorbonne Universit\'e, 98 bis boulevard Arago, 75014 Paris, France\label{aff141}
\and
Institute of Cosmology and Gravitation, University of Portsmouth, Portsmouth PO1 3FX, UK\label{aff142}
\and
Department of Computer Science, Aalto University, PO Box 15400, Espoo, FI-00 076, Finland\label{aff143}
\and
Universidad de La Laguna, Dpto. Astrof\'\i sica, E-38206 La Laguna, Tenerife, Spain\label{aff144}
\and
Ruhr University Bochum, Faculty of Physics and Astronomy, Astronomical Institute (AIRUB), German Centre for Cosmological Lensing (GCCL), 44780 Bochum, Germany\label{aff145}
\and
Department of Physics and Astronomy, Vesilinnantie 5, University of Turku, 20014 Turku, Finland\label{aff146}
\and
Finnish Centre for Astronomy with ESO (FINCA), Quantum, Vesilinnantie 5, University of Turku, 20014 Turku, Finland\label{aff147}
\and
Serco for European Space Agency (ESA), Camino bajo del Castillo, s/n, Urbanizacion Villafranca del Castillo, Villanueva de la Ca\~nada, 28692 Madrid, Spain\label{aff148}
\and
ARC Centre of Excellence for Dark Matter Particle Physics, Melbourne, Australia\label{aff149}
\and
Centre for Astrophysics \& Supercomputing, Swinburne University of Technology,  Hawthorn, Victoria 3122, Australia\label{aff150}
\and
Department of Physics and Astronomy, University of the Western Cape, Bellville, Cape Town, 7535, South Africa\label{aff151}
\and
DAMTP, Centre for Mathematical Sciences, Wilberforce Road, Cambridge CB3 0WA, UK\label{aff152}
\and
Kavli Institute for Cosmology Cambridge, Madingley Road, Cambridge, CB3 0HA, UK\label{aff153}
\and
Department of Physics, Centre for Extragalactic Astronomy, Durham University, South Road, Durham, DH1 3LE, UK\label{aff154}
\and
IRFU, CEA, Universit\'e Paris-Saclay 91191 Gif-sur-Yvette Cedex, France\label{aff155}
\and
INAF-Osservatorio Astrofisico di Arcetri, Largo E. Fermi 5, 50125, Firenze, Italy\label{aff156}
\and
Centro de Astrof\'{\i}sica da Universidade do Porto, Rua das Estrelas, 4150-762 Porto, Portugal\label{aff157}
\and
Instituto de Astrof\'isica e Ci\^encias do Espa\c{c}o, Universidade do Porto, CAUP, Rua das Estrelas, PT4150-762 Porto, Portugal\label{aff158}
\and
HE Space for European Space Agency (ESA), Camino bajo del Castillo, s/n, Urbanizacion Villafranca del Castillo, Villanueva de la Ca\~nada, 28692 Madrid, Spain\label{aff159}
\and
Centre National d'Etudes Spatiales -- Centre spatial de Toulouse, 18 avenue Edouard Belin, 31401 Toulouse Cedex 9, France\label{aff160}
\and
University of Applied Sciences and Arts of Northwestern Switzerland, School of Computer Science, 5210 Windisch, Switzerland\label{aff161}
\and
INAF - Osservatorio Astronomico d'Abruzzo, Via Maggini, 64100, Teramo, Italy\label{aff162}
\and
Theoretical astrophysics, Department of Physics and Astronomy, Uppsala University, Box 516, 751 37 Uppsala, Sweden\label{aff163}
\and
Mathematical Institute, University of Leiden, Einsteinweg 55, 2333 CA Leiden, The Netherlands\label{aff164}
\and
Institute of Astronomy, University of Cambridge, Madingley Road, Cambridge CB3 0HA, UK\label{aff165}
\and
Univ. Lille, CNRS, Centrale Lille, UMR 9189 CRIStAL, 59000 Lille, France\label{aff166}
\and
Center for Astrophysics and Cosmology, University of Nova Gorica, Nova Gorica, Slovenia\label{aff167}
\and
Institute for Particle Physics and Astrophysics, Dept. of Physics, ETH Zurich, Wolfgang-Pauli-Strasse 27, 8093 Zurich, Switzerland\label{aff168}
\and
Fakult\"at f\"ur Physik, Universit\"at Bielefeld, Postfach 100131, 33501 Bielefeld, Germany\label{aff169}
\and
Space physics and astronomy research unit, University of Oulu, Pentti Kaiteran katu 1, FI-90014 Oulu, Finland\label{aff170}
\and
International Centre for Theoretical Physics (ICTP), Strada Costiera 11, 34151 Trieste, Italy\label{aff171}
\and
Center for Computational Astrophysics, Flatiron Institute, 162 5th Avenue, 10010, New York, NY, USA\label{aff172}}    

%% file: aa60445-26.bbl
\begin{thebibliography}{80}
\expandafter\ifx\csname natexlab\endcsname\relax\def\natexlab#1{#1}\fi

\bibitem[{{Abazajian} {et~al.}(2009){Abazajian}, {Adelman-McCarthy}, {Ag{\"u}eros}, {Allam}, {Allende Prieto}, {An}, {Anderson}, {Anderson}, {Annis}, {Bahcall}, {Bailer-Jones}, {Barentine}, {Bassett}, {Becker}, {Beers}, {Bell}, {Belokurov}, {Berlind}, {Berman}, {Bernardi}, {Bickerton}, {Bizyaev}, {Blakeslee}, {Blanton}, {Bochanski}, {Boroski}, {Brewington}, {Brinchmann}, {Brinkmann}, {Brunner}, {Budav{\'a}ri}, {Carey}, {Carliles}, {Carr}, {Castander}, {Cinabro}, {Connolly}, {Csabai}, {Cunha}, {Czarapata}, {Davenport}, {de Haas}, {Dilday}, {Doi}, {Eisenstein}, {Evans}, {Evans}, {Fan}, {Friedman}, {Frieman}, {Fukugita}, {G{\"a}nsicke}, {Gates}, {Gillespie}, {Gilmore}, {Gonzalez}, {Gonzalez}, {Grebel}, {Gunn}, {Gy{\"o}ry}, {Hall}, {Harding}, {Harris}, {Harvanek}, {Hawley}, {Hayes}, {Heckman}, {Hendry}, {Hennessy}, {Hindsley}, {Hoblitt}, {Hogan}, {Hogg}, {Holtzman}, {Hyde}, {Ichikawa}, {Ichikawa}, {Im}, {Ivezi{\'c}}, {Jester}, {Jiang}, {Johnson}, {Jorgensen}, {Juri{\'c}}, {Kent}, {Kessler}, {Kleinman}, {Knapp},
  {Konishi}, {Kron}, {Krzesinski}, {Kuropatkin}, {Lampeitl}, {Lebedeva}, {Lee}, {Lee}, {French Leger}, {L{\'e}pine}, {Li}, {Lima}, {Lin}, {Long}, {Loomis}, {Loveday}, {Lupton}, {Magnier}, {Malanushenko}, {Malanushenko}, {Mandelbaum}, {Margon}, {Marriner}, {Mart{\'\i}nez-Delgado}, {Matsubara}, {McGehee}, {McKay}, {Meiksin}, {Morrison}, {Mullally}, {Munn}, {Murphy}, {Nash}, {Nebot}, {Neilsen}, {Newberg}, {Newman}, {Nichol}, {Nicinski}, {Nieto-Santisteban}, {Nitta}, {Okamura}, {Oravetz}, {Ostriker}, {Owen}, {Padmanabhan}, {Pan}, {Park}, {Pauls}, {Peoples}, {Percival}, {Pier}, {Pope}, {Pourbaix}, {Price}, {Purger}, {Quinn}, {Raddick}, {Re Fiorentin}, {Richards}, {Richmond}, {Riess}, {Rix}, {Rockosi}, {Sako}, {Schlegel}, {Schneider}, {Scholz}, {Schreiber}, {Schwope}, {Seljak}, {Sesar}, {Sheldon}, {Shimasaku}, {Sibley}, {Simmons}, {Sivarani}, {Allyn Smith}, {Smith}, {Smol{\v{c}}i{\'c}}, {Snedden}, {Stebbins}, {Steinmetz}, {Stoughton}, {Strauss}, {SubbaRao}, {Suto}, {Szalay}, {Szapudi}, {Szkody}, {Tanaka},
  {Tegmark}, {Teodoro}, {Thakar}, {Tremonti}, {Tucker}, {Uomoto}, {Vanden Berk}, {Vandenberg}, {Vidrih}, {Vogeley}, {Voges}, {Vogt}, {Wadadekar}, {Watters}, {Weinberg}, {West}, {White}, {Wilhite}, {Wonders}, {Yanny}, {Yocum}, {York}, {Zehavi}, {Zibetti}, \& {Zucker}}]{Abazajian2009}
{Abazajian}, K.~N., {Adelman-McCarthy}, J.~K., {Ag{\"u}eros}, M.~A., {et~al.} 2009, \apjs, 182, 543

\bibitem[{{Avila} {et~al.}(2020){Avila}, {Gonzalez-Perez}, {Mohammad}, {de Mattia}, {Zhao}, {Raichoor}, {Tamone}, {Alam}, {Bautista}, {Bianchi}, {Burtin}, {Chapman}, {Chuang}, {Comparat}, {Dawson}, {Divers}, {du Mas des Bourboux}, {Gil-Marin}, {Mueller}, {Habib}, {Heitmann}, {Ruhlmann-Kleider}, {Padilla}, {Percival}, {Ross}, {Seo}, {Schneider}, \& {Zhao}}]{Avila2020}
{Avila}, S., {Gonzalez-Perez}, V., {Mohammad}, F.~G., {et~al.} 2020, \mnras, 499, 5486

\bibitem[{{Behroozi} {et~al.}(2019){Behroozi}, {Wechsler}, {Hearin}, \& {Conroy}}]{Behroozi2019}
{Behroozi}, P., {Wechsler}, R.~H., {Hearin}, A.~P., \& {Conroy}, C. 2019, \mnras, 488, 3143

\bibitem[{{Behroozi} {et~al.}(2013){Behroozi}, {Wechsler}, \& {Wu}}]{Behroozi2013}
{Behroozi}, P.~S., {Wechsler}, R.~H., \& {Wu}, H.-Y. 2013, \apj, 762, 109

\bibitem[{{Blanton} {et~al.}(2003){Blanton}, {Hogg}, {Bahcall}, {Brinkmann}, {Britton}, {Connolly}, {Csabai}, {Fukugita}, {Loveday}, {Meiksin}, {Munn}, {Nichol}, {Okamura}, {Quinn}, {Schneider}, {Shimasaku}, {Strauss}, {Tegmark}, {Vogeley}, \& {Weinberg}}]{Blanton2003}
{Blanton}, M.~R., {Hogg}, D.~W., {Bahcall}, N.~A., {et~al.} 2003, \apj, 592, 819

\bibitem[{{Blanton} {et~al.}(2005){Blanton}, {Lupton}, {Schlegel}, {Strauss}, {Brinkmann}, {Fukugita}, \& {Loveday}}]{Blanton2005}
{Blanton}, M.~R., {Lupton}, R.~H., {Schlegel}, D.~J., {et~al.} 2005, \apj, 631, 208

\bibitem[{{Bose} {et~al.}(2019){Bose}, {Eisenstein}, {Hernquist}, {Pillepich}, {Nelson}, {Marinacci}, {Springel}, \& {Vogelsberger}}]{Bose2019}
{Bose}, S., {Eisenstein}, D.~J., {Hernquist}, L., {et~al.} 2019, \mnras, 490, 5693

\bibitem[{{Carretero} {et~al.}(2015){Carretero}, {Castander}, {Gazta{\~n}aga}, {Crocce}, \& {Fosalba}}]{Carretero:15}
{Carretero}, J., {Castander}, F.~J., {Gazta{\~n}aga}, E., {Crocce}, M., \& {Fosalba}, P. 2015, \mnras, 447, 646

\bibitem[{{Carretero} {et~al.}(2017){Carretero}, {Tallada}, {Casals}, {Caubet}, {Castander}, {Blot}, {Alarc{\'o}n}, {Serrano}, {Fosalba}, {Acosta-Silva}, {Tonello}, {Torradeflot}, {Eriksen}, {Neissner}, \& {Delfino}}]{Carretero:17}
{Carretero}, J., {Tallada}, P., {Casals}, J., {et~al.} 2017, in Proceedings of the European Physical Society Conference on High Energy Physics. 5-12 July, 488

\bibitem[{{Contreras} {et~al.}(2019){Contreras}, {Zehavi}, {Padilla}, {Baugh}, {Jim{\'e}nez}, \& {Lacerna}}]{Contreras2019}
{Contreras}, S., {Zehavi}, I., {Padilla}, N., {et~al.} 2019, \mnras, 484, 1133

\bibitem[{{Dahlen} {et~al.}(2005){Dahlen}, {Mobasher}, {Somerville}, {Moustakas}, {Dickinson}, {Ferguson}, \& {Giavalisco}}]{Dahlen2005}
{Dahlen}, T., {Mobasher}, B., {Somerville}, R.~S., {et~al.} 2005, \apj, 631, 126

\bibitem[{{d'Assignies} {et~al.}(2025){d'Assignies}, {Manera}, {Padilla}, {Ilbert}, {Hildebrandt}, {Reynolds}, {Chaves-Montero}, {Wright}, {Tallada-Cresp{\'\i}}, {Eriksen}, {Carretero}, {Roster}, {Kang}, {Naidoo}, {Miquel}, {Altieri}, {Amara}, {Andreon}, {Auricchio}, {Baccigalupi}, {Bagot}, {Baldi}, {Balestra}, {Bardelli}, {Battaglia}, {Biviano}, {Branchini}, {Brescia}, {Camera}, {Capobianco}, {Carbone}, {Cardone}, {Casas}, {Castander}, {Castellano}, {Castignani}, {Cavuoti}, {Chambers}, {Cimatti}, {Colodro-Conde}, {Congedo}, {Conselice}, {Conversi}, {Copin}, {Courbin}, {Courtois}, {Crocce}, {Da Silva}, {Degaudenzi}, {de la Torre}, {De Lucia}, {Douspis}, {Dupac}, {Ealet}, {Escoffier}, {Farina}, {Faustini}, {Ferriol}, {Finelli}, {Fosalba}, {Fotopoulou}, {Frailis}, {Franceschi}, {Fumana}, {Galeotta}, {George}, {Gillis}, {Giocoli}, {G{\'o}mez-Alvarez}, {Gracia-Carpio}, {Grazian}, {Grupp}, {Holmes}, {Hook}, {Hornstrup}, {Jahnke}, {Jhabvala}, {Joachimi}, {Keih{\"a}nen}, {Kermiche}, {Kiessling}, {Kubik},
  {K{\"u}mmel}, {Kunz}, {Kurki-Suonio}, {Lahav}, {Le Brun}, {Ligori}, {Lilje}, {Lindholm}, {Lloro}, {Mainetti}, {Maino}, {Maiorano}, {Mansutti}, {Marcin}, {Marggraf}, {Markovic}, {Martinelli}, {Martinet}, {Marulli}, {Massey}, {Masters}, {Medinaceli}, {Mei}, {Melchior}, {Mellier}, {Meneghetti}, {Merlin}, {Meylan}, {Mora}, {Moresco}, {Moscardini}, {Neissner}, {Niemi}, {Paltani}, {Pasian}, {Pedersen}, {Pettorino}, {Pires}, {Polenta}, {Poncet}, {Popa}, {Pozzetti}, {Raison}, {Rebolo}, {Renzi}, {Rhodes}, {Riccio}, {Romelli}, {Roncarelli}, {Rossetti}, {Saglia}, {Sakr}, {Sapone}, {Sartoris}, {Schewtschenko}, {Schneider}, {Schrabback}, {Secroun}, {Sefusatti}, {Seidel}, {Seiffert}, {Serrano}, {Simon}, {Sirignano}, {Sirri}, {Spurio Mancini}, {Stanco}, {Steinwagner}, {Tavagnacco}, {Taylor}, {Teplitz}, {Tereno}, {Tessore}, {Toft}, {Toledo-Moreo}, {Torradeflot}, {Tsyganov}, {Tutusaus}, {Valenziano}, {Valiviita}, {Vassallo}, {Verdoes Kleijn}, {Wang}, {Weller}, {Zamorani}, {Zucca}, {Bolzonella}, {Burigana}, {Gabarra},
  {Mart{\'\i}n-Fleitas}, {Risso}, {Scottez}, \& {Viel}}]{Euclid_Photoz_Optimiz}
{d'Assignies}, W., {Manera}, M., {Padilla}, C., {et~al.} 2025, \aap, 702, A155

\bibitem[{{DESI Collaboration: Abareshi} {et~al.}(2022){DESI Collaboration: Abareshi}, {Abareshi}, {Aguilar}, {Ahlen}, {Alam}, {Alexander}, {Alfarsy}, {Allen}, {Allende Prieto}, {Alves}, {Ameel}, {Armengaud}, {Asorey}, {Aviles}, {Bailey}, {Balaguera-Antol{\'\i}nez}, {Ballester}, {Baltay}, {Bault}, {Beltran}, {Benavides}, {BenZvi}, {Berti}, {Besuner}, {Beutler}, {Bianchi}, {Blake}, {Blanc}, {Blum}, {Bolton}, {Bose}, {Bramall}, {Brieden}, {Brodzeller}, {Brooks}, {Brownewell}, {Buckley-Geer}, {Cahn}, {Cai}, {Canning}, {Capasso}, {Carnero Rosell}, {Carton}, {Casas}, {Castander}, {Cervantes-Cota}, {Chabanier}, {Chaussidon}, {Chuang}, {Circosta}, {Cole}, {Cooper}, {da Costa}, {Cousinou}, {Cuceu}, {Davis}, {Dawson}, {de la Cruz-Noriega}, {de la Macorra}, {de Mattia}, {Della Costa}, {Demmer}, {Derwent}, {Dey}, {Dey}, {Dhungana}, {Ding}, {Dobson}, {Doel}, {Donald-McCann}, {Donaldson}, {Douglass}, {Duan}, {Dunlop}, {Edelstein}, {Eftekharzadeh}, {Eisenstein}, {Enriquez-Vargas}, {Escoffier}, {Evatt}, {Fagrelius}, {Fan},
  {Fanning}, {Fawcett}, {Ferraro}, {Ereza}, {Flaugher}, {Font-Ribera}, {Forero-Romero}, {Frenk}, {Fromenteau}, {G{\"a}nsicke}, {Garcia-Quintero}, {Garrison}, {Gazta{\~n}aga}, {Gerardi}, {Gil-Mar{\'\i}n}, {Gontcho A Gontcho}, {Gonzalez-Morales}, {Gonzalez-de-Rivera}, {Gonzalez-Perez}, {Gordon}, {Graur}, {Green}, {Grove}, {Gruen}, {Gutierrez}, {Guy}, {Hahn}, {Harris}, {Herrera}, {Herrera-Alcantar}, {Honscheid}, {Howlett}, {Huterer}, {Ir{\v{s}}i{\v{c}}}, {Ishak}, {Jelinsky}, {Jiang}, {Jimenez}, {Jing}, {Joyce}, {Jullo}, {Juneau}, {Kara{\c{c}}ayl{\i}}, {Karamanis}, {Karcher}, {Karim}, {Kehoe}, {Kent}, {Kirkby}, {Kisner}, {Kitaura}, {Koposov}, {Kov{\'a}cs}, {Kremin}, {Krolewski}, {L'Huillier}, {Lahav}, {Lambert}, {Lamman}, {Lan}, {Landriau}, {Lane}, {Lang}, {Lange}, {Lasker}, {Le Guillou}, {Leauthaud}, {Le Van Suu}, {Levi}, {Li}, {Magneville}, {Manera}, {Manser}, {Marshall}, {Martini}, {McCollam}, {McDonald}, {Meisner}, {Mena-Fern{\'a}ndez}, {Meneses-Rizo}, {Mezcua}, {Miller}, {Miquel}, {Montero-Camacho}, {Moon},
  {Moustakas}, {Mueller}, {Mu{\~n}oz-Guti{\'e}rrez}, {Myers}, {Nadathur}, {Najita}, {Napolitano}, {Neilsen}, {Newman}, {Nie}, {Ning}, {Niz}, {Norberg}, {Noriega}, {O'Brien}, {Obuljen}, {Palanque-Delabrouille}, {Palmese}, {Zhiwei}, {Pappalardo}, {PENG}, {Percival}, {Perruchot}, {Pogge}, {Poppett}, {Porredon}, {Prada}, {Prochaska}, {Pucha}, {P{\'e}rez-Fern{\'a}ndez}, {P{\'e}rez-R{\`a}fols}, {Rabinowitz}, {Raichoor}, {Ramirez-Solano}, {Ram{\'\i}rez-P{\'e}rez}, {Ravoux}, {Reil}, {Rezaie}, {Rocher}, {Rockosi}, {Roe}, {Roodman}, {Ross}, {Rossi}, {Ruggeri}, {Ruhlmann-Kleider}, {Sabiu}, {Gaines}, {Said}, {Saintonge}, {Salas Catonga}, {Samushia}, {Sanchez}, {Saulder}, {Schaan}, {Schlafly}, {Schlegel}, {Schmoll}, {Scholte}, {Schubnell}, {Secroun}, {Seo}, {Serrano}, {Sharples}, {Sholl}, {Silber}, {Silva}, {Sirk}, {Siudek}, {Smith}, {Sprayberry}, {Staten}, {Stupak}, {Tan}, {Tarl{\'e}}, {Tie}, {Tojeiro}, {Ure{\~n}a-L{\'o}pez}, {Valdes}, {Valenzuela}, {Valluri}, {Vargas-Maga{\~n}a}, {Verde}, {Walther}, {Wang}, {Wang},
  {Weaver}, {Weaverdyck}, {Wechsler}, {Wilson}, {Yang}, {Yu}, {Yuan}, {Y{\`e}che}, {Zhang}, {Zhang}, {Zhao}, {Zhou}, {Zhou}, {Zou}, {Zou}, {Zou}, \& {Zu}}]{DESI2022}
{DESI Collaboration: Abareshi}, {Abareshi}, B., {Aguilar}, J., {et~al.} 2022, \aj, 164, 207

\bibitem[{{DESI Collaboration: Aghamousa} {et~al.}(2016){DESI Collaboration: Aghamousa}, {Aghamousa}, {Aguilar}, {Ahlen}, {Alam}, {Allen}, {Allende Prieto}, {Annis}, {Bailey}, {Balland}, {Ballester}, {Baltay}, {Beaufore}, {Bebek}, {Beers}, {Bell}, {Bernal}, {Besuner}, {Beutler}, {Blake}, {Bleuler}, {Blomqvist}, {Blum}, {Bolton}, {Briceno}, {Brooks}, {Brownstein}, {Buckley-Geer}, {Burden}, {Burtin}, {Busca}, {Cahn}, {Cai}, {Cardiel-Sas}, {Carlberg}, {Carton}, {Casas}, {Castander}, {Cervantes-Cota}, {Claybaugh}, {Close}, {Coker}, {Cole}, {Comparat}, {Cooper}, {Cousinou}, {Crocce}, {Cuby}, {Cunningham}, {Davis}, {Dawson}, {de la Macorra}, {De Vicente}, {Delubac}, {Derwent}, {Dey}, {Dhungana}, {Ding}, {Doel}, {Duan}, {Ealet}, {Edelstein}, {Eftekharzadeh}, {Eisenstein}, {Elliott}, {Escoffier}, {Evatt}, {Fagrelius}, {Fan}, {Fanning}, {Farahi}, {Farihi}, {Favole}, {Feng}, {Fernandez}, {Findlay}, {Finkbeiner}, {Fitzpatrick}, {Flaugher}, {Flender}, {Font-Ribera}, {Forero-Romero}, {Fosalba}, {Frenk}, {Fumagalli},
  {Gaensicke}, {Gallo}, {Garcia-Bellido}, {Gaztanaga}, {Pietro Gentile Fusillo}, {Gerard}, {Gershkovich}, {Giannantonio}, {Gillet}, {Gonzalez-de-Rivera}, {Gonzalez-Perez}, {Gott}, {Graur}, {Gutierrez}, {Guy}, {Habib}, {Heetderks}, {Heetderks}, {Heitmann}, {Hellwing}, {Herrera}, {Ho}, {Holland}, {Honscheid}, {Huff}, {Hutchinson}, {Huterer}, {Hwang}, {Illa Laguna}, {Ishikawa}, {Jacobs}, {Jeffrey}, {Jelinsky}, {Jennings}, {Jiang}, {Jimenez}, {Johnson}, {Joyce}, {Jullo}, {Juneau}, {Kama}, {Karcher}, {Karkar}, {Kehoe}, {Kennamer}, {Kent}, {Kilbinger}, {Kim}, {Kirkby}, {Kisner}, {Kitanidis}, {Kneib}, {Koposov}, {Kovacs}, {Koyama}, {Kremin}, {Kron}, {Kronig}, {Kueter-Young}, {Lacey}, {Lafever}, {Lahav}, {Lambert}, {Lampton}, {Landriau}, {Lang}, {Lauer}, {Le Goff}, {Le Guillou}, {Le Van Suu}, {Lee}, {Lee}, {Leitner}, {Lesser}, {Levi}, {L'Huillier}, {Li}, {Liang}, {Lin}, {Linder}, {Loebman}, {Luki{\'c}}, {Ma}, {MacCrann}, {Magneville}, {Makarem}, {Manera}, {Manser}, {Marshall}, {Martini}, {Massey}, {Matheson},
  {McCauley}, {McDonald}, {McGreer}, {Meisner}, {Metcalfe}, {Miller}, {Miquel}, {Moustakas}, {Myers}, {Naik}, {Newman}, {Nichol}, {Nicola}, {Nicolati da Costa}, {Nie}, {Niz}, {Norberg}, {Nord}, {Norman}, {Nugent}, {O'Brien}, {Oh}, {Olsen}, {Padilla}, {Padmanabhan}, {Padmanabhan}, {Palanque-Delabrouille}, {Palmese}, {Pappalardo}, {P{\^a}ris}, {Park}, {Patej}, {Peacock}, {Peiris}, {Peng}, {Percival}, {Perruchot}, {Pieri}, {Pogge}, {Pollack}, {Poppett}, {Prada}, {Prakash}, {Probst}, {Rabinowitz}, {Raichoor}, {Ree}, {Refregier}, {Regal}, {Reid}, {Reil}, {Rezaie}, {Rockosi}, {Roe}, {Ronayette}, {Roodman}, {Ross}, {Ross}, {Rossi}, {Rozo}, {Ruhlmann-Kleider}, {Rykoff}, {Sabiu}, {Samushia}, {Sanchez}, {Sanchez}, {Schlegel}, {Schneider}, {Schubnell}, {Secroun}, {Seljak}, {Seo}, {Serrano}, {Shafieloo}, {Shan}, {Sharples}, {Sholl}, {Shourt}, {Silber}, {Silva}, {Sirk}, {Slosar}, {Smith}, {Smoot}, {Som}, {Song}, {Sprayberry}, {Staten}, {Stefanik}, {Tarle}, {Sien Tie}, {Tinker}, {Tojeiro}, {Valdes}, {Valenzuela},
  {Valluri}, {Vargas-Magana}, {Verde}, {Walker}, {Wang}, {Wang}, {Weaver}, {Weaverdyck}, {Wechsler}, {Weinberg}, {White}, {Yang}, {Yeche}, {Zhang}, {Zhao}, {Zheng}, {Zhou}, {Zhou}, {Zhu}, {Zou}, \& {Zu}}]{DESI2016}
{DESI Collaboration: Aghamousa}, {Aghamousa}, A., {Aguilar}, J., {et~al.} 2016, arXiv:1611.00036

\bibitem[{{Despali} {et~al.}(2016){Despali}, {Giocoli}, {Angulo}, {Tormen}, {Sheth}, {Baso}, \& {Moscardini}}]{Despali2016}
{Despali}, G., {Giocoli}, C., {Angulo}, R.~E., {et~al.} 2016, \mnras, 456, 2486

\bibitem[{{Diemer}(2018)}]{Diemer2018}
{Diemer}, B. 2018, \apjs, 239, 35

\bibitem[{{Diemer} \& {Joyce}(2019)}]{Diemer2019}
{Diemer}, B. \& {Joyce}, M. 2019, \apj, 871, 168

\bibitem[{{Dubois} {et~al.}(2014){Dubois}, {Pichon}, {Welker}, {Le Borgne}, {Devriendt}, {Laigle}, {Codis}, {Pogosyan}, {Arnouts}, {Benabed}, {Bertin}, {Blaizot}, {Bouchet}, {Cardoso}, {Colombi}, {de Lapparent}, {Desjacques}, {Gavazzi}, {Kassin}, {Kimm}, {McCracken}, {Milliard}, {Peirani}, {Prunet}, {Rouberol}, {Silk}, {Slyz}, {Sousbie}, {Teyssier}, {Tresse}, {Treyer}, {Vibert}, \& {Volonteri}}]{Dubois2014}
{Dubois}, Y., {Pichon}, C., {Welker}, C., {et~al.} 2014, \mnras, 444, 1453

\bibitem[{{Erben} {et~al.}(2013){Erben}, {Hildebrandt}, {Miller}, {van Waerbeke}, {Heymans}, {Hoekstra}, {Kitching}, {Mellier}, {Benjamin}, {Blake}, {Bonnett}, {Cordes}, {Coupon}, {Fu}, {Gavazzi}, {Gillis}, {Grocutt}, {Gwyn}, {Holhjem}, {Hudson}, {Kilbinger}, {Kuijken}, {Milkeraitis}, {Rowe}, {Schrabback}, {Semboloni}, {Simon}, {Smit}, {Toader}, {Vafaei}, {van Uitert}, \& {Velander}}]{Erben2013}
{Erben}, T., {Hildebrandt}, H., {Miller}, L., {et~al.} 2013, \mnras, 433, 2545

\bibitem[{{Euclid Collaboration: Cardone, V.~F.} {et~al.}(2025){Euclid Collaboration: Cardone, V.~F.}, {Joudaki}, {Blot}, {Bonici}, {Camera}, {Ca{\~n}as-Herrera}, {Carrilho}, {Casas}, {Davini}, {Di Domizio}, {Farrens}, {Goh}, {Gouyou Beauchamps}, {Ili{\'c}}, {Keil}, {Le Brun}, {Martinelli}, {Moretti}, {Pettorino}, {Pezzotta}, {S{\'a}nchez}, {Sakr}, {Sciotti}, {Tanidis}, {Tutusaus}, {Ajani}, {Crocce}, {Giocoli}, {Legrand}, {Lembo}, {Lesci}, {Navarro Girones}, {Nouri-Zonoz}, {Pamuk}, {Tsedrik}, {Bel}, {Carbone}, {Duncan}, {Kilbinger}, {Lacasa}, {Lattanzi}, {Sapone}, {Sellentin}, {Taylor}, {Aghanim}, {Altieri}, {Amendola}, {Andreon}, {Auricchio}, {Aussel}, {Baccigalupi}, {Baldi}, {Bardelli}, {Battaglia}, {Biviano}, {Branchini}, {Brescia}, {Brinchmann}, {Capobianco}, {Carretero}, {Castellano}, {Castignani}, {Cavuoti}, {Chambers}, {Cimatti}, {Colodro-Conde}, {Congedo}, {Conselice}, {Conversi}, {Copin}, {Courbin}, {Courtois}, {Cropper}, {Da Silva}, {Degaudenzi}, {De Lucia}, {Di Giorgio}, {Douspis}, {Dubath},
  {Dupac}, {Dusini}, {Ealet}, {Escoffier}, {Farina}, {Farinelli}, {Faustini}, {Ferriol}, {Finelli}, {Fosalba}, {Fotopoulou}, {Frailis}, {Franceschi}, {Fumana}, {Galeotta}, {Gillis}, {G{\'o}mez-Alvarez}, {Gracia-Carpio}, {Granett}, {Grazian}, {Grupp}, {Guzzo}, {Haugan}, {Hoekstra}, {Holmes}, {Hook}, {Hormuth}, {Hornstrup}, {Jahnke}, {Jhabvala}, {Keih{\"a}nen}, {Kermiche}, {Kiessling}, {Kubik}, {K{\"u}mmel}, {Kunz}, {Kurki-Suonio}, {Lahav}, {Liebing}, {Lilje}, {Lindholm}, {Lloro}, {Mainetti}, {Maino}, {Maiorano}, {Mansutti}, {Marcin}, {Marggraf}, {Martinet}, {Marulli}, {Massey}, {Maurogordato}, {Medinaceli}, {Mei}, {Mellier}, {Meneghetti}, {Merlin}, {Meylan}, {Mora}, {Moresco}, {Moscardini}, {Nakajima}, {Neissner}, {Niemi}, {Padilla}, {Paltani}, {Pasian}, {Pedersen}, {Percival}, {Pires}, {Polenta}, {Poncet}, {Popa}, {Pozzetti}, {Racca}, {Raison}, {Rebolo}, {Renzi}, {Rhodes}, {Riccio}, {Romelli}, {Roncarelli}, {Saglia}, {Sartoris}, {Scaramella}, {Schewtschenko}, {Schneider}, {Schrabback}, {Secroun}, {Sefusatti},
  {Seidel}, {Serrano}, {Simon}, {Sirignano}, {Sirri}, {Stanco}, {Steinwagner}, {Tallada-Cresp{\'\i}}, {Taylor}, {Tereno}, {Toft}, {Toledo-Moreo}, {Torradeflot}, {Valenziano}, {Valiviita}, {Vassallo}, {Verdoes Kleijn}, {Veropalumbo}, {Wang}, {Weller}, {Zacchei}, {Zamorani}, {Zerbi}, {Zucca}, {Allevato}, {Ballardini}, {Bolzonella}, {Bozzo}, {Burigana}, \& {Cabanac}}]{Cloe}
{Euclid Collaboration: Cardone, V.~F.}, {Joudaki}, S., {Blot}, L., {et~al.} 2025, arXiv:2510.09118

\bibitem[{{Euclid Collaboration: Castander} {et~al.}(2025){Euclid Collaboration: Castander}, {Fosalba}, {Stadel}, {Potter}, {Carretero}, {Tallada-Cresp{\'\i}}, {Pozzetti}, {Bolzonella}, {Mamon}, {Blot}, {Hoffmann}, {Huertas-Company}, {Monaco}, {Gonzalez}, {De Lucia}, {Scarlata}, {Breton}, {Linke}, {Viglione}, {Li}, {Zhai}, {Baghkhani}, {Pardede}, {Neissner}, {Teyssier}, {Crocce}, {Tutusaus}, {Miller}, {Congedo}, {Biviano}, {Hirschmann}, {Pezzotta}, {Aussel}, {Hoekstra}, {Kitching}, {Percival}, {Guzzo}, {Mellier}, {Oesch}, {Bowler}, {Bruton}, {Allevato}, {Gonzalez-Perez}, {Manera}, {Avila}, {Kov{\'a}cs}, {Aghanim}, {Altieri}, {Amara}, {Amendola}, {Andreon}, {Auricchio}, {Baccigalupi}, {Baldi}, {Balestra}, {Bardelli}, {Bender}, {Bernardeau}, {Bodendorf}, {Bonino}, {Branchini}, {Brescia}, {Brinchmann}, {Camera}, {Capobianco}, {Carbone}, {Casas}, {Castellano}, {Castignani}, {Cavuoti}, {Cimatti}, {Colodro-Conde}, {Conselice}, {Conversi}, {Copin}, {Corcione}, {Courbin}, {Courtois}, {Da Silva}, {Degaudenzi}, {Di
  Giorgio}, {Dinis}, {Douspis}, {Dubath}, {Duncan}, {Dupac}, {Dusini}, {Ealet}, {Farina}, {Farrens}, {Ferriol}, {Fotopoulou}, {Fourmanoit}, {Frailis}, {Franceschi}, {Franzetti}, {Galeotta}, {Gillard}, {Gillis}, {Giocoli}, {G{\'o}mez-Alvarez}, {Granett}, {Grazian}, {Grupp}, {Haugan}, {Holliman}, {Holmes}, {Hook}, {Hormuth}, {Hornstrup}, {Hudelot}, {Ili{\'c}}, {Jahnke}, {Jhabvala}, {Joachimi}, {Keih{\"a}nen}, {Kermiche}, {Kiessling}, {Kilbinger}, {Kohley}, {Kubik}, {K{\"u}mmel}, {Kunz}, {Kurki-Suonio}, {Lahav}, {Laureijs}, {Le Mignant}, {Liebing}, {Ligori}, {Lilje}, {Lindholm}, {Lloro}, {Maino}, {Maiorano}, {Mansutti}, {Marcin}, {Marggraf}, {Markovic}, {Martinelli}, {Martinet}, {Marulli}, {Massey}, {Masters}, {Maurogordato}, {McCracken}, {Medinaceli}, {Mei}, {Melchior}, {Meneghetti}, {Merlin}, {Meylan}, {Mohr}, {Moresco}, {Moscardini}, {Munari}, {Nakajima}, {Nichol}, {Niemi}, {Padilla}, {Paech}, {Paltani}, {Pasian}, {Peacock}, {Pedersen}, {Pettorino}, {Pires}, {Polenta}, {Poncet}, {Popa}, {Raison}, {Rebolo},
  {Renzi}, {Rhodes}, {Riccio}, {Romelli}, {Roncarelli}, {Rosset}, {Rossetti}, {Rusholme}, {Saglia}, {Sakr}, {S{\'a}nchez}, {Sapone}, {Schewtschenko}, {Schirmer}, {Schneider}, {Schrabback}, {Scodeggio}, {Secroun}, {Sefusatti}, {Seidel}, {Serrano}, {Sirignano}, {Sirri}, {Stanco}, {Starck}, {Steinwagner}, {Taylor}, \& {Teplitz}}]{Castander2024}
{Euclid Collaboration: Castander}, F., {Fosalba}, P., {Stadel}, J., {et~al.} 2025, \aap, 697, A5

\bibitem[{{Euclid Collaboration: Hoffman, K.} {et~al.}(2026){Euclid Collaboration: Hoffman, K.}, {Paviot}, {Joachimi}, {Tessore}, {Tallada-Cresp{\'\i}}, {Chisari}, {Gonzalez}, {Loureiro}, {Fosalba}, {Blazek}, {Laigle}, {Dubois}, {Pichon}, {Altieri}, {Andreon}, {Auricchio}, {Baccigalupi}, {Baldi}, {Bardelli}, {Bernardeau}, {Biviano}, {Branchini}, {Brescia}, {Camera}, {Ca{\~n}as-Herrera}, {Capobianco}, {Carbone}, {Cardone}, {Carretero}, {Casas}, {Castander}, {Castellano}, {Castignani}, {Cavuoti}, {Chambers}, {Cimatti}, {Colodro-Conde}, {Congedo}, {Conversi}, {Copin}, {Courbin}, {Courtois}, {Da Silva}, {Degaudenzi}, {De Lucia}, {Dole}, {Dubath}, {Duncan}, {Dupac}, {Dusini}, {Escoffier}, {Farina}, {Farinelli}, {Farrens}, {Ferriol}, {Finelli}, {Fourmanoit}, {Frailis}, {Franceschi}, {Fumana}, {Galeotta}, {George}, {Gillis}, {Giocoli}, {Gracia-Carpio}, {Grazian}, {Grupp}, {Haugan}, {Hoekstra}, {Holmes}, {Hormuth}, {Hornstrup}, {Jahnke}, {Jhabvala}, {Keih{\"a}nen}, {Kermiche}, {Kiessling}, {Kilbinger}, {Kubik},
  {K{\"u}mmel}, {Kunz}, {Kurki-Suonio}, {Le Brun}, {Ligori}, {Lilje}, {Lindholm}, {Lloro}, {Mainetti}, {Maino}, {Maiorano}, {Mansutti}, {Marcin}, {Marggraf}, {Martinelli}, {Martinet}, {Marulli}, {Massey}, {Medinaceli}, {Mei}, {Mellier}, {Meneghetti}, {Merlin}, {Meylan}, {Mora}, {Moresco}, {Moscardini}, {Neissner}, {Niemi}, {Padilla}, {Paltani}, {Pasian}, {Pedersen}, {Pettorino}, {Pires}, {Polenta}, {Poncet}, {Popa}, {Pozzetti}, {Raison}, {Rebolo}, {Renzi}, {Rhodes}, {Riccio}, {Romelli}, {Roncarelli}, {Saglia}, {Sakr}, {S{\'a}nchez}, {Sapone}, {Sartoris}, {Schneider}, {Schrabback}, {Secroun}, {Sefusatti}, {Seidel}, {Serrano}, {Simon}, {Sirignano}, {Sirri}, {Spurio Mancini}, {Stanco}, {Steinwagner}, {Taylor}, {Tereno}, {Toft}, {Toledo-Moreo}, {Torradeflot}, {Tutusaus}, {Valenziano}, {Valiviita}, {Vassallo}, {Veropalumbo}, {Wang}, {Weller}, {Zamorani}, {Zerbi}, {Zucca}, {Ballardini}, {Bozzo}, {Burigana}, {Cabanac}, {Calabrese}, {Cappi}, {Di Ferdinando}, {Escartin Vigo}, {Gabarra}, {Hartley}, {Matthew}, {Maturi},
  {Mauri}, {Metcalf}, {Pezzotta}, {P{\"o}ntinen}, {Porciani}, {Risso}, {Scottez}, {Sereno}, {Tenti}, {Viel}, {Wiesmann}, {Akrami}, {Alvi}, {Andika}, {Anselmi}, {Archidiacono}, {Atrio-Barandela}, {Bertacca}, {Bethermin}, {Blanchard}, {Blot}, {Bonici}, {Borgani}, {Brown}, {Bruton}, {Calabro}, {Camacho Quevedo}, {Caro}, {Carvalho}, \& {Castro}}]{Hoffmann2026}
{Euclid Collaboration: Hoffman, K.}, {Paviot}, R., {Joachimi}, B., {et~al.} 2026, arXiv e-prints, arXiv:2601.07785

\bibitem[{{Euclid Collaboration: Mellier} {et~al.}(2025){Euclid Collaboration: Mellier}, {Abdurro'uf}, {Acevedo Barroso}, {Ach{\'u}carro}, {Adamek}, {Adam}, {Addison}, {Aghanim}, {Aguena}, {Ajani}, {Akrami}, {Al-Bahlawan}, {Alavi}, {Albuquerque}, {Alestas}, {Alguero}, {Allaoui}, {Allen}, {Allevato}, {Alonso-Tetilla}, {Altieri}, {Alvarez-Candal}, {Alvi}, {Amara}, {Amendola}, {Amiaux}, {Andika}, {Andreon}, {Andrews}, {Angora}, {Angulo}, {Annibali}, {Anselmi}, {Anselmi}, {Arcari}, {Archidiacono}, {Aric{\`o}}, {Arnaud}, {Arnouts}, {Asgari}, {Asorey}, {Atayde}, {Atek}, {Atrio-Barandela}, {Aubert}, {Aubourg}, {Auphan}, {Auricchio}, {Aussel}, {Aussel}, {Avelino}, {Avgoustidis}, {Avila}, {Awan}, {Azzollini}, {Baccigalupi}, {Bachelet}, {Bacon}, {Baes}, {Bagley}, {Bahr-Kalus}, {Balaguera-Antolinez}, {Balbinot}, {Balcells}, {Baldi}, {Baldry}, {Balestra}, {Ballardini}, {Ballester}, {Balogh}, {Ba{\~n}ados}, {Barbier}, {Bardelli}, {Baron}, {Barreiro}, {Barrena}, {Barriere}, {Barros}, {Barthelemy}, {Bartolo}, {Basset},
  {Battaglia}, {Battisti}, {Baugh}, {Baumont}, {Bazzanini}, {Beaulieu}, {Beckmann}, {Belikov}, {Bel}, {Bellagamba}, {Bella}, {Bellini}, {Benabed}, {Bender}, {Benevento}, {Bennett}, {Benson}, {Bergamini}, {Bermejo-Climent}, {Bernardeau}, {Bertacca}, {Berthe}, {Berthier}, {Bethermin}, {Beutler}, {Bevillon}, {Bhargava}, {Bhatawdekar}, {Bianchi}, {Bisigello}, {Biviano}, {Blake}, {Blanchard}, {Blazek}, {Blot}, {Bosco}, {Bodendorf}, {Boenke}, {B{\"o}hringer}, {Boldrini}, {Bolzonella}, {Bonchi}, {Bonici}, {Bonino}, {Bonino}, {Bonvin}, {Bon}, {Booth}, {Borgani}, {Borlaff}, {Borsato}, {Bose}, {Botticella}, {Boucaud}, {Bouche}, {Boucher}, {Boutigny}, {Bouvard}, {Bouwens}, {Bouy}, {Bowler}, {Bozza}, {Bozzo}, {Branchini}, {Brando}, {Brau-Nogue}, {Brekke}, {Bremer}, {Brescia}, {Breton}, {Brinchmann}, {Brinckmann}, {Brockley-Blatt}, {Brodwin}, {Brouard}, {Brown}, {Bruton}, {Bucko}, {Buddelmeijer}, {Buenadicha}, {Buitrago}, {Burger}, {Burigana}, {Busillo}, {Busonero}, {Cabanac}, {Cabayol-Garcia}, {Cagliari}, {Caillat},
  {Caillat}, {Calabrese}, {Calabro}, {Calderone}, {Calura}, {Camacho Quevedo}, {Camera}, {Campos}, {Ca{\~n}as-Herrera}, {Candini}, {Cantiello}, {Capobianco}, {Cappellaro}, {Cappelluti}, {Cappi}, {Caputi}, {Cara}, {Carbone}, {Cardone}, {Carella}, {Carlberg}, {Carle}, {Carminati}, {Caro}, {Carrasco}, {Carretero}, {Carrilho}, {Carron Duque}, \& {Carry}}]{Mellier2025}
{Euclid Collaboration: Mellier}, {Abdurro'uf}, {Acevedo Barroso}, J.~A., {et~al.} 2025, \aap, 697, A1

\bibitem[{{Euclid Collaboration: Monaco, P.} {et~al.}(2025){Euclid Collaboration: Monaco, P.}, {Parimbelli}, {Elkhashab}, {Salvalaggio}, {Castro}, {Lepinzan}, {Sarpa}, {Sefusatti}, {Stanco}, {Tornatore}, {Addison}, {Bruton}, {Carbone}, {Castander}, {Carretero}, {de la Torre}, {Fosalba}, {Lavaux}, {Lee}, {Markovic}, {McCarthy}, {Passalacqua}, {Percival}, {Risso}, {Scarlata}, {Tallada-Cresp{\'\i}}, {Viel}, {Wang}, {Altieri}, {Andreon}, {Auricchio}, {Baccigalupi}, {Baldi}, {Bardelli}, {Battaglia}, {Bernardeau}, {Biviano}, {Branchini}, {Brescia}, {Brinchmann}, {Camera}, {Ca{\~n}as-Herrera}, {Capobianco}, {Cardone}, {Casas}, {Castellano}, {Castignani}, {Cavuoti}, {Cimatti}, {Colodro-Conde}, {Congedo}, {Conselice}, {Conversi}, {Copin}, {Courbin}, {Courtois}, {Da Silva}, {Degaudenzi}, {De Lucia}, {Di Giorgio}, {Dubath}, {Ducret}, {Duncan}, {Dupac}, {Dusini}, {Ealet}, {Escoffier}, {Farina}, {Farinelli}, {Farrens}, {Ferriol}, {Finelli}, {Fourmanoit}, {Frailis}, {Franceschi}, {Fumana}, {Galeotta}, {George}, {Gillis},
  {Giocoli}, {Gracia-Carpio}, {Grazian}, {Grupp}, {Guzzo}, {Haugan}, {Holmes}, {Hormuth}, {Hornstrup}, {Jahnke}, {Jhabvala}, {Joachimi}, {Keih{\"a}nen}, {Kermiche}, {Kubik}, {K{\"u}mmel}, {Kunz}, {Kurki-Suonio}, {Le Brun}, {Ligori}, {Lilje}, {Lindholm}, {Lloro}, {Maino}, {Maiorano}, {Mansutti}, {Marggraf}, {Martinelli}, {Martinet}, {Marulli}, {Massey}, {Medinaceli}, {Mei}, {Melchior}, {Mellier}, {Meneghetti}, {Merlin}, {Meylan}, {Mora}, {Moresco}, {Moscardini}, {Munari}, {Nakajima}, {Neissner}, {Niemi}, {Padilla}, {Paltani}, {Pasian}, {Pedersen}, {Pettorino}, {Pires}, {Polenta}, {Poncet}, {Popa}, {Pozzetti}, {Raison}, {Renzi}, {Rhodes}, {Riccio}, {Rizzo}, {Romelli}, {Roncarelli}, {Saglia}, {Sakr}, {S{\'a}nchez}, {Sapone}, {Sartoris}, {Schneider}, {Schrabback}, {Scodeggio}, {Secroun}, {Seidel}, {Seiffert}, {Serrano}, {Simon}, {Sirignano}, {Sirri}, {Steinwagner}, {Tavagnacco}, {Taylor}, {Tereno}, {Tessore}, {Toft}, {Toledo-Moreo}, {Torradeflot}, {Tutusaus}, {Valenziano}, {Valiviita}, {Vassallo}, {Verdoes
  Kleijn}, {Veropalumbo}, {Weller}, {Zamorani}, {Zucca}, {Allevato}, {Ballardini}, {Burigana}, {Cabanac}, {Calabrese}, {Cappi}, {Di Ferdinando}, {Escartin Vigo}, {Fabbian}, {Gabarra}, {Mart{\'\i}n-Fleitas}, {Matthew}, {Mauri}, {Metcalf}, {Pezzotta}, {P{\"o}ntinen}, {Porciani}, {Scottez}, {Sereno}, {Tenti}, {Wiesmann}, {Akrami}, {Alvi}, {Andika}, {Anselmi}, \& {Archidiacono}}]{Monaco2025}
{Euclid Collaboration: Monaco, P.}, {Parimbelli}, G., {Elkhashab}, M.~Y., {et~al.} 2025, arXiv:2507.12116

\bibitem[{{Euclid Collaboration: Pocino, A.} {et~al.}(2021){Euclid Collaboration: Pocino, A.}, {Tutusaus}, {Castander}, {Fosalba}, {Crocce}, {Porredon}, {Camera}, {Cardone}, {Casas}, {Kitching}, {Lacasa}, {Martinelli}, {Pourtsidou}, {Sakr}, {Andreon}, {Auricchio}, {Baccigalupi}, {Balaguera-Antol{\'\i}nez}, {Baldi}, {Balestra}, {Bardelli}, {Bender}, {Biviano}, {Bodendorf}, {Bonino}, {Boucaud}, {Bozzo}, {Branchini}, {Brescia}, {Brinchmann}, {Burigana}, {Cabanac}, {Capobianco}, {Cappi}, {Carvalho}, {Castellano}, {Castignani}, {Cavuoti}, {Cimatti}, {Cledassou}, {Colodro-Conde}, {Congedo}, {Conselice}, {Conversi}, {Copin}, {Corcione}, {Costille}, {Coupon}, {Courtois}, {Cropper}, {Cuby}, {Da Silva}, {de la Torre}, {Di Ferdinando}, {Dubath}, {Duncan}, {Dupac}, {Dusini}, {Farrens}, {Ferreira}, {Ferrero}, {Finelli}, {Fotopoulou}, {Frailis}, {Franceschi}, {Galeotta}, {Garilli}, {Gillard}, {Gillis}, {Giocoli}, {Gozaliasl}, {Graci{\'a}-Carpio}, {Grupp}, {Guzzo}, {Holmes}, {Hormuth}, {Jahnke}, {Keihanen}, {Kermiche},
  {Kiessling}, {Kirkpatrick}, {Kunz}, {Kurki-Suonio}, {Ligori}, {Lilje}, {Lloro}, {Maino}, {Maiorano}, {Mansutti}, {Marggraf}, {Martinet}, {Marulli}, {Massey}, {Maurogordato}, {Medinaceli}, {Mei}, {Meneghetti}, {Benton Metcalf}, {Meylan}, {Moresco}, {Morin}, {Moscardini}, {Munari}, {Nakajima}, {Neissner}, {Nichol}, {Niemi}, {Nightingale}, {Padilla}, {Paltani}, {Pasian}, {Patrizii}, {Pedersen}, {Percival}, {Pettorino}, {Pires}, {Polenta}, {Poncet}, {Popa}, {Potter}, {Pozzetti}, {Raison}, {Renzi}, {Rhodes}, {Riccio}, {Romelli}, {Roncarelli}, {Rossetti}, {Saglia}, {S{\'a}nchez}, {Sapone}, {Scaramella}, {Schneider}, {Scottez}, {Secroun}, {Seidel}, {Serrano}, {Sirignano}, {Sirri}, {Stanco}, {Sureau}, {Taylor}, {Tenti}, {Tereno}, {Teyssier}, {Toledo-Moreo}, {Tramacere}, {Valentijn}, {Valenziano}, {Valiviita}, {Vassallo}, {Viel}, {Wang}, {Welikala}, {Whittaker}, {Zacchei}, {Zamorani}, {Zoubian}, \& {Zucca}}]{Pocino2021}
{Euclid Collaboration: Pocino, A.}, {Tutusaus}, I., {Castander}, F.~J., {et~al.} 2021, \aap, 655, A44

\bibitem[{{Euclid Collaboration: Scaramella, R.} {et~al.}(2022){Euclid Collaboration: Scaramella, R.}, {Amiaux}, {Mellier}, {Burigana}, {Carvalho}, {Cuillandre}, {Da Silva}, {Derosa}, {Dinis}, {Maiorano}, {Maris}, {Tereno}, {Laureijs}, {Boenke}, {Buenadicha}, {Dupac}, {Gaspar Venancio}, {G{\'o}mez-{\'A}lvarez}, {Hoar}, {Lorenzo Alvarez}, {Racca}, {Saavedra-Criado}, {Schwartz}, {Vavrek}, {Schirmer}, {Aussel}, {Azzollini}, {Cardone}, {Cropper}, {Ealet}, {Garilli}, {Gillard}, {Granett}, {Guzzo}, {Hoekstra}, {Jahnke}, {Kitching}, {Maciaszek}, {Meneghetti}, {Miller}, {Nakajima}, {Niemi}, {Pasian}, {Percival}, {Pottinger}, {Sauvage}, {Scodeggio}, {Wachter}, {Zacchei}, {Aghanim}, {Amara}, {Auphan}, {Auricchio}, {Awan}, {Balestra}, {Bender}, {Bodendorf}, {Bonino}, {Branchini}, {Brau-Nogue}, {Brescia}, {Candini}, {Capobianco}, {Carbone}, {Carlberg}, {Carretero}, {Casas}, {Castander}, {Castellano}, {Cavuoti}, {Cimatti}, {Cledassou}, {Congedo}, {Conselice}, {Conversi}, {Copin}, {Corcione}, {Costille}, {Courbin},
  {Degaudenzi}, {Douspis}, {Dubath}, {Duncan}, {Dusini}, {Farrens}, {Ferriol}, {Fosalba}, {Fourmanoit}, {Frailis}, {Franceschi}, {Franzetti}, {Fumana}, {Gillis}, {Giocoli}, {Grazian}, {Grupp}, {Haugan}, {Holmes}, {Hormuth}, {Hudelot}, {Kermiche}, {Kiessling}, {Kilbinger}, {Kohley}, {Kubik}, {K{\"u}mmel}, {Kunz}, {Kurki-Suonio}, {Lahav}, {Ligori}, {Lilje}, {Lloro}, {Mansutti}, {Marggraf}, {Markovic}, {Marulli}, {Massey}, {Maurogordato}, {Melchior}, {Merlin}, {Meylan}, {Mohr}, {Moresco}, {Morin}, {Moscardini}, {Munari}, {Nichol}, {Padilla}, {Paltani}, {Peacock}, {Pedersen}, {Pettorino}, {Pires}, {Poncet}, {Popa}, {Pozzetti}, {Raison}, {Rebolo}, {Rhodes}, {Rix}, {Roncarelli}, {Rossetti}, {Saglia}, {Schneider}, {Schrabback}, {Secroun}, {Seidel}, {Serrano}, {Sirignano}, {Sirri}, {Skottfelt}, {Stanco}, {Starck}, {Tallada-Cresp{\'\i}}, {Tavagnacco}, {Taylor}, {Teplitz}, {Toledo-Moreo}, {Torradeflot}, {Trifoglio}, {Valentijn}, {Valenziano}, {Verdoes Kleijn}, {Wang}, {Welikala}, {Weller}, {Wetzstein}, {Zamorani},
  {Zoubian}, {Andreon}, {Baldi}, {Bardelli}, {Boucaud}, {Camera}, {Di Ferdinando}, {Fabbian}, {Farinelli}, {Galeotta}, {Graci{\'a}-Carpio}, {Maino}, {Medinaceli}, {Mei}, {Neissner}, {Polenta}, {Renzi}, {Romelli}, {Rosset}, {Sureau}, {Tenti}, {Vassallo}, {Zucca}, {Baccigalupi}, {Balaguera-Antol{\'\i}nez}, {Battaglia}, {Biviano}, {Borgani}, {Bozzo}, {Cabanac}, \& {Cappi}}]{Scaramella2022}
{Euclid Collaboration: Scaramella, R.}, {Amiaux}, J., {Mellier}, Y., {et~al.} 2022, \aap, 662, A112

\bibitem[{{Euclid Collaboration: Serrano} {et~al.}(2024){Euclid Collaboration: Serrano}, {Hudelot}, {Seidel}, {Pollack}, {Jullo}, {Torradeflot}, {Benielli}, {Fahed}, {Auphan}, {Carretero}, {Aussel}, {Casenove}, {Castander}, {Davies}, {Fourmanoit}, {Huot}, {Kara}, {Keih{\"a}nen}, {Kermiche}, {Okumura}, {Zoubian}, {Ealet}, {Boucaud}, {Bretonni{\`e}re}, {Casas}, {Cl{\'e}ment}, {Duncan}, {George}, {Kiiveri}, {Kurki-Suonio}, {K{\"u}mmel}, {Laugier}, {Mainetti}, {Mohr}, {Montoro}, {Neissner}, {Rosset}, {Schirmer}, {Tallada-Cresp{\'\i}}, {Tonello}, {Venhola}, {Verderi}, {Zacchei}, {Aghanim}, {Altieri}, {Amara}, {Andreon}, {Auricchio}, {Azzollini}, {Baccigalupi}, {Baldi}, {Bardelli}, {Basset}, {Battaglia}, {Bernardeau}, {Bodendorf}, {Bonino}, {Branchini}, {Brescia}, {Brinchmann}, {Camera}, {Candini}, {Capobianco}, {Carbone}, {Casas}, {Castellano}, {Castignani}, {Cavuoti}, {Cimatti}, {Cledassou}, {Colodro-Conde}, {Congedo}, {Conselice}, {Conversi}, {Copin}, {Corcione}, {Courbin}, {Courtois}, {Crocce}, {Cropper}, {Da
  Silva}, {Degaudenzi}, {De Lucia}, {Di Giorgio}, {Dinis}, {Dubath}, {Dupac}, {Dusini}, {Farina}, {Farrens}, {Ferriol}, {Frailis}, {Franceschi}, {Franzetti}, {Galeotta}, {Garilli}, {Gillard}, {Gillis}, {Giocoli}, {Granett}, {Grazian}, {Grupp}, {Guzzo}, {Haugan}, {Hoar}, {Hoekstra}, {Holmes}, {Hook}, {Hormuth}, {Hornstrup}, {Jahnke}, {Joachimi}, {Kiessling}, {Kitching}, {Kohley}, {Kunz}, {Le Boulc'h}, {Liebing}, {Ligori}, {Lilje}, {Lindholm}, {Lloro}, {Maino}, {Maiorano}, {Mansutti}, {Marcin}, {Marggraf}, {Markovic}, {Martinelli}, {Martinet}, {Marulli}, {Massey}, {Maurogordato}, {Medinaceli}, {Mei}, {Melchior}, {Mellier}, {Meneghetti}, {Merlin}, {Meylan}, {Moresco}, {Morris}, {Moscardini}, {Munari}, {Nakajima}, {Niemi}, {Nutma}, {Padilla}, {Paltani}, {Pasian}, {Pedersen}, {Percival}, {Pettorino}, {Pires}, {Polenta}, {Poncet}, {Popa}, {Pozzetti}, {Raison}, {Rebolo}, {Renzi}, {Rhodes}, {Riccio}, {Romelli}, {Roncarelli}, {Rossetti}, {Rusholme}, {Saglia}, {Sakr}, {S{\'a}nchez}, {Sapone}, {Sartoris}, {Sauvage},
  {Schneider}, {Schrabback}, {Scodeggio}, {Secroun}, {Sirignano}, {Sirri}, {Skottfelt}, {Stanco}, {Starck}, {Steinwagner}, {Taylor}, {Teplitz}, {Tereno}, {Toledo-Moreo}, {Tutusaus}, {Valentijn}, {Valenziano}, {Vassallo}, {Veropalumbo}, {Wang}, {Weller}, {Zamorani}, {Zucca}, {Biviano}, {Bozzo}, \& {Di Ferdinando}}]{Serrano2024}
{Euclid Collaboration: Serrano}, S., {Hudelot}, P., {Seidel}, G., {et~al.} 2024, \aap, 690, A103

\bibitem[{{Foreman-Mackey} {et~al.}(2013){Foreman-Mackey}, {Hogg}, {Lang}, \& {Goodman}}]{Foreman2013}
{Foreman-Mackey}, D., {Hogg}, D.~W., {Lang}, D., \& {Goodman}, J. 2013, \pasp, 125, 306

\bibitem[{{Grogin} {et~al.}(2012){Grogin}, {Rajan}, {Donley}, {Kartaltepe}, {Koekemoer}, {Lucas}, {Rosario}, \& {Salvato}}]{Grogin2012}
{Grogin}, N.~A., {Rajan}, A., {Donley}, J.~L., {et~al.} 2012, in American Astronomical Society Meeting Abstracts, Vol. 220, American Astronomical Society Meeting Abstracts \#220, 335.23

\bibitem[{{Guzzo} \& {VIPERS Team}(2013)}]{Guzzo:2013}
{Guzzo}, L. \& {VIPERS Team}. 2013, The Messenger, 151, 41

\bibitem[{{Haslbauer} {et~al.}(2022){Haslbauer}, {Banik}, {Kroupa}, {Wittenburg}, \& {Javanmardi}}]{Haslbauer2022}
{Haslbauer}, M., {Banik}, I., {Kroupa}, P., {Wittenburg}, N., \& {Javanmardi}, B. 2022, \apj, 925, 183

\bibitem[{{Hatfield} {et~al.}(2019){Hatfield}, {Laigle}, {Jarvis}, {Devriendt}, {Davidzon}, {Ilbert}, {Pichon}, \& {Dubois}}]{Hatfield2019}
{Hatfield}, P.~W., {Laigle}, C., {Jarvis}, M.~J., {et~al.} 2019, \mnras, 490, 5043

\bibitem[{{Hearin} {et~al.}(2016){Hearin}, {Zentner}, {van den Bosch}, {Campbell}, \& {Tollerud}}]{Hearin2016}
{Hearin}, A.~P., {Zentner}, A.~R., {van den Bosch}, F.~C., {Campbell}, D., \& {Tollerud}, E. 2016, \mnras, 460, 2552

\bibitem[{{Ivezi{\'c}} {et~al.}(2019){Ivezi{\'c}}, {Kahn}, {Tyson}, {Abel}, {Acosta}, {Allsman}, {Alonso}, {AlSayyad}, {Anderson}, {Andrew}, {Angel}, {Angeli}, {Ansari}, {Antilogus}, {Araujo}, {Armstrong}, {Arndt}, {Astier}, {Aubourg}, {Auza}, {Axelrod}, {Bard}, {Barr}, {Barrau}, {Bartlett}, {Bauer}, {Bauman}, {Baumont}, {Bechtol}, {Bechtol}, {Becker}, {Becla}, {Beldica}, {Bellavia}, {Bianco}, {Biswas}, {Blanc}, {Blazek}, {Blandford}, {Bloom}, {Bogart}, {Bond}, {Booth}, {Borgland}, {Borne}, {Bosch}, {Boutigny}, {Brackett}, {Bradshaw}, {Brandt}, {Brown}, {Bullock}, {Burchat}, {Burke}, {Cagnoli}, {Calabrese}, {Callahan}, {Callen}, {Carlin}, {Carlson}, {Chandrasekharan}, {Charles-Emerson}, {Chesley}, {Cheu}, {Chiang}, {Chiang}, {Chirino}, {Chow}, {Ciardi}, {Claver}, {Cohen-Tanugi}, {Cockrum}, {Coles}, {Connolly}, {Cook}, {Cooray}, {Covey}, {Cribbs}, {Cui}, {Cutri}, {Daly}, {Daniel}, {Daruich}, {Daubard}, {Daues}, {Dawson}, {Delgado}, {Dellapenna}, {de Peyster}, {de Val-Borro}, {Digel}, {Doherty}, {Dubois},
  {Dubois-Felsmann}, {Durech}, {Economou}, {Eifler}, {Eracleous}, {Emmons}, {Fausti Neto}, {Ferguson}, {Figueroa}, {Fisher-Levine}, {Focke}, {Foss}, {Frank}, {Freemon}, {Gangler}, {Gawiser}, {Geary}, {Gee}, {Geha}, {Gessner}, {Gibson}, {Gilmore}, {Glanzman}, {Glick}, {Goldina}, {Goldstein}, {Goodenow}, {Graham}, {Gressler}, {Gris}, {Guy}, {Guyonnet}, {Haller}, {Harris}, {Hascall}, {Haupt}, {Hernandez}, {Herrmann}, {Hileman}, {Hoblitt}, {Hodgson}, {Hogan}, {Howard}, {Huang}, {Huffer}, {Ingraham}, {Innes}, {Jacoby}, {Jain}, {Jammes}, {Jee}, {Jenness}, {Jernigan}, {Jevremovi{\'c}}, {Johns}, {Johnson}, {Johnson}, {Jones}, {Juramy-Gilles}, {Juri{\'c}}, {Kalirai}, {Kallivayalil}, {Kalmbach}, {Kantor}, {Karst}, {Kasliwal}, {Kelly}, {Kessler}, {Kinnison}, {Kirkby}, {Knox}, {Kotov}, {Krabbendam}, {Krughoff}, {Kub{\'a}nek}, {Kuczewski}, {Kulkarni}, {Ku}, {Kurita}, {Lage}, {Lambert}, {Lange}, {Langton}, {Le Guillou}, {Levine}, {Liang}, {Lim}, {Lintott}, {Long}, {Lopez}, {Lotz}, {Lupton}, {Lust}, {MacArthur}, {Mahabal},
  {Mandelbaum}, {Markiewicz}, {Marsh}, {Marshall}, {Marshall}, {May}, {McKercher}, {McQueen}, {Meyers}, {Migliore}, {Miller}, {Mills}, {Miraval}, {Moeyens}, {Moolekamp}, {Monet}, {Moniez}, {Monkewitz}, {Montgomery}, {Morrison}, {Mueller}, {Muller}, {Mu{\~n}oz Arancibia}, {Neill}, {Newbry}, {Nief}, {Nomerotski}, {Nordby}, {O'Connor}, {Oliver}, {Olivier}, {Olsen}, {O'Mullane}, {Ortiz}, {Osier}, {Owen}, {Pain}, {Palecek}, {Parejko}, {Parsons}, {Pease}, {Peterson}, {Peterson}, {Petravick}, {Libby Petrick}, {Petry}, {Pierfederici}, {Pietrowicz}, {Pike}, {Pinto}, {Plante}, {Plate}, {Plutchak}, {Price}, {Prouza}, {Radeka}, {Rajagopal}, {Rasmussen}, {Regnault}, {Reil}, {Reiss}, {Reuter}, {Ridgway}, {Riot}, {Ritz}, {Robinson}, {Roby}, {Roodman}, {Rosing}, {Roucelle}, {Rumore}, {Russo}, {Saha}, {Sassolas}, {Schalk}, {Schellart}, {Schindler}, {Schmidt}, {Schneider}, {Schneider}, {Schoening}, {Schumacher}, {Schwamb}, {Sebag}, {Selvy}, {Sembroski}, {Seppala}, {Serio}, {Serrano}, {Shaw}, {Shipsey}, {Sick}, {Silvestri},
  {Slater}, {Smith}, {Smith}, {Sobhani}, {Soldahl}, {Storrie-Lombardi}, {Stover}, {Strauss}, {Street}, {Stubbs}, {Sullivan}, {Sweeney}, {Swinbank}, {Szalay}, {Takacs}, {Tether}, {Thaler}, {Thayer}, {Thomas}, {Thornton}, {Thukral}, {Tice}, {Trilling}, {Turri}, {Van Berg}, {Vanden Berk}, {Vetter}, {Virieux}, {Vucina}, {Wahl}, {Walkowicz}, {Walsh}, {Walter}, {Wang}, {Wang}, {Warner}, {Wiecha}, {Willman}, {Winters}, {Wittman}, {Wolff}, {Wood-Vasey}, {Wu}, {Xin}, {Yoachim}, \& {Zhan}}]{Ivezic2019}
{Ivezi{\'c}}, {\v{Z}}., {Kahn}, S.~M., {Tyson}, J.~A., {et~al.} 2019, \apj, 873, 111

\bibitem[{{Jarvis} {et~al.}(2013){Jarvis}, {H{\"a}u{\ss}ler}, \& {McAlpine}}]{Jarvis2013}
{Jarvis}, M.~J., {H{\"a}u{\ss}ler}, B., \& {McAlpine}, K. 2013, The Messenger, 154, 26

\bibitem[{{Jing} \& {Suto}(2002)}]{Jing2002}
{Jing}, Y.~P. \& {Suto}, Y. 2002, \apj, 574, 538

\bibitem[{{Johnston} {et~al.}(2019){Johnston}, {Georgiou}, {Joachimi}, {Hoekstra}, {Chisari}, {Farrow}, {Fortuna}, {Heymans}, {Joudaki}, {Kuijken}, \& {Wright}}]{Johnston2019}
{Johnston}, H., {Georgiou}, C., {Joachimi}, B., {et~al.} 2019, \aap, 624, A30

\bibitem[{{Kaviraj} {et~al.}(2017){Kaviraj}, {Laigle}, {Kimm}, {Devriendt}, {Dubois}, {Pichon}, {Slyz}, {Chisari}, \& {Peirani}}]{Kaviraj2017}
{Kaviraj}, S., {Laigle}, C., {Kimm}, T., {et~al.} 2017, \mnras, 467, 4739

\bibitem[{{Kerscher} {et~al.}(2000){Kerscher}, {Szapudi}, \& {Szalay}}]{Kerscher2000}
{Kerscher}, M., {Szapudi}, I., \& {Szalay}, A.~S. 2000, \apjl, 535, L13

\bibitem[{{Kim} {et~al.}(2016){Kim}, {Agertz}, {Teyssier}, {Butler}, {Ceverino}, {Choi}, {Feldmann}, {Keller}, {Lupi}, {Quinn}, {Revaz}, {Wallace}, {Gnedin}, {Leitner}, {Shen}, {Smith}, {Thompson}, {Turk}, {Abel}, {Arraki}, {Benincasa}, {Chakrabarti}, {DeGraf}, {Dekel}, {Goldbaum}, {Hopkins}, {Hummels}, {Klypin}, {Li}, {Madau}, {Mandelker}, {Mayer}, {Nagamine}, {Nickerson}, {O'Shea}, {Primack}, {Roca-F{\`a}brega}, {Semenov}, {Shimizu}, {Simpson}, {Todoroki}, {Wadsley}, {Wise}, \& {AGORA Collaboration}}]{Kim2016}
{Kim}, J.-h., {Agertz}, O., {Teyssier}, R., {et~al.} 2016, \apj, 833, 202

\bibitem[{{Koekemoer} {et~al.}(2011){Koekemoer}, {Faber}, {Ferguson}, {Grogin}, {Kocevski}, {Koo}, {Lai}, {Lotz}, {Lucas}, {McGrath}, {Ogaz}, {Rajan}, {Riess}, {Rodney}, {Strolger}, {Casertano}, {Castellano}, {Dahlen}, {Dickinson}, {Dolch}, {Fontana}, {Giavalisco}, {Grazian}, {Guo}, {Hathi}, {Huang}, {van der Wel}, {Yan}, {Acquaviva}, {Alexander}, {Almaini}, {Ashby}, {Barden}, {Bell}, {Bournaud}, {Brown}, {Caputi}, {Cassata}, {Challis}, {Chary}, {Cheung}, {Cirasuolo}, {Conselice}, {Roshan Cooray}, {Croton}, {Daddi}, {Dav{\'e}}, {de Mello}, {de Ravel}, {Dekel}, {Donley}, {Dunlop}, {Dutton}, {Elbaz}, {Fazio}, {Filippenko}, {Finkelstein}, {Frazer}, {Gardner}, {Garnavich}, {Gawiser}, {Gruetzbauch}, {Hartley}, {H{\"a}ussler}, {Herrington}, {Hopkins}, {Huang}, {Jha}, {Johnson}, {Kartaltepe}, {Khostovan}, {Kirshner}, {Lani}, {Lee}, {Li}, {Madau}, {McCarthy}, {McIntosh}, {McLure}, {McPartland}, {Mobasher}, {Moreira}, {Mortlock}, {Moustakas}, {Mozena}, {Nandra}, {Newman}, {Nielsen}, {Niemi}, {Noeske}, {Papovich},
  {Pentericci}, {Pope}, {Primack}, {Ravindranath}, {Reddy}, {Renzini}, {Rix}, {Robaina}, {Rosario}, {Rosati}, {Salimbeni}, {Scarlata}, {Siana}, {Simard}, {Smidt}, {Snyder}, {Somerville}, {Spinrad}, {Straughn}, {Telford}, {Teplitz}, {Trump}, {Vargas}, {Villforth}, {Wagner}, {Wandro}, {Wechsler}, {Weiner}, {Wiklind}, {Wild}, {Wilson}, {Wuyts}, \& {Yun}}]{Koekemoer2011}
{Koekemoer}, A.~M., {Faber}, S.~M., {Ferguson}, H.~C., {et~al.} 2011, \apjs, 197, 36

\bibitem[{{Komatsu} {et~al.}(2011){Komatsu}, {Smith}, {Dunkley}, {Bennett}, {Gold}, {Hinshaw}, {Jarosik}, {Larson}, {Nolta}, {Page}, {Spergel}, {Halpern}, {Hill}, {Kogut}, {Limon}, {Meyer}, {Odegard}, {Tucker}, {Weiland}, {Wollack}, \& {Wright}}]{Komatsu2011}
{Komatsu}, E., {Smith}, K.~M., {Dunkley}, J., {et~al.} 2011, \apjs, 192, 18

\bibitem[{{Kravtsov} {et~al.}(2004){Kravtsov}, {Berlind}, {Wechsler}, {Klypin}, {Gottl{\"o}ber}, {Allgood}, \& {Primack}}]{Kravtsov2004}
{Kravtsov}, A.~V., {Berlind}, A.~A., {Wechsler}, R.~H., {et~al.} 2004, \apj, 609, 35

\bibitem[{{Landy} \& {Szalay}(1993)}]{Landy1993}
{Landy}, S.~D. \& {Szalay}, A.~S. 1993, \apj, 412, 64

\bibitem[{{Laureijs} {et~al.}(2011){Laureijs}, {Amiaux}, {Arduini}, {Augu{\`e}res}, {Brinchmann}, {Cole}, {Cropper}, {Dabin}, {Duvet}, {Ealet}, {Garilli}, {Gondoin}, {Guzzo}, {Hoar}, {Hoekstra}, {Holmes}, {Kitching}, {Maciaszek}, {Mellier}, {Pasian}, {Percival}, {Rhodes}, {Saavedra Criado}, {Sauvage}, {Scaramella}, {Valenziano}, {Warren}, {Bender}, {Castander}, {Cimatti}, {Le F{\`e}vre}, {Kurki-Suonio}, {Levi}, {Lilje}, {Meylan}, {Nichol}, {Pedersen}, {Popa}, {Rebolo Lopez}, {Rix}, {Rottgering}, {Zeilinger}, {Grupp}, {Hudelot}, {Massey}, {Meneghetti}, {Miller}, {Paltani}, {Paulin-Henriksson}, {Pires}, {Saxton}, {Schrabback}, {Seidel}, {Walsh}, {Aghanim}, {Amendola}, {Bartlett}, {Baccigalupi}, {Beaulieu}, {Benabed}, {Cuby}, {Elbaz}, {Fosalba}, {Gavazzi}, {Helmi}, {Hook}, {Irwin}, {Kneib}, {Kunz}, {Mannucci}, {Moscardini}, {Tao}, {Teyssier}, {Weller}, {Zamorani}, {Zapatero Osorio}, {Boulade}, {Foumond}, {Di Giorgio}, {Guttridge}, {James}, {Kemp}, {Martignac}, {Spencer}, {Walton}, {Bl{\"u}mchen}, {Bonoli},
  {Bortoletto}, {Cerna}, {Corcione}, {Fabron}, {Jahnke}, {Ligori}, {Madrid}, {Martin}, {Morgante}, {Pamplona}, {Prieto}, {Riva}, {Toledo}, {Trifoglio}, {Zerbi}, {Abdalla}, {Douspis}, {Grenet}, {Borgani}, {Bouwens}, {Courbin}, {Delouis}, {Dubath}, {Fontana}, {Frailis}, {Grazian}, {Koppenh{\"o}fer}, {Mansutti}, {Melchior}, {Mignoli}, {Mohr}, {Neissner}, {Noddle}, {Poncet}, {Scodeggio}, {Serrano}, {Shane}, {Starck}, {Surace}, {Taylor}, {Verdoes-Kleijn}, {Vuerli}, {Williams}, {Zacchei}, {Altieri}, {Escudero Sanz}, {Kohley}, {Oosterbroek}, {Astier}, {Bacon}, {Bardelli}, {Baugh}, {Bellagamba}, {Benoist}, {Bianchi}, {Biviano}, {Branchini}, {Carbone}, {Cardone}, {Clements}, {Colombi}, {Conselice}, {Cresci}, {Deacon}, {Dunlop}, {Fedeli}, {Fontanot}, {Franzetti}, {Giocoli}, {Garcia-Bellido}, {Gow}, {Heavens}, {Hewett}, {Heymans}, {Holland}, {Huang}, {Ilbert}, {Joachimi}, {Jennins}, {Kerins}, {Kiessling}, {Kirk}, {Kotak}, {Krause}, {Lahav}, {van Leeuwen}, {Lesgourgues}, {Lombardi}, {Magliocchetti}, {Maguire},
  {Majerotto}, {Maoli}, {Marulli}, {Maurogordato}, {McCracken}, {McLure}, {Melchiorri}, {Merson}, {Moresco}, {Nonino}, {Norberg}, {Peacock}, {Pello}, {Penny}, {Pettorino}, {Di Porto}, {Pozzetti}, {Quercellini}, {Radovich}, {Rassat}, {Roche}, {Ronayette}, {Rossetti}, {Sartoris}, {Schneider}, {Semboloni}, {Serjeant}, {Simpson}, {Skordis}, {Smadja}, {Smartt}, {Spano}, {Spiro}, {Sullivan}, {Tilquin}, {Trotta}, {Verde}, {Wang}, {Williger}, {Zhao}, {Zoubian}, \& {Zucca}}]{Laureijs2011}
{Laureijs}, R., {Amiaux}, J., {Arduini}, S., {et~al.} 2011, arXiv:1110.3193

\bibitem[{{LSST Science Collaboration: Abell} {et~al.}(2009){LSST Science Collaboration: Abell}, {Abell}, {Allison}, {Anderson}, {Andrew}, {Angel}, {Armus}, {Arnett}, {Asztalos}, {Axelrod}, {Bailey}, {Ballantyne}, {Bankert}, {Barkhouse}, {Barr}, {Barrientos}, {Barth}, {Bartlett}, {Becker}, {Becla}, {Beers}, {Bernstein}, {Biswas}, {Blanton}, {Bloom}, {Bochanski}, {Boeshaar}, {Borne}, {Bradac}, {Brandt}, {Bridge}, {Brown}, {Brunner}, {Bullock}, {Burgasser}, {Burge}, {Burke}, {Cargile}, {Chandrasekharan}, {Chartas}, {Chesley}, {Chu}, {Cinabro}, {Claire}, {Claver}, {Clowe}, {Connolly}, {Cook}, {Cooke}, {Cooray}, {Covey}, {Culliton}, {de Jong}, {de Vries}, {Debattista}, {Delgado}, {Dell'Antonio}, {Dhital}, {Di Stefano}, {Dickinson}, {Dilday}, {Djorgovski}, {Dobler}, {Donalek}, {Dubois-Felsmann}, {Durech}, {Eliasdottir}, {Eracleous}, {Eyer}, {Falco}, {Fan}, {Fassnacht}, {Ferguson}, {Fernandez}, {Fields}, {Finkbeiner}, {Figueroa}, {Fox}, {Francke}, {Frank}, {Frieman}, {Fromenteau}, {Furqan}, {Galaz}, {Gal-Yam},
  {Garnavich}, {Gawiser}, {Geary}, {Gee}, {Gibson}, {Gilmore}, {Grace}, {Green}, {Gressler}, {Grillmair}, {Habib}, {Haggerty}, {Hamuy}, {Harris}, {Hawley}, {Heavens}, {Hebb}, {Henry}, {Hileman}, {Hilton}, {Hoadley}, {Holberg}, {Holman}, {Howell}, {Infante}, {Ivezic}, {Jacoby}, {Jain}, {R}, {Jedicke}, {Jee}, {Garrett Jernigan}, {Jha}, {Johnston}, {Jones}, {Juric}, {Kaasalainen}, {Styliani}, {Kafka}, {Kahn}, {Kaib}, {Kalirai}, {Kantor}, {Kasliwal}, {Keeton}, {Kessler}, {Knezevic}, {Kowalski}, {Krabbendam}, {Krughoff}, {Kulkarni}, {Kuhlman}, {Lacy}, {Lepine}, {Liang}, {Lien}, {Lira}, {Long}, {Lorenz}, {Lotz}, {Lupton}, {Lutz}, {Macri}, {Mahabal}, {Mandelbaum}, {Marshall}, {May}, {McGehee}, {Meadows}, {Meert}, {Milani}, {Miller}, {Miller}, {Mills}, {Minniti}, {Monet}, {Mukadam}, {Nakar}, {Neill}, {Newman}, {Nikolaev}, {Nordby}, {O'Connor}, {Oguri}, {Oliver}, {Olivier}, {Olsen}, {Olsen}, {Olszewski}, {Oluseyi}, {Padilla}, {Parker}, {Pepper}, {Peterson}, {Petry}, {Pinto}, {Pizagno}, {Popescu}, {Prsa}, {Radcka},
  {Raddick}, {Rasmussen}, {Rau}, {Rho}, {Rhoads}, {Richards}, {Ridgway}, {Robertson}, {Roskar}, {Saha}, {Sarajedini}, {Scannapieco}, {Schalk}, {Schindler}, {Schmidt}, {Schmidt}, {Schneider}, {Schumacher}, {Scranton}, {Sebag}, {Seppala}, {Shemmer}, {Simon}, {Sivertz}, {Smith}, {Allyn Smith}, {Smith}, {Spitz}, {Stanford}, {Stassun}, {Strader}, {Strauss}, {Stubbs}, {Sweeney}, {Szalay}, {Szkody}, {Takada}, {Thorman}, {Trilling}, {Trimble}, {Tyson}, {Van Berg}, {Vanden Berk}, {VanderPlas}, {Verde}, {Vrsnak}, {Walkowicz}, {Wandelt}, {Wang}, {Wang}, {Warner}, {Wechsler}, {West}, {Wiecha}, {Williams}, {Willman}, {Wittman}, {Wolff}, {Wood-Vasey}, {Wozniak}, {Young}, {Zentner}, \& {Zhan}}]{LSST2009}
{LSST Science Collaboration: Abell}, {Abell}, P.~A., {Allison}, J., {et~al.} 2009, arXiv:0912.0201

\bibitem[{{Lu} {et~al.}(2014){Lu}, {Wechsler}, {Somerville}, {Croton}, {Porter}, {Primack}, {Behroozi}, {Ferguson}, {Koo}, {Guo}, {Safarzadeh}, {Finlator}, {Castellano}, {White}, {Sommariva}, \& {Moody}}]{Lu2014}
{Lu}, Y., {Wechsler}, R.~H., {Somerville}, R.~S., {et~al.} 2014, \apj, 795, 123

\bibitem[{{Mahony} {et~al.}(2025){Mahony}, {Contreras}, {Angulo}, {Alonso}, {Georgiou}, \& {Dvornik}}]{Mahony2025}
{Mahony}, C., {Contreras}, S., {Angulo}, R.~E., {et~al.} 2025, arXiv:2507.01601

\bibitem[{{Mamon} {et~al.}(2019){Mamon}, {Cava}, {Biviano}, {Moretti}, {Poggianti}, \& {Bettoni}}]{Mamon2019}
{Mamon}, G.~A., {Cava}, A., {Biviano}, A., {et~al.} 2019, \aap, 631, A131

\bibitem[{{Marulli} {et~al.}(2013){Marulli}, {Bolzonella}, {Branchini}, {Davidzon}, {de la Torre}, {Granett}, {Guzzo}, {Iovino}, {Moscardini}, {Pollo}, {Abbas}, {Adami}, {Arnouts}, {Bel}, {Bottini}, {Cappi}, {Coupon}, {Cucciati}, {De Lucia}, {Fritz}, {Franzetti}, {Fumana}, {Garilli}, {Ilbert}, {Krywult}, {Le Brun}, {Le F{\`e}vre}, {Maccagni}, {Ma{\l}ek}, {McCracken}, {Paioro}, {Polletta}, {Schlagenhaufer}, {Scodeggio}, {Tasca}, {Tojeiro}, {Vergani}, {Zanichelli}, {Burden}, {Di Porto}, {Marchetti}, {Marinoni}, {Mellier}, {Nichol}, {Peacock}, {Percival}, {Phleps}, {Wolk}, \& {Zamorani}}]{Marulli2013}
{Marulli}, F., {Bolzonella}, M., {Branchini}, E., {et~al.} 2013, \aap, 557, A17

\bibitem[{{Navarro} {et~al.}(1997){Navarro}, {Frenk}, \& {White}}]{Navarro97}
{Navarro}, J.~F., {Frenk}, C.~S., \& {White}, S. D.~M. 1997, \apj, 490, 493

\bibitem[{{Nelson} {et~al.}(2024){Nelson}, {Pillepich}, {Ayromlou}, {Lee}, {Lehle}, {Rohr}, \& {Truong}}]{Nelson2024}
{Nelson}, D., {Pillepich}, A., {Ayromlou}, M., {et~al.} 2024, \aap, 686, A157

\bibitem[{{Peacock} \& {Smith}(2000)}]{Peacock2000}
{Peacock}, J.~A. \& {Smith}, R.~E. 2000, \mnras, 318, 1144

\bibitem[{{Peebles} \& {Hauser}(1974)}]{Peebles1974}
{Peebles}, P.~J.~E. \& {Hauser}, M.~G. 1974, \apjs, 28, 19

\bibitem[{{Perez} {et~al.}(2024){Perez}, {Pereyra}, {Coldwell}, {Rodriguez}, {Alfaro}, \& {Ruiz}}]{Perez2024}
{Perez}, N.~R., {Pereyra}, L.~A., {Coldwell}, G., {et~al.} 2024, \mnras, 528, 3186

\bibitem[{{Porredon} {et~al.}(2021){Porredon}, {Crocce}, {Fosalba}, {Elvin-Poole}, {Carnero Rosell}, {Cawthon}, {Eifler}, {Fang}, {Ferrero}, {Krause}, {MacCrann}, {Weaverdyck}, {Abbott}, {Aguena}, {Allam}, {Amon}, {Avila}, {Bacon}, {Bertin}, {Bhargava}, {Bridle}, {Brooks}, {Carrasco Kind}, {Carretero}, {Castander}, {Choi}, {Costanzi}, {da Costa}, {Pereira}, {De Vicente}, {Desai}, {Diehl}, {Doel}, {Drlica-Wagner}, {Eckert}, {Fert{\'e}}, {Flaugher}, {Frieman}, {Garc{\'\i}a-Bellido}, {Gaztanaga}, {Gerdes}, {Giannantonio}, {Gruen}, {Gruendl}, {Gschwend}, {Gutierrez}, {Hartley}, {Hinton}, {Hollowood}, {Honscheid}, {Hoyle}, {James}, {Jarvis}, {Kuehn}, {Kuropatkin}, {Maia}, {Marshall}, {Menanteau}, {Miquel}, {Morgan}, {Palmese}, {Pandey}, {Paz-Chinch{\'o}n}, {Plazas}, {Rodriguez-Monroy}, {Roodman}, {Samuroff}, {Sanchez}, {Scarpine}, {Serrano}, {Sevilla-Noarbe}, {Smith}, {Soares-Santos}, {Suchyta}, {Swanson}, {Tarle}, {To}, {Varga}, {Weller}, {Wilkinson}, \& {DES Collaboration}}]{Porredon2021}
{Porredon}, A., {Crocce}, M., {Fosalba}, P., {et~al.} 2021, \prd, 103, 043503

\bibitem[{{Potter} \& {Stadel}(2016)}]{Potter2016}
{Potter}, D. \& {Stadel}, J. 2016, {PKDGRAV3: Parallel gravity code}, Astrophysics Source Code Library, record ascl:1609.016

\bibitem[{{Ramakrishnan} {et~al.}(2025{\natexlab{a}}){Ramakrishnan}, {Castander}, {Gonzalez}, {Eriksen}, {Baghkhani}, {Fosalba}, {Carretero}, {Parimbelli}, \& {Tallada-Cresp{\'\i}}}]{Ramakrishnan2025b}
{Ramakrishnan}, S., {Castander}, F.~J., {Gonzalez}, E.~J., {et~al.} 2025{\natexlab{a}}, arXiv:2512.13801

\bibitem[{{Ramakrishnan} {et~al.}(2025{\natexlab{b}}){Ramakrishnan}, {Gonzalez-Perez}, {Parimbelli}, \& {Yepes}}]{Ramakrishnan2025}
{Ramakrishnan}, S., {Gonzalez-Perez}, V., {Parimbelli}, G., \& {Yepes}, G. 2025{\natexlab{b}}, \aap, 697, A70

\bibitem[{{Reid} {et~al.}(2014){Reid}, {Seo}, {Leauthaud}, {Tinker}, \& {White}}]{Reid2014}
{Reid}, B.~A., {Seo}, H.-J., {Leauthaud}, A., {Tinker}, J.~L., \& {White}, M. 2014, \mnras, 444, 476

\bibitem[{{Rodr{\'\i}guez-Monroy} {et~al.}(2022){Rodr{\'\i}guez-Monroy}, {Weaverdyck}, {Elvin-Poole}, {Crocce}, {Carnero Rosell}, {Andrade-Oliveira}, {Avila}, {Bechtol}, {Bernstein}, {Blazek}, {Camacho}, {Cawthon}, {De Vicente}, {DeRose}, {Dodelson}, {Everett}, {Fang}, {Ferrero}, {Fert{\'e}}, {Friedrich}, {Gaztanaga}, {Giannini}, {Gruendl}, {Hartley}, {Herner}, {Huff}, {Jarvis}, {Krause}, {MacCrann}, {Mena-Fern{\'a}ndez}, {Muir}, {Pandey}, {Park}, {Porredon}, {Prat}, {Rosenfeld}, {Ross}, {Rozo}, {Rykoff}, {Sanchez}, {Sanchez Cid}, {Sevilla-Noarbe}, {Tabbutt}, {To}, {Wagoner}, {Wechsler}, {Aguena}, {Allam}, {Amon}, {Annis}, {Bacon}, {Baxter}, {Bertin}, {Bhargava}, {Brooks}, {Burke}, {Carrasco Kind}, {Carretero}, {Castander}, {Choi}, {Conselice}, {Costanzi}, {da Costa}, {Pereira}, {Desai}, {Diehl}, {Flaugher}, {Fosalba}, {Frieman}, {Garc{\'\i}a-Bellido}, {Giannantonio}, {Gruen}, {Gschwend}, {Gutierrez}, {Hinton}, {Hollowood}, {Honscheid}, {Huterer}, {Jain}, {James}, {Kuehn}, {Kuropatkin}, {Lima}, {Maia},
  {March}, {Marshall}, {Melchior}, {Menanteau}, {Miller}, {Miquel}, {Mohr}, {Morgan}, {Palmese}, {Paz-Chinch{\'o}n}, {Pieres}, {Plazas Malag{\'o}n}, {Roodman}, {Scarpine}, {Serrano}, {Smith}, {Soares-Santos}, {Suchyta}, {Tarle}, {Thomas}, {Varga}, \& {DES Collaboration}}]{DES2022}
{Rodr{\'\i}guez-Monroy}, M., {Weaverdyck}, N., {Elvin-Poole}, J., {et~al.} 2022, \mnras, 511, 2665

\bibitem[{{Roster} {et~al.}(2026){Roster}, {Wright}, {Hildebrandt}, {Reischke}, {Ilbert}, {d'Assignies D.}, {Manera}, {Bolzonella}, {Masters}, {Paltani}, {Hartley}, {Kang}, {Hoekstra}, {Altieri}, {Amara}, {Andreon}, {Auricchio}, {Baccigalupi}, {Baldi}, {Balestra}, {Bardelli}, {Battaglia}, {Bender}, {Biviano}, {Branchini}, {Brescia}, {Camera}, {Ca{\~n}as-Herrera}, {Capobianco}, {Carbone}, {Cardone}, {Carretero}, {Casas}, {Casas}, {Castander}, {Castellano}, {Castignani}, {Cavuoti}, {Chambers}, {Cimatti}, {Colodro-Conde}, {Congedo}, {Conselice}, {Conversi}, {Copin}, {Costille}, {Courbin}, {Courtois}, {Cropper}, {Da Silva}, {Degaudenzi}, {de la Torre}, {De Lucia}, {Dubath}, {Duncan}, {Dupac}, {Dusini}, {Escoffier}, {Farina}, {Farinelli}, {Farrens}, {Faustini}, {Ferriol}, {Finelli}, {Fosalba}, {Fourmanoit}, {Frailis}, {Franceschi}, {Fumana}, {Galeotta}, {George}, {Gillard}, {Gillis}, {Giocoli}, {Gracia-Carpio}, {Grazian}, {Grupp}, {Haugan}, {Holmes}, {Hormuth}, {Hornstrup}, {Hudelot}, {Jahnke}, {Jhabvala},
  {Joachimi}, {Keih{\"a}nen}, {Kermiche}, {Kubik}, {Kurki-Suonio}, {Le Brun}, {Le Mignant}, {Ligori}, {Lilje}, {Lindholm}, {Lloro}, {Maino}, {Maiorano}, {Mansutti}, {Marggraf}, {Martinelli}, {Martinet}, {Marulli}, {Massey}, {Medinaceli}, {Mei}, {Melchior}, {Mellier}, {Meneghetti}, {Merlin}, {Meylan}, {Mora}, {Moresco}, {Moscardini}, {Nakajima}, {Neissner}, {Niemi}, {Padilla}, {Pasian}, {Pedersen}, {Pettorino}, {Pires}, {Polenta}, {Poncet}, {Popa}, {Pozzetti}, {Raison}, {Rebolo}, {Renzi}, {Rhodes}, {Riccio}, {Romelli}, {Roncarelli}, {Rosset}, {Rossetti}, {Saglia}, {Sakr}, {Sapone}, {Sartoris}, {Schirmer}, {Schneider}, {Schrabback}, {Scodeggio}, {Secroun}, {Sefusatti}, {Seidel}, {Serrano}, {Simon}, {Sirignano}, {Sirri}, {Skottfelt}, {Stanco}, {Steinwagner}, {Tallada-Cresp{\'\i}}, {Taylor}, {Teplitz}, {Tereno}, {Tessore}, {Toft}, {Toledo-Moreo}, {Torradeflot}, {Tutusaus}, {Valenziano}, {Valiviita}, {Vassallo}, {Verdoes Kleijn}, {Veropalumbo}, {Wang}, {Weller}, {Zamorani}, {Zerbi}, {Zucca}, {Burigana}, {Gabarra},
  {Porciani}, {Scottez}, \& {Sereno}}]{Euclid_Photoz_SOM}
{Roster}, W., {Wright}, A.~H., {Hildebrandt}, H., {et~al.} 2026, \aap, 707, A277

\bibitem[{{Schechter}(1976)}]{Schechter1976}
{Schechter}, P. 1976, \apj, 203, 297

\bibitem[{{Spergel} {et~al.}(2015){Spergel}, {Gehrels}, {Baltay}, {Bennett}, {Breckinridge}, {Donahue}, {Dressler}, {Gaudi}, {Greene}, {Guyon}, {Hirata}, {Kalirai}, {Kasdin}, {Macintosh}, {Moos}, {Perlmutter}, {Postman}, {Rauscher}, {Rhodes}, {Wang}, {Weinberg}, {Benford}, {Hudson}, {Jeong}, {Mellier}, {Traub}, {Yamada}, {Capak}, {Colbert}, {Masters}, {Penny}, {Savransky}, {Stern}, {Zimmerman}, {Barry}, {Bartusek}, {Carpenter}, {Cheng}, {Content}, {Dekens}, {Demers}, {Grady}, {Jackson}, {Kuan}, {Kruk}, {Melton}, {Nemati}, {Parvin}, {Poberezhskiy}, {Peddie}, {Ruffa}, {Wallace}, {Whipple}, {Wollack}, \& {Zhao}}]{Spergel2015}
{Spergel}, D., {Gehrels}, N., {Baltay}, C., {et~al.} 2015, arXiv:1503.03757

\bibitem[{{Tallada-Crespi} {et~al.}(2024){Tallada-Crespi}, {Carretero}, {Serrano}, {Castander}, {Cesar}, {Eriksen}, {Fosalba}, {Gaztanaga}, {Merino}, {Neissner}, {Tonello}, \& {Torradeflot}}]{Tallada2024}
{Tallada-Crespi}, P., {Carretero}, J., {Serrano}, S., {et~al.} 2024, in Astronomical Society of the Pacific Conference Series, Vol. 535, Astromical Data Analysis Software and Systems XXXI, ed. B.~V. {Hugo}, R.~{Van Rooyen}, \& O.~M. {Smirnov}, 381

\bibitem[{{Teyssier}(2002)}]{Teyssier2002}
{Teyssier}, R. 2002, \aap, 385, 337

\bibitem[{{The Dark Energy Survey Collaboration}(2005)}]{DES2005}
{The Dark Energy Survey Collaboration}. 2005, arXiv:astro-ph/0510346

\bibitem[{{Tinker} {et~al.}(2008){Tinker}, {Kravtsov}, {Klypin}, {Abazajian}, {Warren}, {Yepes}, {Gottl{\"o}ber}, \& {Holz}}]{Tinker2008}
{Tinker}, J., {Kravtsov}, A.~V., {Klypin}, A., {et~al.} 2008, \apj, 688, 709

\bibitem[{{Tutusaus} {et~al.}(2025){Tutusaus}, {Fosalba}, {Blot}, {Tallada-Cresp{\'\i}}, {Carretero}, {Castander}, {Gonzalez}, \& {Alarcon}}]{Tutusaus2025}
{Tutusaus}, I., {Fosalba}, P., {Blot}, L., {et~al.} 2025, arXiv:2507.23451

\bibitem[{{Vale} \& {Ostriker}(2004)}]{Vale2004}
{Vale}, A. \& {Ostriker}, J.~P. 2004, \mnras, 353, 189

\bibitem[{{Villaescusa-Navarro} {et~al.}(2021){Villaescusa-Navarro}, {Angl{\'e}s-Alc{\'a}zar}, {Genel}, {Spergel}, {Somerville}, {Dave}, {Pillepich}, {Hernquist}, {Nelson}, {Torrey}, {Narayanan}, {Li}, {Philcox}, {La Torre}, {Maria Delgado}, {Ho}, {Hassan}, {Burkhart}, {Wadekar}, {Battaglia}, {Contardo}, \& {Bryan}}]{Villaescusa2021}
{Villaescusa-Navarro}, F., {Angl{\'e}s-Alc{\'a}zar}, D., {Genel}, S., {et~al.} 2021, \apj, 915, 71

\bibitem[{{Vogelsberger} {et~al.}(2020){Vogelsberger}, {Marinacci}, {Torrey}, \& {Puchwein}}]{Vogelsberger2020}
{Vogelsberger}, M., {Marinacci}, F., {Torrey}, P., \& {Puchwein}, E. 2020, Nature Reviews Physics, 2, 42

\bibitem[{{Watson} {et~al.}(2013){Watson}, {Iliev}, {D'Aloisio}, {Knebe}, {Shapiro}, \& {Yepes}}]{Watson2013}
{Watson}, W.~A., {Iliev}, I.~T., {D'Aloisio}, A., {et~al.} 2013, \mnras, 433, 1230

\bibitem[{{Weaver} {et~al.}(2022){Weaver}, {Kauffmann}, {Ilbert}, {McCracken}, {Moneti}, {Toft}, {Brammer}, {Shuntov}, {Davidzon}, {Hsieh}, {Laigle}, {Anastasiou}, {Jespersen}, {Vinther}, {Capak}, {Casey}, {McPartland}, {Milvang-Jensen}, {Mobasher}, {Sanders}, {Zalesky}, {Arnouts}, {Aussel}, {Dunlop}, {Faisst}, {Franx}, {Furtak}, {Fynbo}, {Gould}, {Greve}, {Gwyn}, {Kartaltepe}, {Kashino}, {Koekemoer}, {Kokorev}, {Le F{\`e}vre}, {Lilly}, {Masters}, {Magdis}, {Mehta}, {Peng}, {Riechers}, {Salvato}, {Sawicki}, {Scarlata}, {Scoville}, {Shirley}, {Silverman}, {Sneppen}, {Smolc̆i{\'c}}, {Steinhardt}, {Stern}, {Tanaka}, {Taniguchi}, {Teplitz}, {Vaccari}, {Wang}, \& {Zamorani}}]{COSMOS2020}
{Weaver}, J.~R., {Kauffmann}, O.~B., {Ilbert}, O., {et~al.} 2022, \apjs, 258, 11

\bibitem[{{York} {et~al.}(2000){York}, {Adelman}, {Anderson}, {Anderson}, {Annis}, {Bahcall}, {Bakken}, {Barkhouser}, {Bastian}, {Berman}, {Boroski}, {Bracker}, {Briegel}, {Briggs}, {Brinkmann}, {Brunner}, {Burles}, {Carey}, {Carr}, {Castander}, {Chen}, {Colestock}, {Connolly}, {Crocker}, {Csabai}, {Czarapata}, {Davis}, {Doi}, {Dombeck}, {Eisenstein}, {Ellman}, {Elms}, {Evans}, {Fan}, {Federwitz}, {Fiscelli}, {Friedman}, {Frieman}, {Fukugita}, {Gillespie}, {Gunn}, {Gurbani}, {de Haas}, {Haldeman}, {Harris}, {Hayes}, {Heckman}, {Hennessy}, {Hindsley}, {Holm}, {Holmgren}, {Huang}, {Hull}, {Husby}, {Ichikawa}, {Ichikawa}, {Ivezi{\'c}}, {Kent}, {Kim}, {Kinney}, {Klaene}, {Kleinman}, {Kleinman}, {Knapp}, {Korienek}, {Kron}, {Kunszt}, {Lamb}, {Lee}, {Leger}, {Limmongkol}, {Lindenmeyer}, {Long}, {Loomis}, {Loveday}, {Lucinio}, {Lupton}, {MacKinnon}, {Mannery}, {Mantsch}, {Margon}, {McGehee}, {McKay}, {Meiksin}, {Merelli}, {Monet}, {Munn}, {Narayanan}, {Nash}, {Neilsen}, {Neswold}, {Newberg}, {Nichol}, {Nicinski},
  {Nonino}, {Okada}, {Okamura}, {Ostriker}, {Owen}, {Pauls}, {Peoples}, {Peterson}, {Petravick}, {Pier}, {Pope}, {Pordes}, {Prosapio}, {Rechenmacher}, {Quinn}, {Richards}, {Richmond}, {Rivetta}, {Rockosi}, {Ruthmansdorfer}, {Sandford}, {Schlegel}, {Schneider}, {Sekiguchi}, {Sergey}, {Shimasaku}, {Siegmund}, {Smee}, {Smith}, {Snedden}, {Stone}, {Stoughton}, {Strauss}, {Stubbs}, {SubbaRao}, {Szalay}, {Szapudi}, {Szokoly}, {Thakar}, {Tremonti}, {Tucker}, {Uomoto}, {Vanden Berk}, {Vogeley}, {Waddell}, {Wang}, {Watanabe}, {Weinberg}, {Yanny}, {Yasuda}, \& {SDSS Collaboration}}]{York2000}
{York}, D.~G., {Adelman}, J., {Anderson}, John~E., J., {et~al.} 2000, \aj, 120, 1579

\bibitem[{{Yuan} {et~al.}(2022){Yuan}, {Garrison}, {Eisenstein}, \& {Wechsler}}]{Yuan2022}
{Yuan}, S., {Garrison}, L.~H., {Eisenstein}, D.~J., \& {Wechsler}, R.~H. 2022, \mnras, 515, 871

\bibitem[{{Zehavi} {et~al.}(2011){Zehavi}, {Zheng}, {Weinberg}, {Blanton}, {Bahcall}, {Berlind}, {Brinkmann}, {Frieman}, {Gunn}, {Lupton}, {Nichol}, {Percival}, {Schneider}, {Skibba}, {Strauss}, {Tegmark}, \& {York}}]{Zehavi2011}
{Zehavi}, I., {Zheng}, Z., {Weinberg}, D.~H., {et~al.} 2011, \apj, 736, 59

\bibitem[{{Zhai} {et~al.}(2019){Zhai}, {Tinker}, {Becker}, {DeRose}, {Mao}, {McClintock}, {McLaughlin}, {Rozo}, \& {Wechsler}}]{Zhai2019}
{Zhai}, Z., {Tinker}, J.~L., {Becker}, M.~R., {et~al.} 2019, \apj, 874, 95

\bibitem[{{Zheng} {et~al.}(2005){Zheng}, {Berlind}, {Weinberg}, {Benson}, {Baugh}, {Cole}, {Dav{\'e}}, {Frenk}, {Katz}, \& {Lacey}}]{Zheng2005}
{Zheng}, Z., {Berlind}, A.~A., {Weinberg}, D.~H., {et~al.} 2005, \apj, 633, 791

\bibitem[{{Zheng} {et~al.}(2007){Zheng}, {Coil}, \& {Zehavi}}]{Zheng2007}
{Zheng}, Z., {Coil}, A.~L., \& {Zehavi}, I. 2007, \apj, 667, 760

\end{thebibliography}
